\documentclass[twocolumn,showpacs,preprintnumbers,amsmath,amssymb,superscriptaddress]{revtex4}

\usepackage{graphicx}
\usepackage{dcolumn}
\usepackage{bm}
\usepackage{epsfig}
\usepackage{amsmath}
\usepackage{color}
\usepackage{longtable}
\usepackage[OT2,T1]{fontenc}

\newcommand{\beq}{\begin{equation}}
\newcommand{\eeq}{\end{equation}}
\newcommand{\beqa}{\begin{eqnarray}}
\newcommand{\eeqa}{\end{eqnarray}}
\newcommand\lsim{\mathrel{\rlap{\lower4pt\hbox{\hskip1pt$\sim$}}
        \raise1pt\hbox{$<$}}}
\newcommand\gsim{\mathrel{\rlap{\lower4pt\hbox{\hskip1pt$\sim$}}
        \raise1pt\hbox{$>$}}}
\def\thetaB{\mbox{\boldmath$\hat\theta$}}

\newcommand{\white}[1]{{\textcolor{white}#1}}
\newcommand{\nhat}{\hat{\bf n}}

\begin{document}


\title{Correlation of CMB with large-scale structure: I. ISW Tomography and Cosmological Implications}

\author{Shirley Ho}
\email{shirley@astro.princeton.edu}
\affiliation{Department of Astrophysical Sciences, Princeton University, NJ 08544, USA}
\author{Christopher Hirata}
\affiliation{Caltech M/C 130-33, Pasadena, CA 91125, USA}
\author{Nikhil Padmanabhan}
\affiliation{Lawrence Berkeley National Laboratory, Berkeley, CA 94720, USA}
\author{Uros Seljak}
\affiliation{Institute for Theoretical Physics, Zurich University, Zurich 8057, Switzerland}
\affiliation{Department of Physics, University of California at 
Berkeley, Berkeley, CA 94704, USA}
\author{Neta Bahcall}
\affiliation{Department of Astrophysical Sciences, Princeton University, NJ 08544, USA}

\date{January 3, 2008}

\begin{abstract}
We cross-correlate large scale structure (LSS) observations from a number of surveys 
with cosmic microwave background (CMB) anisotropies from the Wilkinson Microwave Anisotropy Probe (WMAP) 
to investigate the Integrated Sachs-Wolfe (ISW) effect as a function of redshift, 
covering $z \sim 0.1 - 2.5$.  
Our main goal is to go beyond reporting detections towards developing a reliable 
likelihood analysis that allows one to determine cosmological constraints from ISW 
observations. With this in mind 
we spend a considerable amount of effort in determining
 the redshift-dependent bias and redshift distribution ($b(z)\times dN/dz$) of these samples 
by matching with spectroscopic observations where available, and analyzing 
auto-power spectra and cross-power spectra between the samples. 
Due to wide redshift distributions of some of the data sets 
 we do not assume a constant bias model, in contrast to previous work on this subject. 
We only use the LSS data sets for which we can extract such information reliably 
and as a result the data sets we use are 
2-Micron All Sky Survey (2MASS) samples, Sloan Digital Sky Survey (SDSS) photometric
Luminous Red Galaxies, SDSS photometric quasars and NRAO VLA Sky Survey (NVSS) radio 
sources.  
We make a joint 
analysis of all samples constructing a full covariance matrix, which we subsequently use  
for cosmological parameter fitting.
We report a 3.7$\sigma$ detection of ISW combining all the datasets.
We do not find significant evidence for an ISW signal at $z>1$, in agreement with theoretical 
expectation in $\Lambda$CDM model. 
We combine the ISW likelihood function with weak lensing of CMB (hereafter Paper II \cite{paperII}) and CMB power spectrum
to constrain the equation of state of dark energy and the curvature of the Universe.
While ISW does not significantly improve the constraints in the simplest 6-parameter flat $\Lambda$CDM model,
it improves constraints on 7-parameter models with curvature by a factor of 3.2 (relative to WMAP alone)
to $\Omega_K=-0.004^{+0.014}_{-0.020}$,
and with dark energy equation of state by 15\% to $w=-1.01^{+0.30}_{-0.40}$
[posterior median with ``$1\sigma$'' (16th--84th percentile) range].
A software package for calculating the ISW likelihood function 
can be downloaded at
{\tt http://www.astro.princeton.edu/\~{}shirley/ISW\_WL.html}.

\end{abstract}


\maketitle

\section{\label{sec:intro}Introduction}

The Cosmic Microwave Background (CMB) has provided us with a wealth of cosmological information.
The large-scale anisotropies were first discovered by the Differential Microwave Radiometer (DMR)  
on Cosmic Background Explorer (COBE) satellite \cite{smoot92}, 
and the smaller-scale CMB anisotropies were subsequently measured by various
ground-based/balloon-borne experiments.
More recently, the Wilkinson Microwave Anisotropy Probe 
(WMAP) satellite \cite{bennett03, jarosik07} produced a cosmic variance limited
map of CMB anisotropies down to $l\sim 400$. 
The structure of the angular power spectrum when combined with other
cosmological probes (such as Sloan Digital Sky Survey,  
\cite{tegmark06}, Hubble Key Project \cite{freedman94} and 2dF Galaxy Redshift Survey \cite{cole05}), allows
extremely precise measurements
of the cosmological parameters of the $\Lambda$CDM model. 
While most of the fluctuations seen by WMAP and other CMB experiments were generated
at the last surface of scattering, structures formed at low redshift also 
leave imprints on the CMB.  These anisotropies, such as the thermal 
Sunyaev-Zeldovich (tSZ) \cite{sunyaev80a} and kinetic Sunyaev Zeldovich effects (kSZ) 
\cite{sunyaev80b}, the Integrated Sachs-Wolfe (ISW) effect \cite{sachs67}, and gravitational 
lensing, contribute only slightly to the CMB power spectrum on scales measured by WMAP, but they can be detected 
by cross-correlating the CMB with suitable tracers of the large scale structure. 

This is the first of two papers that measure the Integrated Sachs-Wolfe effect and 
gravitational lensing (Paper II) in cross-correlation.
In this paper, we focus on large scale galaxy-temperature correlations and their 
large scale cosmological source, the Integrated Sachs-Wolfe (ISW) effect.  The 
ISW effect results from the red- or blue-shifting of the CMB photons as they 
propagate through gravitational potential wells. As the potential wells of the
Universe (i.e., the spatial metric) evolve, the energy gained by photons falling into the potential well
does not cancel out the energy loss as photons climb out of the well. This is 
important at late times when the Universe is not matter dominated and the 
gravitational potential is time dependent. It is only 
significant on large scales, since 
on small scales the amount of time spent 
by the photon in each coherence region of the gravitational 
potential is small and any small scale fluctuations will be 
smoothed out as the photon go through numerous potential wells along the line
of sight. 

To measure the above effect, we cross-correlate the CMB temperature anisotropies
with maps of galaxies from the Two Micron All Sky Survey (2MASS), luminous red
galaxies (LRGs) and quasars from the Sloan Digital Sky Survey, and radio sources
from the NRAO VLA Sky Survey (NVSS). This incorporates most of the LSS tracers used by 
previous efforts \cite{boughn98,fosalba03,scranton03,afshordi04,boughn04,fosalba04,nolta04,padmanabhan05ISW,gaztanaga05,cabre06,giannantonio06,vielva06, pietrobon06,mcewen07,rassat07}
to detect the ISW effect. Our goal in this work
extends this previous literature by going beyond detecting the ISW effect to
measuring its redshift evolution and using that to constrain different cosmological
models (e.g. the ISW effect due to spatial curvature occurs at significantly higher
redshifts than that due to a cosmological constant). 
We therefore require a large redshift range ($z \sim 0$ to $2.5$) but with sufficient redshift
resolution to unambiguously discern any redshift evolution of the signal.
In addition, to draw robust cosmological conclusions from an observed redshift evolution,
we must constrain both the redshift distribution and evolution of the bias with redshift
for each of the samples; the simple assumption of constant bias is in most cases no longer sufficient.
These considerations drive our survey selections; we discuss these in more
detail in Sec.~\ref{sec:discussion}.
Our final product is a likelihood code that can be applied to any cosmological 
model. In addition to providing complementary constraints on standard cosmological parameters,
we expect it can be a strong discriminator of the modified 
gravity models, which have very distinctive ISW predictions \cite{song07}.

We review the theory behind the ISW effect in Sec.~\ref{sec:theory}.
The CMB and LSS data sets used are described in Sec.~\ref{sec:data};
the results of cross-correlating the two are in Sec.~\ref{sec:cross}.
Sec.~\ref{sec:dndz} and ~\ref{sec:sys} constrain the redshift distributions of the samples,
and possible systematic contamination of the cross-correlations. 
Sec.~\ref{sec:cosmo} presents the cosmological implications of these results,
and Sec.~\ref{sec:discussion} summarizes our conclusions.
The companion paper (Paper II) uses the same data sets to detect the  weak lensing 
of the CMB. 
All of the theoretical predictions are made with WMAP 3 year parameters 
($\Omega_b h^2$=$0.0223$,
$\Omega_c h^2$ = $0.128$, $\Omega_K=0$, $h=0.732$, $\sigma_8 =0.761$)
except in Section~\ref{sec:dndz} or otherwise stated.


\section{\label{sec:theory} Theory}

We briefly review the ISW effect and its cross-correlation
with the galaxy density
(see also Refs.~\citep{peiris00,cooray02,
afshordi04}).
%
The temperature anisotropy due to the ISW effect is expressed as an integral of the
time derivative of the gravitational potential $\phi$ over conformal time $\eta$,
\begin{equation}
\Delta T_{\rm ISW} (\thetaB)=2\int^{\eta_0}_{\eta_r}\!\!\!d\eta\, \frac{\partial\phi}{\partial\eta},
\label{deltaTisw}
\end{equation}
where $\eta_r$ and $\eta_0$ are the conformal time at recombination and today, respectively, 
and we ignored the effect of Thomson scattering suppression, 
which is negligible for the redshift range of interest here.  
For scales sufficiently within the horizon, the gravitational potential $\phi$
is related to the mass fluctuation $\delta = \delta\rho/\bar\rho$ in Fourier space by the Poisson equation:
\begin{equation}
\phi({\bf k},z)=-\frac{3}{2}\frac{H_0^2}{c^2}\Omega_m (1+z)\frac{\delta({\bf k},z)}{k^2},
\end{equation}
where $\Omega_m$ is the ratio of the matter density to the critical density today,
$H_0$ is the Hubble constant today, $c$ is the speed of light, $z$ is
the redshift, and $k$ is the comoving wave number.
On large scales where the mass fluctuation $\delta \ll 1$, the
perturbations grow according to linear theory
$\delta(k, z)$ = $\delta(k,0) D(z)/D(0)$.

We are interested in cross--correlating the temperature anisotropies, $\delta_T$, with the observed projected galaxy overdensity $g$.  
The intrinsic angular galaxy fluctuations are given by:
\begin{equation}
g(\thetaB)= \int dz \,b(z)\Pi(z) \delta(\chi(z)\thetaB,z),
\label{dg}
\end{equation}
where $b(z)$ is an assumed scale-independent bias factor relating the galaxy overdensity to
the mass overdensity, i.e. $\delta_g =b\,\delta$,
$\Pi(z)$ is the normalized selection function, and $\chi(z)$ is the comoving distance to redshift $z$.
We focus on the cross-spectrum of the galaxies with the CMB temperature fluctuation:
\begin{equation}
\label{cc1}
C^{gT}_\ell=\frac{2}{\pi} \int k^2 dk P(k) [g]_\ell (k) [T]_\ell(k)
\end{equation}
where $P(k)$ is the matter power spectrum today as a function of the wave number $k$,
and the functions $[g]_\ell$ and $[T]_\ell$ are 
\begin{equation}
\label{cc2}
\left[g\right]_\ell(k)=\int dz \, b_i(z) \Pi(z) D(z)j_\ell(k\chi(z))
\end{equation}
and  
\begin{eqnarray}
\left[T \right]_\ell(k)&=&3\frac{H_0^2}{c^2}\Omega_mT_{\rm CMB} \nonumber \\
&&\times\int dz \frac{d}{dz}\left[D(z)(1+z)\right]\frac{j_\ell(k\chi(z))}{k^2}.
\label{cc4}
\end{eqnarray}
The Limber approximation, which is quite accurate when $\ell$ is not
too small ($\ell \gsim 10$),
can be obtained from Eq.~(\ref{cc1}) by setting $P(k) = P(k=(\ell+1/2)/\chi(z))$ and using the
asymptotic formula that $(2/\pi)\int k^2 dk j_\ell (k\chi) j_\ell (k\chi') = (1/\chi^2) \delta (\chi-\chi')$ (when $\ell \gg 1$). We find that the 
substitution 
$k=(\ell+1/2)/\chi(z)$ is a better approximation to the exact expressions than $k=\ell/\chi(z)$.
This gives
\begin{eqnarray}
C_\ell^{gT} &=& {3\Omega_m H_0^2 T_{\rm CMB}\over c^2} {1\over ( \ell+1/2)^2} \nonumber \\
&&\times \int dz b(z) \Pi(z) {H(z) \over c} D(z){d\over dz}[D(z)(1+z)]
\nonumber \\
&&\times  P\left(\frac{\ell+1/2}\chi\right).
\end{eqnarray}

%

The above discussion ignores the effects of gravitational lensing, which alters
the expected signal through two competing effects -- changing the flux limit of 
the survey as well as the observed galaxy density. Both of these effects can
be thought of as altering the redshift distribution of the tracers, and so
we defer the discussion to Sec.~\ref{sec:dndz}.

\section{Data}
\label{sec:data}

We describe the CMB and galaxy data sets used in our analysis below; these are summarized in Table~\ref{tab:lss}.
The data sets not used in this paper are discussed further in the Sec.~\ref{sec:discussion}, 
where we provide detailed explanations for the choices made.  All large scale structure data were pixelized in the HEALPix system with the resolution 
and sky coverage shown in Table~\ref{tab:lss}.

\begin{table*}
\caption{\label{tab:lss}The large-scale structure data sets used.  The effective bias $b_{\rm eff}$ and bias-weighted redshift $\langle z\rangle_b$ 
are given here for the purpose of qualitatively illustrating which redshift ranges are probed by each sample.
They are computed for the fiducial WMAP cosmology as $b_{\rm eff}=\int f(z)\,dz$ and $\langle z\rangle_b=\int zf(z)\,dz/b_{\rm eff}$, respectively; the 
redshift distributions $f(z)$ will be computed in Sec.~\ref{sec:dndz}.
The data are pixelized using HEALPix \cite{gorski05} at the resolutions listed in the table.
}
\begin{tabular}{lcccccccccccccccc}
\hline\hline
Sample  (its notation in paper) &   & Area & & Density & & Number of & & HEALPix & &Number of & & $b_{\rm eff}$ & & $\langle z\rangle_b$ \\
 & & deg$^2$ & & deg$^{-2}$ & & galaxies & & resolution & &HEALPix Pixels & & & & \\
\hline
2MASS, $12.0<K_s<12.5$ (2MASS0) & & 27$\,$191 & & \white{0}1.84 & & \white{0}$\,$\white{0}50$\,$096 & & \white{0}9 & & $2\,$073$\,$457 & & 1.63 & & 0.06 \\
2MASS, $12.5<K_s<13.0$ (2MASS1) & & 27$\,$191 & & \white{0}3.79 & & \white{0}$\,$103$\,$060 & & \white{0}9  & & $2\,$073$\,$457 & & 1.52 & & 0.07 \\
2MASS, $13.0<K_s<13.5$ (2MASS2) & & 27$\,$191 & & \white{0}7.85 & & \white{0}$\,$213$\,$516 & & \white{0}9  & & $2\,$073$\,$457 & &1.54 & & 0.10 \\
2MASS, $13.5<K_s<14.0$ (2MASS3) & & 27$\,$191 & & 16.0\white{0} & & \white{0}$\,$435$\,$570 & & \white{0}9  & & $2\,$073$\,$457 & &1.65 & & 0.12 \\
SDSS, LRG, low-$z$     (LRG0)& & \white{0}6$\,$641 & & 35.1\white{0} & & \white{0}$\,$232$\,$888 & & 10 & & $2\,$025$\,$731 & &1.97 & & 0.31 \\
SDSS, LRG, high-$z$    (LRG1)& & \white{0}6$\,$641 & & 93.8\white{0} & & \white{0}$\,$622$\,$646 & & 10 & & $2\,$025$\,$731 & &1.98 & & 0.53 \\
SDSS, QSO, low-$z$     (QSO0)& & \white{0}6$\,$039 & & 20.8\white{0} & & \white{0}$\,$125$\,$407 & & 10 & & $1\,$842$\,$044 & &2.36 & & 1.29 \\
SDSS, QSO, high-$z$    (QSO1)& & \white{0}6$\,$039 & & 18.3\white{0} & & \white{0}$\,$110$\,$528 & & 10 & & $1\,$842$\,$044 & &2.75 & & 1.67 \\
NVSS point sources     (NVSS)& & 27$\,$361 & & 40.3\white{0} & & 1$\,$104$\,$983 & & \white{0}8 & & \white{0}521$\,$594 & & 1.98 & & 1.43 \\
\hline\hline
\end{tabular}
\end{table*}

\subsection{\label{sec:cmb}CMB temperature from WMAP} 

The WMAP mission \cite{bennett03, jarosik07} measured the all-sky maps of the 
CMB at multipoles up to $\ell \sim $ several
hundred.  We use the second public
data release of the WMAP data with the first
three years of observations.
The all-sky CMB maps are constructed  
in the following bands: K (23 GHz), Ka (33 GHz), Q (41 GHz), V (61 GHz)
and W (94 GHz).
These maps are pixelized in the HEALPix \cite{gorski05}
resolution 9 format with $3\,145\,728$ pixels, each 47.2 sq. arcmin in area. 
These maps are not beam-deconvolved and this, with the scan strategy of WMAP,
results in nearly uncorrelated Gaussian uncertainties on the temperature in 
each pixel \cite{jarosik07}.
We limit our analysis to Ka through W band as the K-band is heavily contaminated 
by the Galactic emission.
We trim all masks with the
WMAP Kp0 mask and point source mask to remove regions contaminated by
Galactic emission and point sources, leaving 76.8\% ($2\,414\,613$ resolution 9 HEALPix pixels)
of the sky for the ISW analysis.
We choose not to use either the WMAP ``Internal Linear Combination'' (ILC) map or the 
foreground cleaned map to avoid a number of practical difficulties as these 
maps lose frequency dependence of the original maps and have complicated pixel-pixel
noise correlations. 


\subsection{Two Micron All Sky Survey (2MASS)}
\label{ss:2mass}

We use galaxies from the Two Micron All Sky Survey (2MASS) Extended Source Catalog (XSC)
\cite{skrutskie97, jarrett00, skrutskie06} as mass tracers of the low redshift Universe. The median redshift
of these objects is $\sim 0.1$. 
We use $K_{20}$, the $K_s$-band isophotal magnitude measured inside a circular isophote 
with surface brightness of 20 mag arcsec$^{-2}$, as our default flux measure.
We extinction correct the magnitudes from the catalog using the reddening maps \cite{schlegel98}:
\begin{equation}
K_{20} = K_{20,raw}-A_K,
\end{equation}
where $A_K = 0.367  E(B-V)$ \citep{afshordi04}.  
Note that we ignore changes to
the isophotal radius due to extinction.
We remove regions with $A_K > 0.05$ in the dataset as the galaxy density starts to drop drastically. 
We visually inspects how the galaxy density changes with $A_K$ and decide to cut 
with $A_K> 0.05$ as there is a drastic drop. 
There are $1\,586\,854$ galaxies in the 2MASS XSC after removing
known artifacts and sources in close proximity to a large galaxy ($cc\_flag\neq$'a' and 'z')
and requiring $use\_src=1$ (which rejects duplicate observations of the same part of the sky).
The 2MASS XSC can miss objects near bright stars or overlapping artifacts, and so we used the
XSC coverage map \cite{jarrett00} and masked out pixels with $< 98\%$ coverage, thus $\sim8\%$ of the sky.

We divided the 2MASS sample into 4 flux bins: 
$12.0<K_{20}<12.5$, $12.5<K_{20}<13.0$, $13.0<K_{20}<13.5$, $13.5<K_{20}<14.0$.
Note that the 
redshift distribution of these 4 bins actually overlap significantly.
Our sample selection for 2MASS is similar to \citet{afshordi04} except 
the pixelization.

\subsection{Data from Sloan Digital Sky Survey (SDSS)}

The Sloan Digital Sky Survey has taken $ugriz$ CCD images of $10^4$ deg$^2$ of the high-latitude sky \citep{york00}. A dedicated 2.5m telescope 
\citep{gunn98,gunn06} at Apache Point Observatory images the sky in photometric conditions \citep{hogg01} in five bands ($ugriz$) 
\citep{fukugita96,smith02} using a drift-scanning, mosaic CCD camera \citep{gunn98}. All the data processing are done by completely automated 
pipelines, including astrometry, source identification, photometry \citep{lupton01,pier03}, calibration \citep{tucker06,padmanabhan07a}, spectroscopic 
target selection \citep{eisenstein01,strauss02,richards02}, and spectroscopic fiber placement \citep{blanton03}. The SDSS is well underway, and has 
produced seven major releases \citep{stoughton02,abazajian03,abazajian04,abazajian05,adel06,adel07a,adel07b}.

In addition to constructing LRG and quasar maps, we constructed three additional maps that we use to reject region sheavily affected by poor seeing or 
stellar contamination.  These include (i) a map of the full width at half-maximum (FWHM) of the point-spread 
function (PSF) in $r$ band; (ii) a map of stellar density ($18.0<r<18.5$ stars, smoothed with a 2 degree FMHM Gaussian); and (iii) a similar map using 
only the red stars ($g-r>1.4$).

All SDSS magnitudes used here are extinction-corrected using the maps of Ref.~\cite{schlegel98}.  We use SDSS model magnitudes for the LRGs, and PSF 
magnitudes for the quasars and stars.

\subsubsection{Luminous Red Galaxies}

We use the photometric Luminous Red Galaxies (LRGs) from Sloan Digital Sky Survey (SDSS)
constructed as described in \cite{padmanabhan05}. 
The LRGs have been very useful as a cosmological probe since they are typically the most luminous 
galaxies in the Universe, thus probing a larger volume than most other tracers. On top of this,
they also have very regular spectral energy distributions and a prominent 4000\AA\ break, making photo-$z$ acquisition much 
easier than the other galaxies. 
We will not be repeat our selection criteria here as it is thoroughly described in \cite{padmanabhan05}.
We only accept sky regions with $E(B-V) \le 0.08$ (almost identical to $A_r\le 0.2$ as in \cite{padmanabhan05})
and an $r$ band FWHM$<2.0$ arcsec.

Furthermore, there are a few regions in SDSS that have $\ge$60\% more red stars than typical for their galactic latitude; we suspect photometric 
problems and rejected these regions. The red star cut removed $427$ deg$^2$ in 
assorted parts of the sky.

We slice our LRG sample into two redshift bins for the ISW analysis: $0.2\le z_{\rm photo}\le 0.4$ and  $0.4\le z_{\rm photo}\le 0.6$.

 
\begin{figure*}
\includegraphics[width=7.0in]{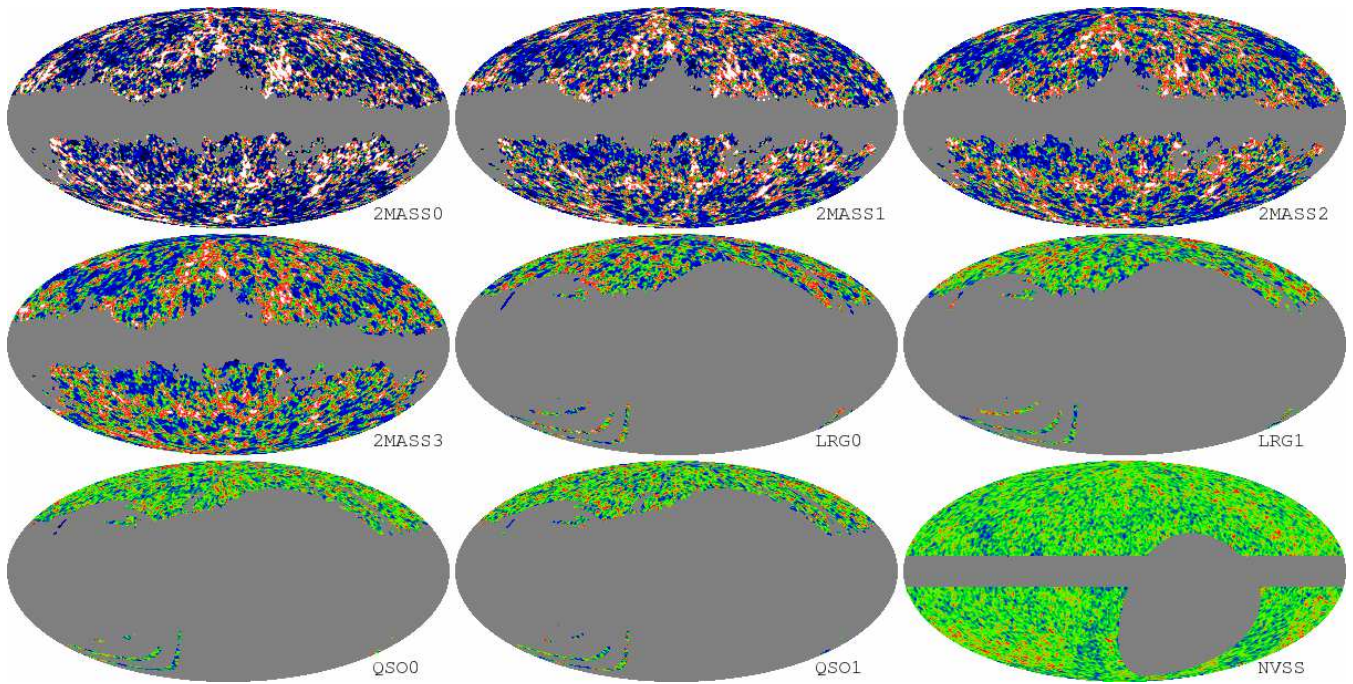} 
\caption{The overdensity maps of various tracer samples in Galactic coordinates.  The scale runs from $g=-1$ (black, no galaxies) to $g=-0.25$ (blue), 
$g=0$ (green), $g=+0.25$ (red), and $g=+1$ (white, $\ge 2\times$ mean density).}
\label{fig:mollew}
\end{figure*}

\subsubsection{Photometric quasars}

We select quasars photometrically from the Sloan Digital Sky Survey by first generating a candidate 
quasar catalog consisting of UVX objects \cite{richards04}. These are
point sources with excess UV flux (i.e. $u-g < 1.0$)
observed $g$ magnitudes fainter than 14.5 (to avoid saturation problems), extinction corrected $g$ magnitudes brighter than $21.0$, and 
u-band error less than $0.5$ mag ($\ge 2\sigma$ detection in $u$).
We call this the ALL-UVX catalog.
We also have the public catalog of photometric quasars from Data Release 3 (DR3) 
generated by Ref.~\cite{richards06}, which we will call DR3-QSO objects.
We also construct a UVX object list from only DR3 data, denoted DR3-UVX.
This catalog is used to extend the selection and photometric redshifts from the DR3 region 
to the ALL region. Ideally the catalog would have been based on running the algorithm of Ref.~\cite{richards06} on the ALL region but this option was 
not available at the time we constructed the quasar catalog.

We first match the DR3-UVX objects to the DR3-QSO objects and then assign the photometric redshifts 
from the DR3-QSO objects to the matched DR3-UVX object. 
For objects that are in DR3-UVX catalog, but not in the DR3-QSO catalog, we 
mark them as rejects.
We now have a DR3-UVX catalog with every object either assigned a redshift or marked as a reject.
The reject rate for DR3-UVX (ALL-UVX) is $89\%$ ($93\%$).
Then, we lay down the DR3-UVX catalog in color$^4$  ($u-g$,$g-r$,$r-i$,$i-z$) space, and 
then for each ALL-UVX object, we find its nearest neighbor in this color$^4$ space, then
assigning it the same ``redshift'' as its matched DR3-UVX neighbor. 
If the DR3-UVX object has a redshift (not a reject), then the ALL-UVX object is classified as a
quasar with the same redshift (photo-$z$ only), otherwise it is rejected.
This procedure generates a photometric catalog of quasars in the full survey area, based on the matching against DR3 
quasars in color$^4$ space. 
However, this catalog only has the photometric redshifts, but not the actual redshift 
distribution. The actual redshift distribution will be discussed in Sec.~\ref{sec:sys}. 
The average color offsets of the quasar candidate to its match for $u-g$, $g-r$, $r-i$ and $i-z$ are 
$0.0018$, $0.0056$, $0.0075$ and $0.0045$, while
the typical errors on the colors of the candidates are $0.11$ ($u-g$), $0.13$ ($g-r$), $0.14$ ($r-i$) and 
$0.17$ ($i-z$). As the color differences between the match and the candidate are well within the 
error of the colors,  we conclude that the quasar candidates are matched with
high accuracy. 

We then cut the catalog according to $E(B-V) <0.05$ and FWHM$< 2.0$ arcsec. These cuts are determined when we look at the variation of the quasar 
number overdensity over a range of extinction and seeing. Also, since quasars are more sensitive than LRGs to extinction (as a result of the importance 
of the $u$ filter in selecting quasars), we cut the catalog at a lower $E(B-V)$.  We also imposed a cut rejecting regions with more than twice average 
stellar density, i.e. we require $n_{\rm star}<564$ stars/deg$^2$.

We further divide the sample into two redshift (photo-z) bins: $0.65<z_{\rm photo}<1.45$ (low-$z$) and
$1.45<z_{\rm photo}<2.0$ (high-$z$). 
This division of sample is due to the fact that there are strong emission lines (e.g. Mg$\,${\sc ii}) that redshift from one filter into the next 
around the redshifts of 
$0.65$, $1.45$ and $2.0$, causing these two redshift bins to be relatively free of cross-contamination.  
However, as we will see, they do contain significant 
contamination from redshifts below $0.65$ and above $2.0$.  
We therefore constrain their redshift distribution by cross-correlating these with 
auxiliary data sets; we discuss this further in Sec.~\ref{sec:dndz}.

\begin{figure*}
\begin{center}
\includegraphics[angle=-90.0,width=6.2in]{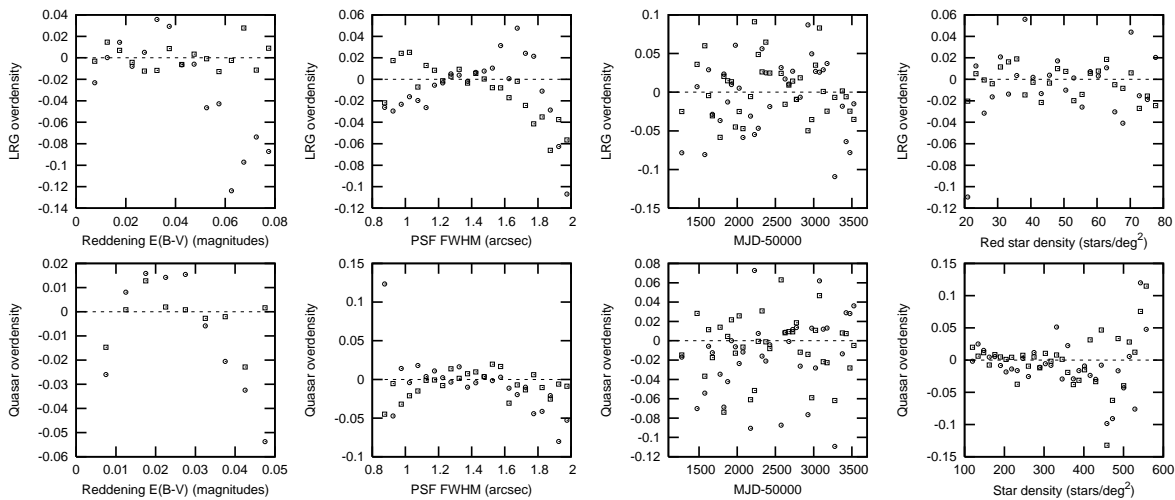}
\end{center}
\caption{LRG and QSO overdensity vs various quantities such as reddening, PSF FWHM ($r$ band), observing time (MJD),
red star density, and star density.  In each panel the circles show the low-redshift sample and the squares show the high-redshift sample.  The 
Modified Julian Date (MJD) of the DR3 ending date is 52821.  Note that there are very few accepted pixels at the extremes of reddening, PSF FWHM, and 
stellar density, resulting in the large fluctuations seen in the figure.}
\label{fig:syst}
\end{figure*}

The construction of the full sample using the DR3 catalog as described above introduces one potentially worrying systematic, namely the possibility that 
regions of the sky observed after DR3 would have a different density of sources than DR3 regions as a result of the nearest-neighbor method misbehaving 
in low-density regions of color$^4$ space.  This would provide a spurious feature in the quasar maps that resembles the DR3 coverage map.
In order to check for this problem, we look for correlations between  
observing dates (if the ALL sample is misbehaving, it will be different from DR3 sample)
with galaxy overdensity, and we do not find any significant correlations (Fig.~\ref{fig:syst}). 
We also look at the correlation between quasar overdensity and the stellar number density
to see if there is significant stellar contamination, we do not find any either (Fig.~\ref{fig:syst}). 

\subsection{NRAO VLA Sky Survey (NVSS)}
\label{sec:nvss_data}
The NRAO VLA Sky Survey (NVSS) is a 1.4 GHz continuum survey covering the entire sky north of 
$-40^\circ$ declination using the compact D and DnC configurations of the Very Large Array (VLA) \cite{condon98}.
The images all have 45 arcsec FWHM resolution and nearly uniform sensitivity and yield
a catalog of almost $2\times 10^6$ discrete sources stronger than $\sim 2$ mJy. 

This survey has several potentially major artifacts: 
Galactic synchrotron emission, spurious power from bright sources and 
a declination-dependent striping problem. 
All of these have to be treated properly before one can claim that the power coming from 
the cross-/auto-correlation is not due to some spurious issues.
The Galactic synchrotron emission can in principle be an issue because it contributes significantly to the noise temperature of the VLA, and for 
realistic number counts, increased 
noise temperature could change the number of sources with measured flux above some threshold.  (As an interferometer the VLA is not directly sensitive 
to the diffuse synchrotron foreground.)
This issue is treated by incorporating a template -- the Haslam map \cite{haslam81} -- in 
the cross-correlation analysis and projecting out the power that are correlated to 
this template. Even though 
the Haslam map is at 408 MHz, the frequency dependence of the galactic synchrotron emission
is fairly flat, allowing us to use it as a template of the Galactic synchroton radiation.
The bright sources are problematic since the VLA has a finite dynamic range ($\sim 1000$ in snapshot mode with limited $uv$-plane coverage) and thus
the identification of faint sources in fields with a bright source is unreliable.
This issue is mitigated by masking out all the bright sources. 
Striping is a known systematic effect in NVSS \citep{blake02}: the galaxy density has a systematic 
dependence on declination, which can mimic long-wavelength modes in the galaxy field.
To deal with the above potential problems, we first impose a flux limit of 2.5 mJy (where NVSS is 50\% complete), mask out a 0.6 degree radius around 
all the bright sources ($>2.5$ Jy).  Then to reduce striping, we also include templates to project out the synchrotron and declination-striping modes.
The implementation of this projection of spurious power will be further discussed in Sec.~\ref{sec:cross}.

\section{\label{sec:cross} Cross-correlation power spectrum analysis}

\subsection{Methodology}
\label{ss:method}

We start by organizing the temperature fluctuations
and the galaxy overdensities into a single data vector,
\begin{equation}
\mathbf{x} = (\mathbf{x}_{B,T},\mathbf{x}_{g}) \,\,,
\end{equation}
where $\mathbf{x}_{B,T}$ is a vector with the measured
CMB temperature (with the monopole and dipole subtracted)
in band $B$ at every HEALPix pixel;
analogously, $\mathbf{x}_{g}$ is the tracer number overdensity.  The vector $\mathbf{x}$ has a total length $N_{\rm pix,CMB}+N_{\rm pix,LSS}$ where 
$N_{\rm pix,CMB}$ and $N_{\rm pix,LSS}$ are the number of accepted pixels for the CMB and LSS maps respectively.
We suppress the band
subscript for simplicity, with the implicit understanding
that we always refer to the cross correlation of
a single WMAP band with the tracer overdensity.
The covariance matrix of $\mathbf{x}$ is,
\begin{equation}
\mathbf{C} = \mathbf{C}_{diag} + \left(
\begin{array}{cc} {\bf 0} & {\bf C}^{gT\dagger} \\ {\bf C}^{gT} & {\bf 0} \end{array}\right) \,\,,
\label{eq:cdef}
\end{equation}
where $\mathbf{C}_{diag}$ is given by,
\begin{equation}
\mathbf{C}_{diag} = \left(\begin{array}{cc} {\bf C}^{TT}+{\bf N}^{TT} & {\bf 0} \\ {\bf 0} & {\bf C}^{gg} + {\bf N}^{gg}
\end{array}\right) \,\,,
\label{eq:cdiagdef}
\end{equation}
where ${\bf N}^{xx}$ is the noise matrix.
The submatrices ${\bf C}^{TT}$, ${\bf C}^{gg}$ and ${\bf C}^{gT}$ are defined by
\begin{equation}
C^{ab}_{ij} = \sum_{lm} C^{ab}_{l}
Y_{lm}^{*}(\hat{n}_{i}^{a}) Y_{lm}(\hat{n}_{j}^{b}) \,\,,
\end{equation}
where $\hat{n}_{i}^{a}$ is the position (on the sky) of the
$i^{th}$ point of the vector $\mathbf{x}_{a}$. The temperature-temperature,
galaxy-galaxy and galaxy-temperature angular power spectra are denoted
by $C^{TT}_{l}, C^{gg}_{l}$ and $C^{gT}_{l}$ respectively. 

The galaxy power spectrum is first estimated using a pseudo-$C_{l}$
estimator \cite{efstathiou04}, and fit by the non-linear power spectrum of
\cite{smith03}, multiplied by a constant linear bias. We project out
the monopole and dipole of both these power spectra by setting the power in
the $l=0,1$ modes to a value ($10^{-1}$) much greater than the true
power spectrum. 

We parametrize $C^{gT}_{l}$ as a sum of bandpowers,
$\tilde{P}_{i,l}$, with amplitudes $c_{i}$ to be estimated,
\begin{equation}
C^{gT}_{l} = \sum_{i} c_{i} \tilde{P}_{i,l} \,\,.
\end{equation}
We consider
``flat'' bandpowers given by
\begin{equation}
\tilde{P}_{i,l}  =  \left\{ \begin{array}{lcl}
 B(l) & & l_{i,min} \leq l <  l_{i,max} \\
0 & & \rm{otherwise}, \end{array}\right.
\label{eq:bandpowers}
\end{equation}
where $B(l)$ is the product of the beam transfer function
\cite{page03}, and the HEALPix pixel transfer functions at WMAP and LSS resolution. This parametrizes the power 
spectrum
as a sum of step functions and is useful
when the shape of the power spectrum is unknown. 

We estimate the $c_{i}$ by forming quadratic combinations of the data
\cite{tegmark97, seljak98},
\begin{equation}
q_{i} = \frac{1}{2} \mathbf{x}^{t} \mathbf{C}_{diag}^{-1}
\frac{\partial \mathbf{C}}{\partial c_{i}} \mathbf{C}_{diag}^{-1} \mathbf{x} \,\,.
\label{eq:qi}
\end{equation}
These are related to the estimated $\hat{c}_{i}$ by the response matrix $\mathbf{F}$,
\begin{equation}
\hat{c}_{i} = \sum_{j} (\mathbf{F}^{-1})_{ij} q_{j} \,\,,
\label{eq:ci}
\end{equation}
where
\begin{equation}
\mathbf{F}_{ij} = \frac{1}{2} \rm{tr}\left[
\mathbf{C}_{diag}^{-1} \frac{\partial \mathbf{C}}{\partial c_{i}}
\mathbf{C}_{diag}^{-1} \frac{\partial \mathbf{C}}{\partial c_{j}} \right] \,\,.
\end{equation}
If $C^{gT}_{l} \ll \sqrt{C^{gg}_{l} C^{TT}_{l}}$, then the $\hat{c}_{i}$ are 
good approximations to the maximum likelihood estimates of the $c_{i}$.
The covariance matrix of the $\hat{c}_{i}$ is
the inverse of the response matrix, if the
fiducial power spectra and noise used to compute $\mathbf{C}_{diag}^{-1}$ correctly
describe the data (in this case $\mathbf{F}$ is the Fisher matrix, hence the notation).
The matrix $\mathbf{C}_{diag}$ determines the weighting and is often called a ``prior'' in quadratic estimation theory.  Note that this usage has 
nothing to do with Bayesian priors -- in particular, Eq.~(\ref{eq:ci}) is unbiased regardless of the choice of prior (though for bad choices the 
estimator is not minimum variance).
Implementing the above algorithm is complicated by the sizes of the datasets;
the implementation we use is in \citep{padmanabhan03, hirata04, padmanabhan05ISW},
and we refer to the reader to the discussion there.

[In addition to the cross-power spectra in Eq.~(\ref{eq:bandpowers}), in quadratic estimator theory one usually tries to estimate the CMB and galaxy 
auto-power spectra as well.  Because our prior is diagonal, however, these decouple, i.e. the entries in $F_{ij}$ that couple the auto-powers and 
cross-powers are zero.  For this reason we can leave the auto-powers out of the quadratic estimator.]

As mentioned earlier, the NVSS dataset has issues that require additional processing.
Assume a systematic $E$ that we characterize as follows:
\begin{equation}
\mathbf{x^{obs}} = \mathbf{x^{true}} + \lambda E \,\,. 
\end{equation}
If estimate $\hat{c}_{i}$, even if $\mathbf{C}$ is the true covariance,
we will still have a biased answer.
However, the substitution
\begin{equation}
\mathbf{C} = \mathbf{C^{true}} + \zeta E E^{t}
\end{equation}
yields an unbiased estimate of $\hat{c}_{i}$ when $\zeta \rightarrow \infty$.
One can add as many systematic templates $E$ (i.e. modes to project out of the map) as desired.
To immunize the NVSS correlations from possible systematics, we break the NVSS map into 74 declination rings, and for each ring include a
template map $E$ consisting of either $+1$ (for 
pixels within the declination ring) or $0$ (for all other pixels).
This removes the declination-dependent stripes.
We also put in the 408 MHz Haslam map \cite{haslam81} (technically $T_{\rm Haslam}-20\,$K) as a template for the 
Galactic synchrotron radiation.  We experimented with the values of $\zeta$ and found that the cross-spectra are converged with the choice $\zeta=1$ 
for the declination rings and $\zeta=10^{-3}\,$K$^{-2}$ for the synchrotron map.

\subsection{Priors}

To generate the priors $\mathbf{C}_{diag}$ for the cross-correlation power spectrum analysis, we need the approximate autopower spectrum of the 
galaxies. The auto-correlation is done using the same methodology as described in Sec.~\ref{ss:method}.
The resulting autopower spectra must be smoothed, before being used as priors.  
This avoids statistical fluctuations in $C_{\ell}$ over- or under-weighting the corresponding monopoles in the cross-correlation, which could result in 
underestimation of $C_\ell^{gT}$ signal since we would artifically down-weight multipoles that had accidentally high power in galaxies and place more 
weight on multipoles that had little power.
We did the smoothing in two different ways.  For the cases where the redshift distribution was available early enough in 
the analysis (2MASS or LRG), we fit the auto-power spectrum
to the non-linear matter power spectrum \cite{smith03} to get the linear bias.
In other cases (quasars, NVSS) we did not have the redshift distribution at the time the priors 
were created; we created the priors by using a smoothed,
splined auto-power spectrum of the sample as the prior.


In the cases where we did a fit using the nonlinear matter power spectrum, the fit biases are 
1.15, 1.18, 1.20, and 1.22 (2MASS, brightest to faintest); 1.92 (LRG low-$z$); and 1.86 (LRG high-$z$).  After generating the priors, we made several 
modifications to the analysis, including the inclusion of redshift-dependent bias in 2MASS.  Thus while the priors were not updated since they give a 
good fit to the observed autopower spectrum, it should be noted that these bias values are {\em not} used in the cosmological analysis (i.e. for ISW 
prediction purposes).


To generate priors for the CMB, we generate the priors using the theoretical $C_{\ell}$s from WMAP and take into the account of the 
effect of pixelization and beams by convolving with the pixel and beam window functions.

\subsection{Results of cross-correlation}

%

\begin{figure*}
\begin{center}

\includegraphics[width=6.5in]{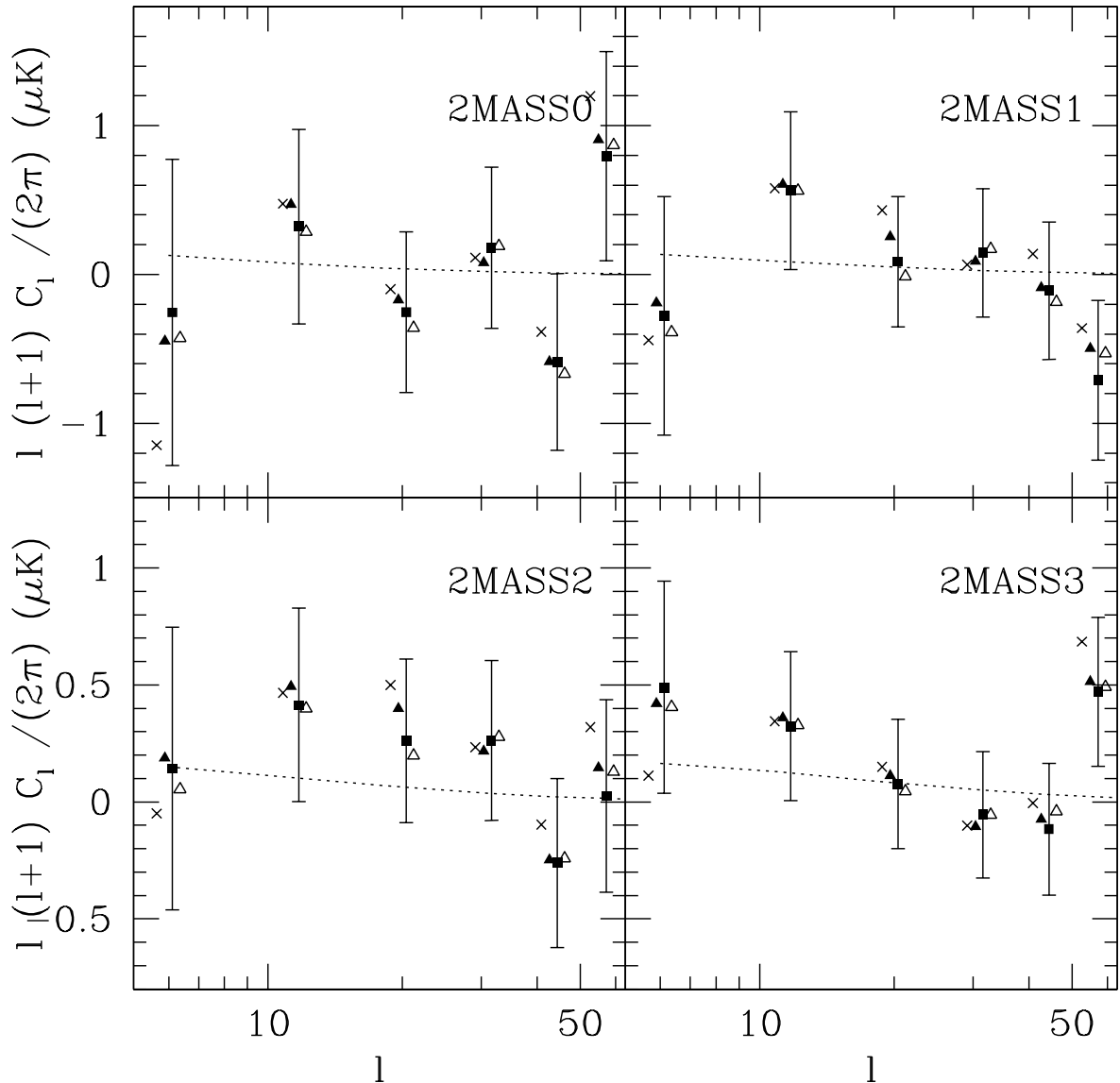}
\end{center}
\caption{Galaxy density correlations with WMAP temperatures (4 bands: Ka (crosses), Q (triangles), V (squares), W (empty triangles), error bars are from the correlations with V-band. 
This contains 2MASS galaxy density correlations with WMAP, 
starting from (from left to right, top to bottom) the brightest sample, to the bottom the dimmest sample.
We shift the points on x-axis for clarity. The dotted line shows the 
predicted signal for the sample with WMAP 3-year parameters and $b dN/dz$ estimated in Sec.~\ref{sec:dndz}.}
\label{fig:cross_2mass}
\end{figure*}

\begin{figure*}
\begin{center}
\includegraphics[width=6.5in]{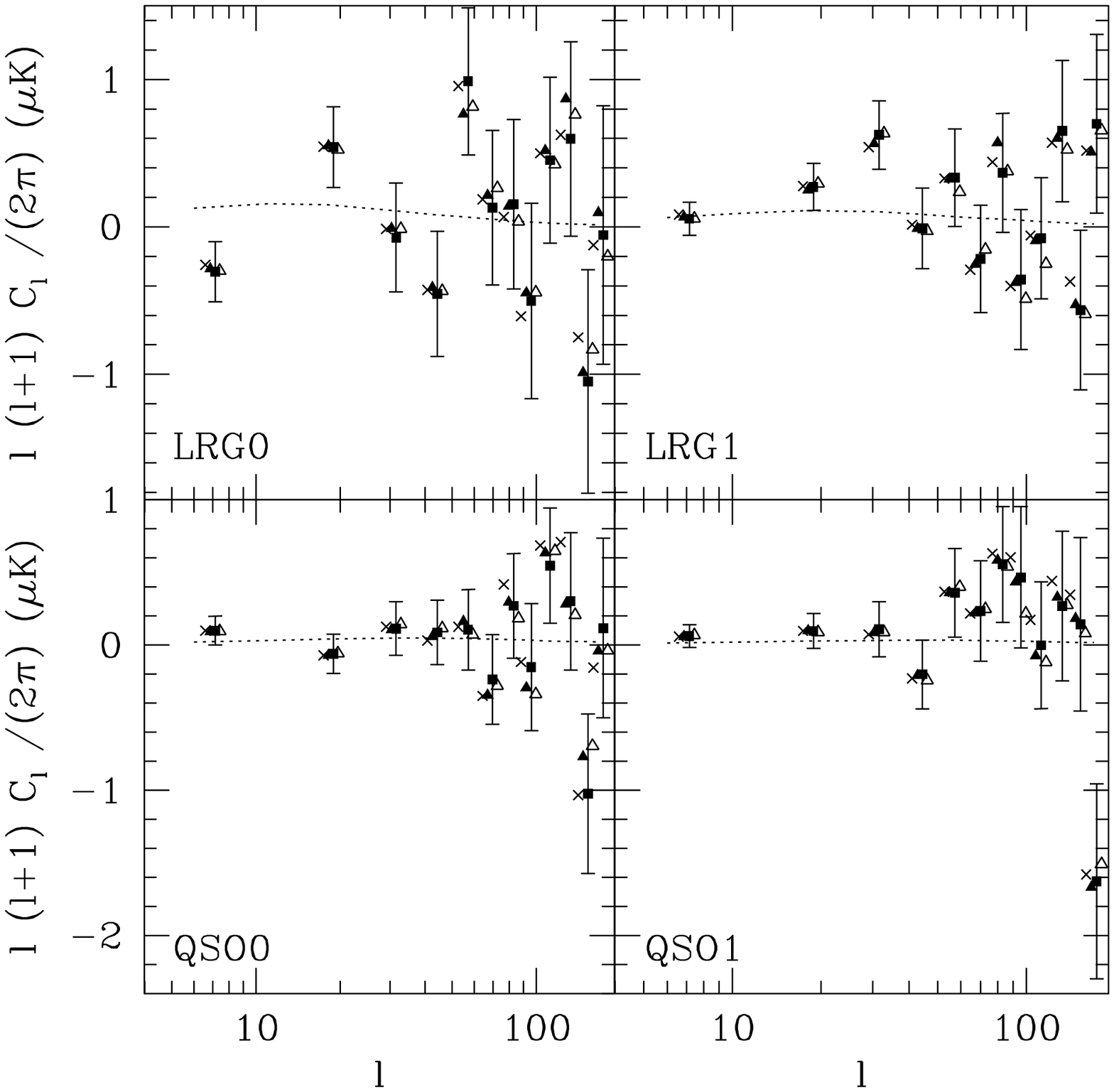}
\end{center}
\caption{Same as Fig.~\ref{fig:cross_2mass} except for the SDSS density maps from (from left to right, top to bottom): low-z LRG, high-z LRG, low-z QSO, high-z QSO.
}
\label{fig:cross_sdss}
\end{figure*}

\begin{figure}
\begin{center}
\includegraphics[width=3.3in]{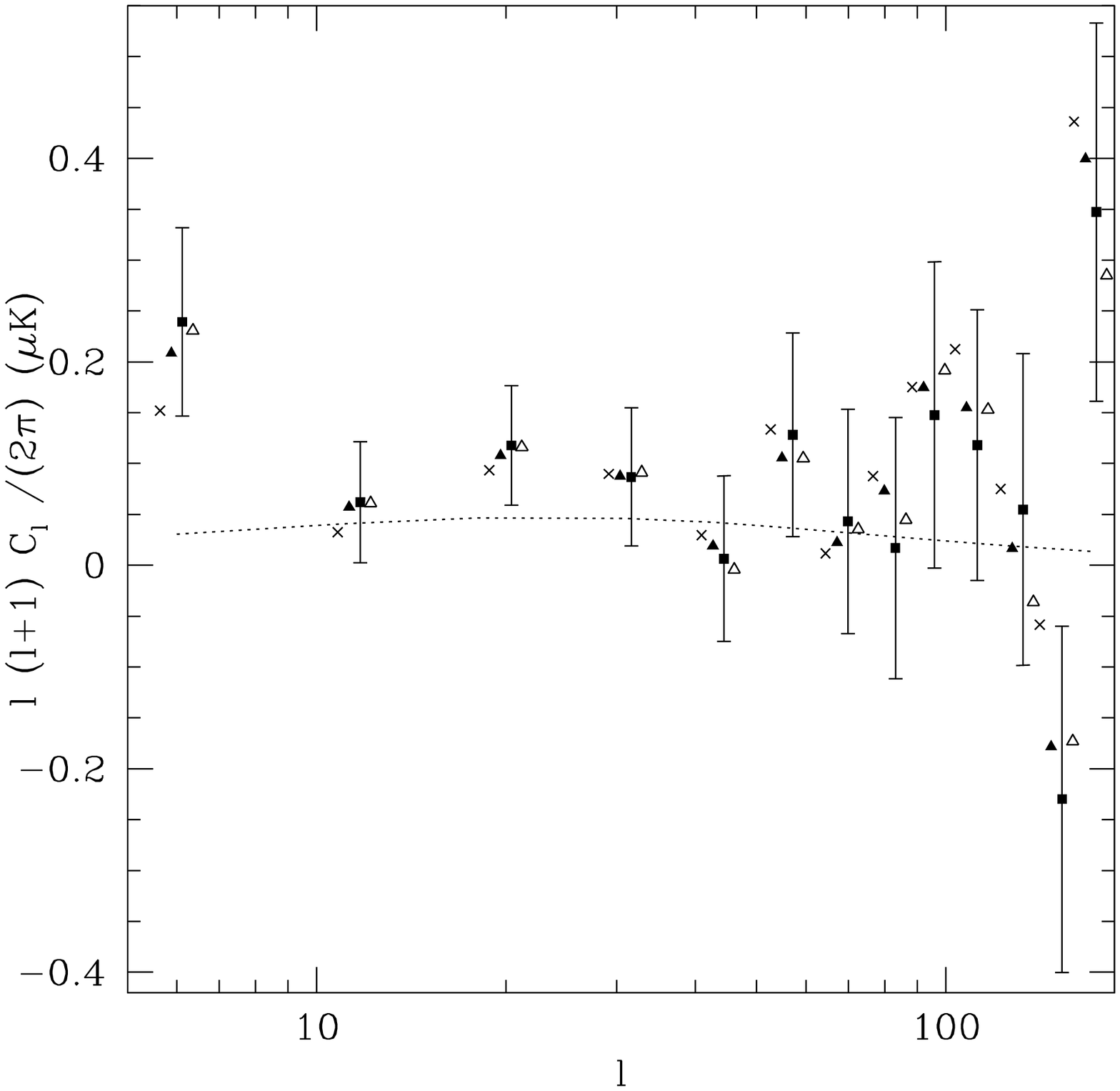}
\end{center}
\caption{Same as Fig.~\ref{fig:cross_2mass} except for the NVSS cross-correlation.
}
\label{fig:nvsscross}
\end{figure}

Figs.~\ref{fig:cross_2mass}, \ref{fig:cross_sdss}, and \ref{fig:nvsscross} plot the 
cross-correlation between WMAP and the 2MASS, SDSS and NVSS samples respectively; the
four different symbols in each of these plots correspond to the four WMAP bands we use.
The observed achromatic nature of the signal is consistent with it being ISW, and is an
important check for frequency dependent systematics. 
The two quasar samples are at the highest redshifts we can probe, 
so if there is an ISW cross-correlation at $z\sim 1$--2, it 
would mean that there is significant gravitational potential 
change at these redshifts. This is not expected in simplest 
$\Lambda$CDM cosmology, but could be present either in models 
where dark energy equation of state is rapidly changing with redshift 
or in models where curvature plays a role. 
The observed lack of a signal for these redshifts therefore strongly constrains 
such models.
Note however that the NVSS cross-correlation
cannot be automatically interpreted as a detection of high redshift ISW, as (see below)
it covers a wide redshift range.

%

\section{Redshift Distributions}
\label{sec:dndz}

The basic problem is to determine for each galaxy sample $i$ and each cosmological model the function $f_i(z)$ that relates the matter density 
$\delta({\bf r})$ to the two-dimensional galaxy overdensity $g_i$:
\beq
g_i(\nhat) = \int_0^\infty f_i(z) \delta[\nhat,\chi(z)] dz.
\label{eq:gi}
\eeq
Eq.~(\ref{eq:gi}) is understood to be valid on scales where the galaxies trace the matter distribution.  In the absence of magnification bias, the function $f_i(z)$ is simply the product of the bias and the redshift distribution: $f_i(z) = b_i(z)\Pi_i(z)$, where $\Pi_i(z)$ is the probability distribution for the galaxy redshift.  In the presence of magnification bias, which is important for the SDSS quasars and possibly the NVSS radio sources, $f_i(z)$ takes on the more complicated form
\beq
f_i(z) = b_i(z)\Pi_i(z) + \int_z^\infty W(z,z')[\alpha(z')-1] \Pi_i(z')dz',
\label{eq:fi}
\eeq
where $\alpha(z')$ is the slope of the number counts of the galaxy density as a function of flux: $N(>F)\propto F^{-\alpha}$.
Here $W(z,z')$ is the lensing window function:
\beqa
W(z,z') &=& \frac32\Omega_mH_0^2\frac{1+z}{cH(z)}\sin_K^2\chi(z)
\nonumber \\ && \times [\cot_K\chi(z)-\cot_K\chi(z')],
\eeqa
where $\chi(z)=\int_0^z dz''/H(z'')$ is the radial comoving distance, $\sin_K\chi$ is the sine like function (equal to $\chi$ in a flat Universe), and $\cot_K\chi=d(\ln\sin_K\chi)/d\chi$ is the cotangent like function (equal to $1/\chi$ in a flat Universe).

It is in fact the function $f_i(z)$ that is required if one is to predict the ISW effect in a given cosmology.  It is this same function that is required to predict the linear-regime angular power spectrum of the galaxies.  This section describes the method by which $f_i(z)$ is obtained for each of the samples.  The methods are quite different due to the different types of information available for each sample.  In particular there are very few spectroscopic redshifts available for NVSS.  Note however that all methods include galaxy clustering data, as this is needed to determine the bias even if the redshift probability distribution $\Pi_i(z)$ is known perfectly.

All of the numbers and plots {\em in this section only} that depend on cosmology are computed using the {\em original} WMAP third-year flat 6-parameter 
$\Lambda$CDM cosmology ($\Omega_bh^2=0.0222$, $\Omega_mh^2=0.1275$, $h=0.727$, $\sigma_8=0.743$, and $n_s=0.948$), i.e. from the first release of 
Ref.~\cite{spergel07}. However in the Markov chain, the functon $f_i(z)$ is re-computed for each cosmological model and used to predict the ISW signal.

\subsection{2MASS}

The 2MASS samples go down to a limiting magnitude of $K_{20}=14$.  At this relatively bright magnitude, almost all objects (97.9\%, after correcting 
for the fiber collisions) have SDSS spectra,
provided of course that they lie within the spectroscopic mask.  In practice there are two subtleties that can occur.  One is that the bias
$b_{\rm 2MASS}$ cannot be obtained to high accuracy from linear theory because even the moderate multipoles ($l\sim 20$) are nonlinear, especially for
the nearest 2MASS slice, and the lowest multipoles suffer from cosmic variance.  The other is that the bias varies with redshift: even though the 2MASS
galaxies cover a narrow range in redshift during which the Universe expands by only $\sim 30$\%, the use of apparent magnitude to define the samples
means that the typical luminosity of a galaxy varies by several magnitudes across the redshift range of interest.  
more biased, this effect shifts the peak of the effective redshift distribution $f(z)$ to higher redshifts than the actual distribution $\Pi(z)$.

\begin{figure}
\includegraphics[angle=-90,width=3.2in]{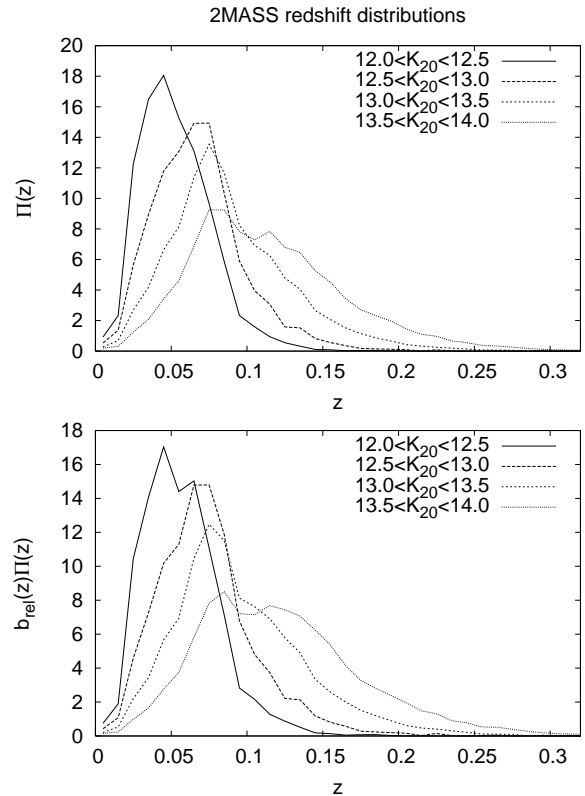}
\caption{\label{fig:2mass-dndz}The 2MASS redshift distribution, binned in units of $\Delta z=0.01$.  The top panel shows the raw measured distribution,
$\Pi(z)$, and the bottom panel is corrected for relative bias $b_{\rm rel}(z)\Pi(z)$.}
\end{figure}

We match the 2MASS galaxies with the SDSS MAIN galaxy sample by first defining the 2MASS sample as discussed in \ref{ss:2mass}, then 
we select 2MASS galaxies only within mask that is more than 90\% complete. 
We then try to match all the 2MASS galaxies with the SDSS MAIN galaxies that are within $3''$ and found that almost all of the objects
from 2MASS sample have SDSS spectra. 
We thus use the spectroscopic redshifts of the matched SDSS galaxies to identify the redshifts of the 2MASS galaxies.
The redshift distribution is binned with $\delta_z = 0.01$.
The redshift distribution for each of the four slices is shown in Fig.~\ref{fig:2mass-dndz}.

The problem of nonlinear evolution is generally very complicated, however for ISW work we only need a solution accurate to a few tens of percent.
Therefore we have used the $Q$-model \cite{cole05}, which relates the galaxy power spectrum to the linear power spectrum via
\beq
P_{\rm gal}(k) = b^2\frac{1+Qk^2}{1+Ak} P_{\rm lin}(k),
\label{eq:Q}
\eeq
where $b$ is the linear bias appearing in Eq.~(\ref{eq:fi}).  Cole et~al. \cite{cole05} found in simulations that this function fits the
galaxy power spectrum in simulations for $A=1.7h^{-1}\,$Mpc, while the required value of $Q$ varies depending on the sample.  Our method is to compute
the theoretical angular galaxy power spectrum $C_\ell^{gg}({\rm th)}$ via the Limber integral, and fit this to the measured $C_\ell^{gg}$ treating $b$ 
and
$Q$ as free parameters.  This procedure can be done either assuming $b$ is constant with redshift, or (better) taking into account the
redshift-dependent bias,
\beq
P_{\rm gal}(k,z) = b_\star^2b_{\rm rel}^2(z)\frac{1+Qk^2}{1+Ak} P_{\rm lin}(k,z),
\label{eq:QQ}
\eeq
where $b_{\rm rel}(z)$ is known and $b_\star$ is a free parameter.  While there is very little evolution in the 2MASS redshift range, the nearby and 
distant galaxies can have very different biases because they correspond to different luminosity ranges.
The results for each are shown in Table~\ref{tab:2mass-bias}.
$b_{\rm rel}(z)$ is based on taking the r-band luminosities of the galaxies and
using $b_{\rm rel}(L)$ from \citet{tegmark06}.
Note that the prominent peak of redshift distribution at $z\sim 0.08$ is a supercluster known 
as the Sloan Great Wall. 
(In principle $Q$ can depend on redshift as well, so one should be careful about 
interpreting the fit value and indeed one can see from Table~\ref{tab:2mass-bias} that $Q$ fit in this way is not stable.  However the $\lsim 1\sigma$ 
changes in $\langle b\rangle$ seen in the table when we restrict to much lower 
$l_{\rm max}$ suggest that this is not a large effect on the bias.)

\begin{table*}
\caption{\label{tab:2mass-bias}The bias of the 2MASS galaxies as determined using the $Q$-model parametrization.
The second column in each line shows the maximum value of $\ell$ used in the main fits (varying or constant $b$).  The first fit (``varying $b$'')
uses Eq.~(\ref{eq:QQ}) and should be viewed as the main result.  For this fit we show the mean bias, i.e. $\langle b\rangle = \int b(z)\Pi(z)\,dz$, as
this is easier to compare with other results than $b_\star$.  The second fit (``constant $b$'') has the bias fixed to a constant value.  The third fit
($l_{\rm max}=24$) has a bias varying according to Eq.~(\ref{eq:QQ}) but the fit is restricted to the region $l<25$ in order to reduce the effect of
nonlinearities.  Note that the biases obtained from the varying-$b$ fits are consistent with each other, while the constant-bias fit finds a lower
value of $b$ by up to $\sim 6$\% depending on the sample.}
\begin{tabular}{ccccrccrccr}
\hline\hline
$K_{20}$ range & $l_{\rm max}$
& & \multicolumn{2}{c}{Varying $b$ fit} & & \multicolumn{2}{c}{Constant $b$ fit} & & \multicolumn{2}{c}{$l_{\rm max}=24$ fit} \\
 & & & $\langle b\rangle$ & $Q\;\;\;$ & & $b$ & $Q\;\;\;$ & & $\langle b\rangle$ & $Q\;\;\;\;$ \\
\hline
12.0--12.5 & 49 & & $1.62\pm 0.08$ & $12\pm 3$ & & $1.54\pm 0.08$ & $12\pm 3$ & & $1.60\pm 0.13$ & $ 12\pm 10$ \\
12.5--13.0 & 61 & & $1.52\pm 0.07$ & $17\pm 3$ & & $1.44\pm 0.06$ & $17\pm 3$ & & $1.57\pm 0.13$ & $  9\pm 15$ \\
13.0--13.5 & 74 & & $1.54\pm 0.05$ & $14\pm 2$ & & $1.45\pm 0.05$ & $14\pm 2$ & & $1.67\pm 0.12$ & $-12\pm 16$ \\
13.5--14.0 & 99 & & $1.65\pm 0.04$ & $ 8\pm 1$ & & $1.55\pm 0.04$ &  $8\pm 1$ & & $1.74\pm 0.10$ & $-32\pm 19$ \\
\hline\hline
\end{tabular}
\end{table*}

\begin{figure}
\includegraphics[angle=-90,width=3.2in]{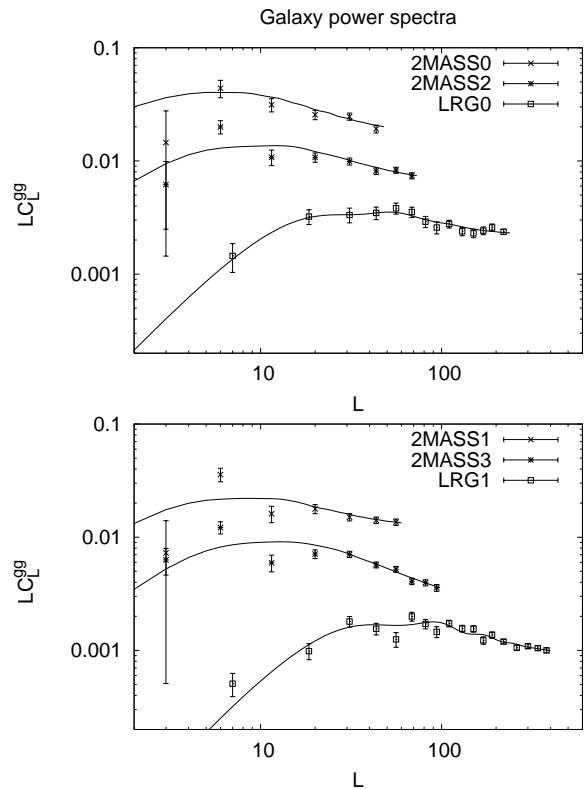}
\caption{\label{fig:loz}The galaxy power spectra for the four 2MASS and two SDSS LRG samples, and the $Q$-model fits.  The solid lines show the range 
of multipoles used in the fit, the dashed lines are extrapolations.  Note that at very small scales the $Q$-model is not a good description of the 
power spectrum.}
\end{figure}

The $Q$-model fits for the 2MASS sample (and the LRGs) are shown in Fig.~\ref{fig:loz}.


\subsection{SDSS LRGs}

\begin{figure}
\includegraphics[angle=-90,width=3.2in]{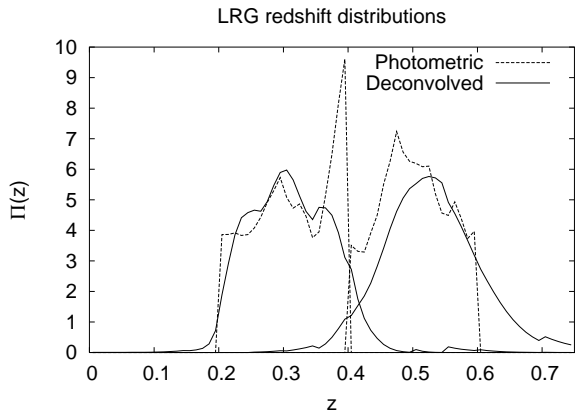}
\caption{\label{fig:lrg-dndz}The redshift distributions of the two LRG samples.  The dashed lines show the probability distribution for the photo-$z$s, where as the solid (``deconvolved'') lines show the smoothed true redshift distribution based on the reconstruction method of Padmanabhan et~al. \cite{padmanabhan05}.}
\end{figure}

Next we consider the photometric LRG sample from SDSS.  The sample is faint enough that spectroscopic redshifts are unavailable for most of the 
objects.  Fortunately, precise photometric redshifts are available for LRGs since they have very uniform spectra whose main broadband feature is a 
break at 400$\,$nm.  This break passes through the SDSS $g$ and $r$ filters in the interesting redshift range, so the $g-r$ and $r-i$ colors of an LRG 
correlate very strongly with its redshift \cite{padmanabhan05}.  The error distribution of the photometric redshifts has been calibrated using 
spectro-$z$s from the 2SLAQ survey\cite{cannon06}; this procedure, and an inversion method used to determine the actual redshift distribution given the photo-$z$ 
distribution, are described in Padmanabhan et~al. \cite{padmanabhan05}.  These methods were applied to determine the redshift probability distribution 
$\Pi_i(z)$ for the LRGs used in this sample.  The redshift distributions so obtained are shown in Fig.~\ref{fig:lrg-dndz}.

The bias is determined by the same $Q$-model fitting procedure as we used for 2MASS.  The maximum values of $\ell$ considered are 240 for the low-$z$
slice and 400 for the high-$z$ slice, which correspond to roughly $k\approx 0.3h\,$Mpc$^{-1}$ at the typical redshifts of these samples.  For the
fiduciual cosmology, the low-$z$ LRG slice gives a bias of $b=1.97\pm 0.05$ and $Q=21.7\pm 2.6$; the high-$z$ slice gives $b=1.98\pm 0.03$ and
$Q=17.1\pm 1.5$.  In order to reduce the possible impact of the nonlinear regime on our results, we also did fits where the maximum value of $\ell$ was
reduced by a factor of 2 or 4.  The results are shown in Table~\ref{tab:lrgbias} and the bias estimates are seen to be consistent with each other.  In
what follows we have used the original ($l_{\rm max}=240,400$) fits for the LRG bias, noting that the remaining uncertainty in $b$ is small compared
to the uncertainty (change in number of sigma detection is: 0.0043 (0.0388) for low-z LRG (high-z LRG)) resulting from statistical error in the ISW signal.  However we note that it is not clear how well the $Q$-model works for
LRGs at small scales, and we recommend more detailed analysis before taking the very small statistical error in $b$ at face value.  The $Q$-model fits 
are shown in Fig.~\ref{fig:loz}.

\begin{table}
\caption{\label{tab:lrgbias}The LRG bias and $Q$-parameter determined using several ranges of $\ell$.  The ``original'' value $l_{\rm orig}$ is 240 for
the low-$z$ slice and 400 for the high-$z$ slice.  The $Q$-values are reported in units of $h^{-2}\,$Mpc$^2$.}
\begin{tabular}{ccccc}
\hline\hline
Value of & \multicolumn{2}{c}{Low-$z$ slice} & \multicolumn{2}{c}{High-$z$ slice} \\
$l_{\rm max}$ & $b$ & $Q$ & $b$ & $Q$ \\
\hline
$l_{\rm orig}$ & $1.97\pm 0.05$ & $21.7\pm 2.6$ & $1.98\pm 0.03$ & $17.1\pm 1.5$ \\
$l_{\rm orig}/2$ & $2.03\pm 0.07$ & $16\pm 8$ & $1.96\pm 0.04$ & $21\pm 5$ \\
$l_{\rm orig}/4$ & $1.99\pm 0.12$ & $33\pm 45$ & $2.00\pm 0.07$ & $-12\pm 24$ \\
\hline\hline
\end{tabular}
\end{table}

For the LRGs -- unlike the 2MASS galaxies -- each of the two photo-$z$ slices covers a narrow redshift range and the threshold luminosity varies slowly
across that range, so we expect the bias to not vary significantly across the redshift range.  This expectation has been confirmed in previous angular
clustering studies which found $\sim 15$\% variation from $z=0.2$ to $z=0.6$ \cite{padmanabhan07}, and also by our own bias analysis which finds
no significant difference between the two bins.  
Thus we conclude that for the purposes of ISW work (where we have a $\sim 1.3 (2.7)$ sigma signal for low-z LRG (high-z LRG) correlation),
variation of the LRG bias within an individual photo-$z$ bin (0.2--0.4 or 0.4--0.6) can be neglected.

We calculate the possible contribution from magnification bias given the redshift distribution of the LRGs and 
also an assumed cosmology. We find that the possible contribution from magnification bias is $100-1000$ times 
(depending on the scale)
smaller than the actual signal.  Therefore magnification bias is not contributing significantly to
our signal. 

\subsection{SDSS quasars}

The function $f_i(z)$ for the quasars is more uncertain than for the LRGs.  This is in part due to the limited spectroscopic coverage available, but also the difficulty of constructing quasar photo-$z$s and the lower clustering amplitude, which leads to noisier estimates of bias parameters.  The basic procedure for obtaining $f_i(z)$ is thus to find a region of sky with as high spectroscopic completeness as possible while still retaining a large area; use this to obtain a preliminary estimate $\Pi(z)$; and then fit for the bias parameters using clustering data, of which several are needed if $\Pi(z)$ is multimodal.  The remainder of this section describes the details of the $f_i(z)$ determination and what possible errors can be introduced by spectroscopic incompleteness, stellar contamination, redshift-dependent bias, and cosmic magnification.

In order to determine the redshift probability distribution, we began by constructing a set of five rectangles that lie within the coverage area of the 
SDSS, 2QZ \cite{croom04}, 6QZ \cite{croom04}, and 2SLAQ \cite{richards05} surveys.  These rectangles lie along the equator (the declination range is 
$-01^\circ00'36''$ to $+00^\circ35'24''$) and cover the five RA ranges 137--143$^\circ$, 150--168$^\circ$, 185--193$^\circ$, 197--214$^\circ$, and 
218--230$^\circ$.  There is a significant amount of area with coverage from all surveys that is rejected as it was found to have lower completeness in 
2SLAQ because there is less plate overlap.  Spectra in SDSS were required to have high confidence (\verb"zConf"$>0.95$) \cite{stoughton02} and those in 
2QZ, 6QZ, and 2SLAQ were required to be of high quality (\verb"quality"$==11$) \cite{croom04}.

\begin{figure}
\includegraphics[angle=-90,width=3.2in]{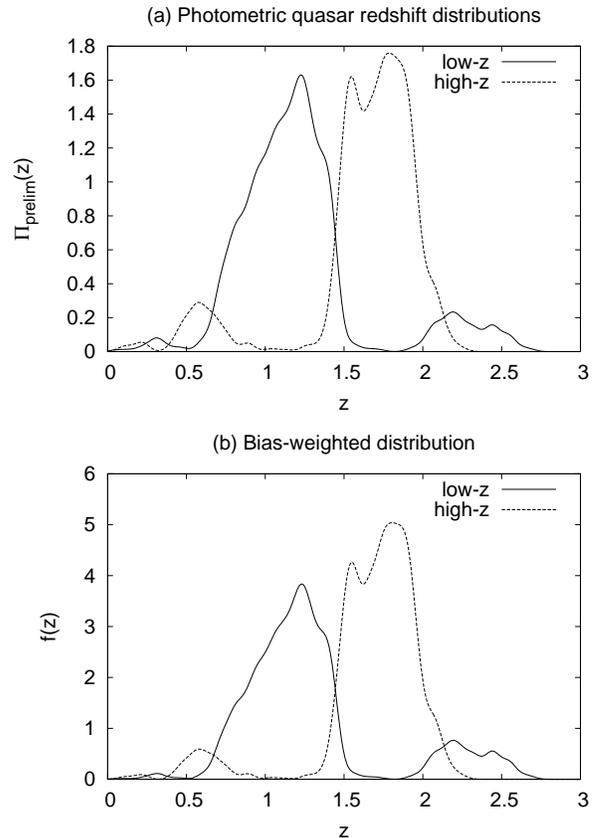}
\caption{\label{fig:qso-dndz}(a) The preliminary quasar redshift distribution, constructed from the successful matches to spectroscopic data. (b) The 
best-fit $f(z)$ for the two quasar samples as described in the text for the WMAP cosmology.}
\end{figure}

Our coverage rectangles contained a total of 1410 low-redshift and 1269 high-redshift photo-$z$ quasars; these numbers are lower than the product of the spectroscopic coverage area and the number density of photo-$z$ quasars because some parts of the latter catalogue were rejected by our stellar density cuts.  Of the low-redshift photo-$z$ quasars, we found that 257 (18\%) had no spectroscopic redshift determination or low quality ones, 58 (4\%) were identified as stars, and the remaining 1095 (78\%) are extragalactic.  For the high-redshift sample these numbers are 208 (16\%), 13 (1\%), and 1048 (83\%) respectively.  From this data we construct a preliminary redshift probability distribution $\Pi_{\rm prelim}(z)$ for each of the photo-$z$ slices using a kernel density estimator,
\beq
\Pi_{\rm prelim}(z) = \frac1{N_{\rm ex}}\sum_{k=1}^{N_{\rm ex}} \frac1{\sqrt{2\pi}\,\sigma}e^{-(z-z_k)^2/2\sigma^2},
\label{eq:prelim}
\eeq
where $N_{\rm ex}$ is the number of matches to extragalactic objects, $z_k$ is the redshift of the $k$th object, and $\sigma$ is the kernel width.  The
estimator is consistent in the limit that the number of objects $N_{\rm ex}\rightarrow\infty$ and $\sigma\rightarrow 0$ at fixed $N_{\rm ex}\sigma$.  
In practice, $\sigma$ must
be chosen to be small compared to the width of any real features in the redshift distribution (otherwise these are artificially smoothed out), and
large enough to smooth out shot noise (and redshift-clustering noise, if significant).  We have used $\sigma=0.04$ (using $\sigma = 0.02$ changes the 
fit bias by only 5\%).  This
preliminary distribution is shown in the top panel of Fig.~\ref{fig:qso-dndz}.
The redshift distributions in the two photo-$z$ quasar slices are multimodal due to the nature of the photo-$z$ error distribution: the quasar spectra redward of Lyman-$\alpha$ are usually characterized by a roughly power-law continuum with superposed emission lines.  This means that quasar colors oscillate as emission lines redshift into and out of the SDSS filters, resulting in an (approximately) self-intersecting locus in color space and many degeneracies in the photo-$z$ solution.

If the quasar bias were constant and magnification bias negligible, then we would have simply $f_i(z)=c\Pi_{\rm prelim}(z)$, with the proportionality
constant $c$ being the product of the bias and the probability for a photo-$z$ quasar to actually be extragalactic.  This constant could then be
determined by fitting the amplitude of the quasar autocorrelation function, as has been done in most past ISW studies.  However, in the real Universe
quasars are known to have an evolving bias, which is potentially significant across the redshift range considered, and at redshifts $z\sim O(1)$
lensing magnification can become significant.  The magnification can be calculated from the slope $\alpha$ of the quasar counts near the $g=21$
magnitude limit, which gives $\alpha=0.82$ for the low-$z$ sample and $\alpha=0.90$ for the high-$z$ sample. 
In principle the cut on the $u$-band magnitude error ($\sigma_u<0.5$)
could have an additional effect since magnification will reduce
$\sigma_u$; however this is not an issue for us since at the $g=21$
threshold, for UVX objects we will have $u<22$ where the typical magnitude
error is $<0.5$ even accounting for extinction ($A_{u,\,\rm max}=0.26$).
Since for these
samples $\alpha-1$ is small, we compute the magnification bias using $\Pi_{\rm prelim}(z)$ in place of the true distribution $\Pi(z)$.  That is, we
replace Eq.~(\ref{eq:fi}) with
\beqa
f_i(z)\!\!&\approx&\!\! b_i(z)\Pi_i(z)
\nonumber \\ &&\!\! + \int_z^\infty W(z,z')[\alpha(z')-1] \Pi_{i,\,\rm prelim}(z')dz'.
\eeqa
This leaves only the problem of constraining the product $b_i(z)\Pi_i(z)$ using the clustering data, i.e. the quasar power spectrum and quasar-LRG
cross-power.  Unfortunately the data is not capable of constraining a full model-independent distribution, so instead we write
\beq
b_i(z)\Pi_i(z)D(z) = A(z)\Pi_{i,\rm prelim}(z),
\eeq
where $D(z)$ is the growth factor, and $A(z)$ is a piecewise constant function of $z$.  This is equivalent to assuming that the clustering amplitude
(divided by spectroscopic completeness) of the quasars is constant in redshift slices, which has been found to be a better approximation than constant
bias in most quasar surveys \cite{myers06}.  For comparison, the empirical ``Model 3'' of Ref.~\cite{2001MNRAS.325..483C} predicts $b(z)D(z)$ to 
change by only 5\% from $z=0.65$ to 1.45, and by 13\% from $z=1.45$ to 2.00.  For the more recent model, Eq.~(15) of Ref.~\cite{2005MNRAS.356..415C}, 
these numbers are 24\% and 15\% respectively.  At higher redshifts ($z\ge 3$) there is a sharp increase in $b(z)D(z)$ \cite{2007AJ....133.2222S} but 
UVX-selected samples do not contain objects from this redshift range.

We constrain $A(z)$ in as many redshift slices as can be constrained using the data.  In particular since
the quasar redshift distributions are multimodal, we would like to be able to fit a different clustering amplitude in each peak.  The treatment of the
two quasar samples is slightly different due to the availability of different information in their redshift ranges, so we now discuss their redshift
distributions separately.  In each case, the autopower spectra were fit to linear theory up to $l=160$ ($k=0.1h\,$Mpc$^{-1}$ at $z=0.6$) and the
quasar-LRG cross-spectra were fit up to $l=140$ ($k=0.1h\,$Mpc$^{-1}$ at $z=0.5$).

\subsubsection{Low-$z$ sample: $0.65<z_{\rm photo}<1.45$}

For the low-$z$ quasar sample, we can only constrain one redshift slice.  An examination of Fig.~\ref{fig:qso-dndz} shows that the distribution is
actually trimodal, with peaks at $z=0.32$, $1.24$, and $2.20$.  A fit assuming a constant $A$ yields $A=1.36\pm 0.10$, with $\chi^2/$dof$=36.32/27$
($p=0.11$).  Almost all of the weight for this comes from the central ($z=1.24$) peak.  We also ran two-slice fits to determine
whether the clustering data constrain the amplitudes of the low- and high-redshift peaks.  The first such fit is of the form
\beq
A(z) = \left\{\begin{array}{lcl} A_1 & & z<0.52 \\ A_2 & & z\ge 0.52 \end{array}\right.,
\eeq
which allows the low-redshift slice to vary ($z=0.52$ is the local minimum of $\Pi_{i,\rm prelim}$).  This fit gives $A_1=4.74\pm 2.12$ and
$A_2=1.35\pm 0.10$, with $\chi^2/$dof$=33.77/26$.  We also tried a two-parameter fit in which the high-redshift slice is allowed to vary:
\beq
A(z) = \left\{\begin{array}{lcl} A'_1 & & z<1.83 \\ A'_2 & & z\ge 1.83 \end{array}\right.
\eeq
(the local minimum of $\Pi_{i,\rm prelim}$ between the main and high-redshift peaks is at $z=1.83$).  This fit gives $A'_1=1.37^{+0.09}_{-0.19}$ and
$A'_2=0.0\pm 8.7$ ($1\sigma$), with $\chi^2/$dof$=36.31/27$.  The errors on $A'_1$ are highly asymmetric in this case because the constraint comes
mainly from the quasar autopower; $A'_1$ and $A'_2$ are then degenerate because one only knows the total power, not how much comes from each redshift
slice.  The shape of the power spectrum breaks this degeneracy in principle, however in practice it is far too noisy.  The fact that the high-redshift
slice cannot give negative power accounts for the ``hard'' upper limit on $A'_1$.

From this exercise we conclude that the clustering data cannot independently measure the bias in either the low- or high-redshift peak.  The reasons
are different in each case.  The low-redshift peak contained only 1.7\% of the spectroscopic identifications, and thus almost certainly contains only a
very small fraction of our quasars.  This peak lies at the same redshift as the low-$z$ SDSS LRGs, and the quasar-LRG cross-correlation is the major
constraint on $A_1$.  Unfortunately this cross-correlation is drowned out by the enormous Poisson noise contributed by the quasars in the other two
peaks, and is detected at only $2.2\sigma$.  On the other hand, the LRGs oversample the cosmic density field on linear scales and cover the same region
of sky as the quasars.  One would thus expect that since the LRG-quasar correlation is only seen at this low significance, and the ISW effect from this
redshift range contributes only a small fraction of the power in the CMB, the contribution of the low-redshift peak to the quasar-ISW correlation would
be statistically insignificant.  
We find that the predicted peak of the quasar-ISW $l(l+1)C_\ell/2\pi$ for only the low-redshift peak quasars is 
lower than the entire sample (high-z QSO)
by $0.015\mu K$, which is significantly smaller than the error on the cross correlation.
This is run using a WMAP-3yr parameters.

The high-redshift peak contains 10\% of the quasars.  Its amplitude must be measured in autocorrelation due to the lack of other samples at that
redshift, which is a serious drawback since only 1\% of quasar pairs come from the high-redshift peak.  An alternative approach to constraining its
amplitude would be cross-correlation against the spectroscopic quasar sample at $2.0<z<2.5$, 
but we did not pursue this approach here.

\subsubsection{High-$z$ sample: $1.45<z_{\rm photo}<2.00$}

The high-$z$ photometric quasar sample also has a trimodal distribution: there is one peak at $z=0.22$, a second at $z=0.58$, and a third at $z=1.80$.
In this case however, it is the highest-redshift peak that contains most of the objects, with the middle peak in second place and only a few objects in
the lowest-redshift peak.  This situation makes it both possible and necessary to fit separate amplitudes for the peaks; in this case we will find that
two amplitudes can be constrained, one for the two low-redshift peaks and one for the main (high-redshift) peak.

As a first step, we attempt to fit all three of the peaks with separate amplitudes,
\beq
A(z) = \left\{\begin{array}{lcl} A_1 & & z<0.33 \\ A_2 & & 0.33\le z<1.18 \\ A_3 & & z\ge 1.18 \end{array}\right..
\eeq
This leads to the results $A_1=8.2\pm 4.5$, $A_2=1.34^{+0.68}_{-0.78}$, and $A_3=1.38^{+0.06}_{-0.14}$ (1$\sigma$), with $\chi^2/$dof$=23.58/25$.  The
large error bar on $A_1$ indicates that this parameter cannot be constrained from the data, so we instead try a two-slice fit in which we fix
$A_1=A_2$. This fit gives the tighter constraints $A_1=A_2=1.59\pm 0.61$ and
$A_3=1.35\pm 0.10$, with $\chi^2/$dof$=25.75/26$, and it is what we use for the rest of the paper.

\subsubsection{Redshift Distribution Summary}

\begin{figure*}
\includegraphics[angle=-90,width=6.5in]{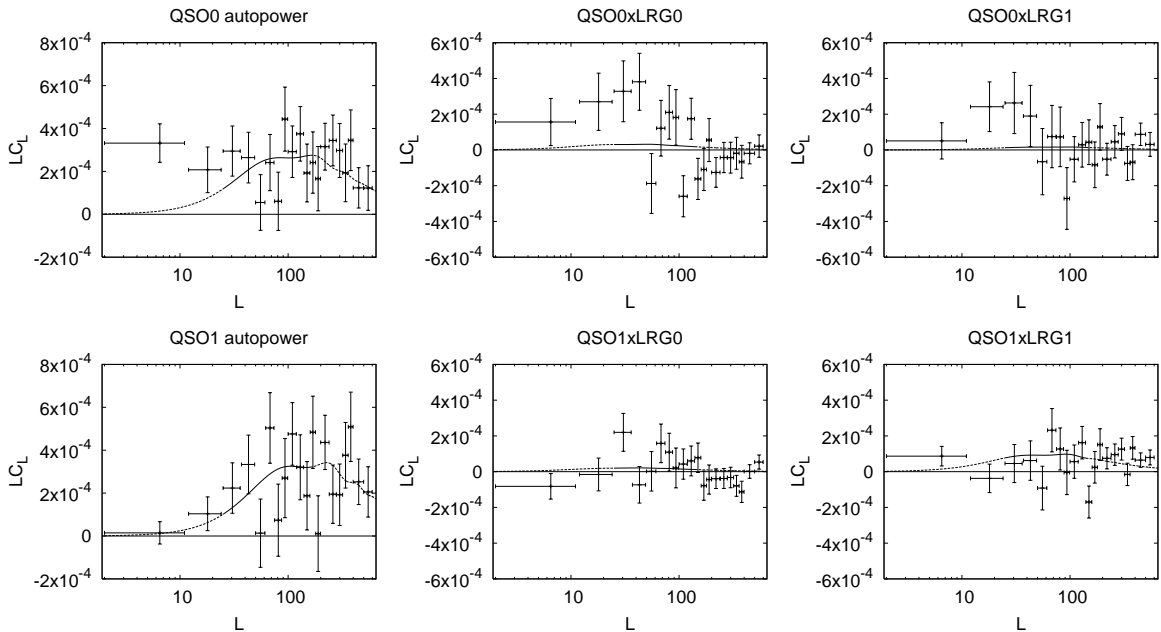}
\caption{\label{fig:fqso}The model fits to the power spectra of the quasars and their cross-correlation with the LRGs.  The low- and high-$z$ quasar
slices are denoted ``QSO0'' and ``QSO1'' respectively, and a similar nomenclature is used for the LRGs.  The model fits using linear theory are shown
with the solid lines over the range of multipoles used in the fit.  The dashed lines show the extension of the model across the remaining range of
multipoles.  Note that for the highest multipoles the linear theory is expected to break down.}
\end{figure*}

The quasar autopower spectra and quasar-LRG cross-spectra, along with the model fits, are shown in Fig.~\ref{fig:fqso}. 
For the QSO0 sample, there is excess power ($\sim 3\sigma$ above the
prediction) in the lowest-$l$ bin, corresponding to a $\sim 2$\% RMS
fluctuation in the number density on scales of $\sim 30$ degrees.  The two
most obvious sources of such power are stellar contamination and
photometric calibration errors.  Given that $\sim 5$\% of the photometric
``quasars'' are actually stars \cite{richards04} and that the
relative photometric calibration across the sky in SDSS is estimated to be
$\sim 2$\% in the $u$ band (the worst band, but one very important for
quasar work) \cite{padmanabhan07}, either of these seems plausible.
In any case, these very low multipoles were not used in fitting the
redshift distribution in either auto- or cross-power.

It is essential to test the robustness of the quasar fits, in particular against the possibility of nonlinear clustering affecting the range of
multipoles used in the fits.  The first way we do this is by repeating our analysis using the nonlinear matter power spectrum of Smith et~al.
\cite{smith03} in place of the linear power spectrum.  In the analysis with the nonlinear spectrum, the amplitude $A$ for the low-$z$
quasar slice increases by $+0.02$, and the amplitudes for the $z<1.18$ and $z\ge 1.18$ parts of the high-$z$ quasar slice increase by $+0.08$ and
$+0.02$, respectively.  If we restrict our attention to the lowest multipoles $l<100$ (instead of cutting at 140 or 160), these changes are $+0.02$,
$-0.14$, and $+0.03$.  In each case the change is very small compared with the error bars.  Thus we do not believe that nonlinear clustering is
affecting our $f_{\rm QSO}(z)$ estimates.

\subsection{NVSS}

The function $f(z)$ for NVSS is the hardest to obtain because there are no spectroscopic samples of NVSS objects that have sufficiently high
completeness to obtain the redshift distribution.  Past ISW analyses \cite{boughn04, nolta04} with the NVSS have been based on
the radio luminosity function $\Phi(L,z)$ of Dunlop \& Peacock \cite{dunlop90}, which itself was fit to a combination of source counts,
redshifts for some of the brightest sources, and the local luminosity function.  A constant bias was then assumed.  The redshift distribution so
obtained is reasonable, however it has three major drawbacks: (i) the redshift probability distribution $\Pi(z)$ for the faint sources (which make up
most of the sample) is constrained only by the functional form used for the luminosity function and not by the data; (ii) it does not give the redshift
dependence of the bias, which could be very important since the redshift range is broad, and the typical luminosity of the sources varies with
redshift; and (iii) the absolute bias $b$ is constrained using the NVSS autopower spectrum, which is known to contain power of instrumental origin
and hence is probably a less reliable constraint than the cross-correlation against other surveys.  The alternative method to measure $f(z)$ is by
cross-correlation against the other samples whose redshift distributions are known.  This method is adopted here, since it does not have any of the
aforementioned problems.  Its main drawback is that the other samples only probe the range out to $z\sim 2.6$, and little data is available to
constrain $f(z)$ above that.  

\subsubsection{Procedure}

\begin{figure*}
\includegraphics[angle=-90,width=6.5in]{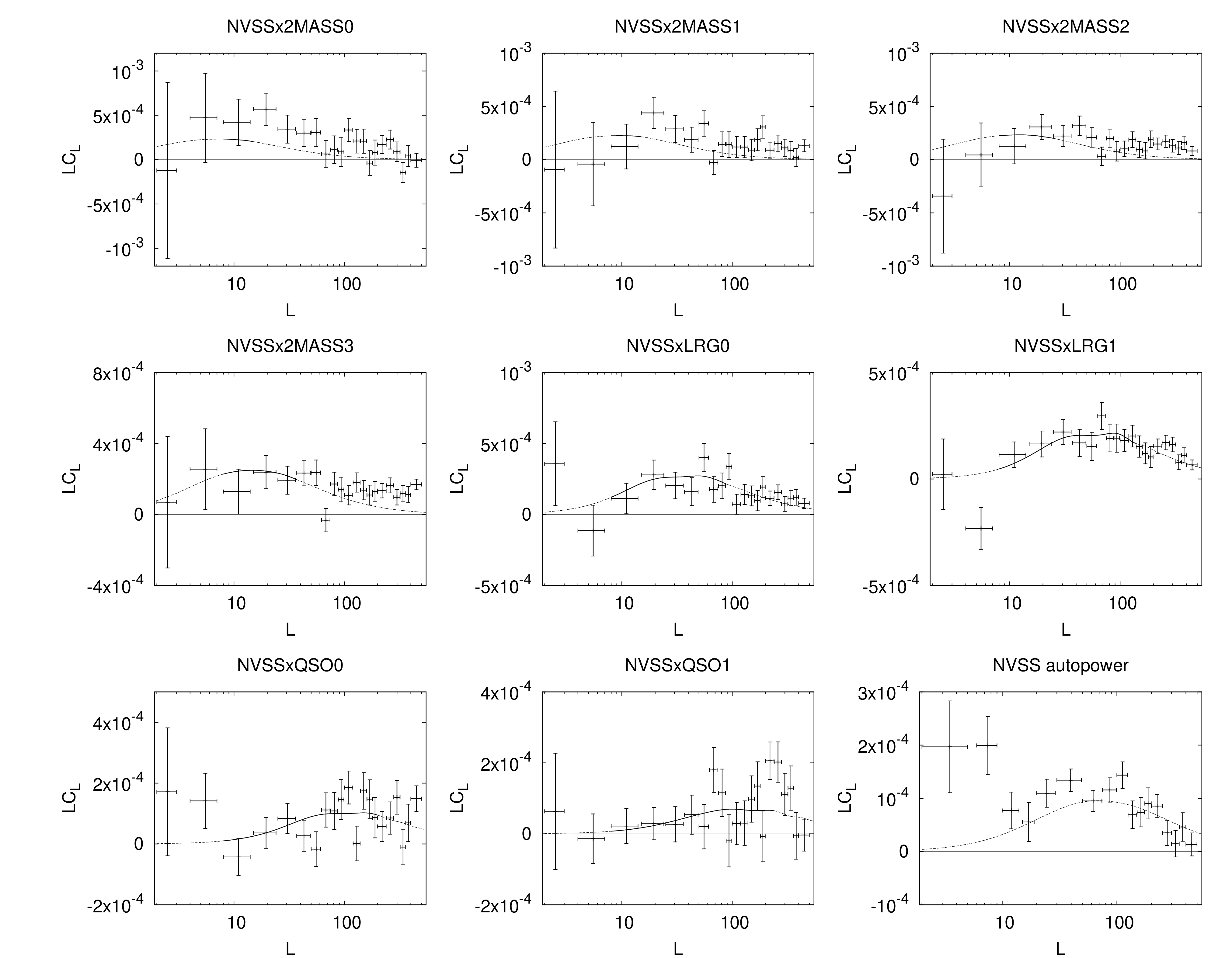}
\caption{\label{fig:xnvss}The cross-spectra of NVSS with the other samples.  The solid lines show the linear theory predictions in the region used for
the fits, and the dashed lines show the extension to higher or lower multipoles.  Note that for the highest multipoles linear theory is not valid.}
\end{figure*}

In order to measure the effective redshift distribution of NVSS, we must first obtain the cross-correlation of NVSS with each of the eight other
samples (the four 2MASS samples, and two samples each of LRGs and quasars).  This is done by using the same angular cross-spectrum estimation method as
was used for the ISW analysis, and the cross-spectra are shown in Fig.~\ref{fig:xnvss}.  The main subtlety that arises is that the cross-spectrum 
$C_\ell^{ij}$ (where
$i$ and $j$ are LSS samples) can actually contain Poisson noise if there are objects that are in both samples.  The Poisson noise term is of the form
\beq
C_\ell^{ij} = C_\ell^{ij}({\rm LSS}) + \frac{\bar n_{ij}}{\bar n_i\bar n_j},
\label{eq:poisson}
\eeq
where $\bar n_i$ is the number of sources per steradian in catalog $i$, and $\bar n_{ij}$ is the number of sources per steradian that appear in both
catalogs. In order to measure $\bar n_{ij}$ we must match the NVSS to each of the other samples.  Note that the
positional errors in NVSS are typically several arc seconds, and consequently there will always be some false matches.  Therefore we estimate the
fraction of matches as
\beq
\frac{\bar n_{i,\rm NVSS}}{\bar n_i}
= \frac{N_{\rm match}}{N_i}-\pi\theta_{\rm max}^2\bar n_{\rm NVSS},
\eeq
where $N_{\rm match}$ is the number of matches within some radius $\theta_{\rm max}$, and $N_i$ is the number of sources in catalog $i$ in the NVSS
mask.  This was estimated for radii $\theta_{\rm max}$ of 40 and 20 arcsec, and the results are shown in Table~\ref{tab:clmatch}.

\begin{table}
\caption{\label{tab:clmatch}Details of the cross-correlation of NVSS with the eight other samples.
The second and third columns show the fraction of objects in each of the samples that match to the NVSS, i.e. $\bar n_{i,\rm NVSS}/\bar n_i$.  Results
are presented for two matching radii, 40 and 20 arcsec.  The final two columns show the range of multipoles used in the cross-correlation.}
\begin{tabular}{ccccccrr}
\hline\hline
Sample & & \multicolumn{3}{c}{$\bar n_{i,\rm NVSS}/\bar n_i$} & & \multicolumn{2}{c}{Multipoles used} \\
 & & $40''$ & & $20''$ & & $l_{\rm min}$ & $l_{\rm max}$ \\
\hline
2MASS $12.0<K_{20}<12.5$ & & $0.1317$ & & $0.1302$ & & 10 &  14 \\
2MASS $12.5<K_{20}<13.0$ & & $0.0802$ & & $0.0787$ & & 10 &  14 \\
2MASS $13.0<K_{20}<13.5$ & & $0.0473$ & & $0.0455$ & & 10 &  24 \\
2MASS $13.5<K_{20}<14.0$ & & $0.0292$ & & $0.0280$ & & 10 &  36 \\
SDSS LRG  low-$z$     & & $0.0450$ & & $0.0425$ & & 10 &  87 \\
SDSS LRG  high-$z$    & & $0.0263$ & & $0.0249$ & & 10 & 139 \\
SDSS QSO  low-$z$     & & $0.0180$ & & $0.0192$ & & 10 & 239 \\
SDSS QSO  high-$z$    & & $0.0189$ & & $0.0207$ & & 10 & 159 \\
\hline\hline
\end{tabular}
\end{table}

We next computed the cross-power spectra between NVSS and each of the other samples.  These spectra (after subtraction of the Poisson term) are shown
in Fig.~\ref{fig:xnvss}.  The redshift distribution was then fit to the cross-power spectra.  In this fit the minimum multipole used is $l_{\rm
min}=10$ (below which there is a large amount of spurious power in the NVSS map) and the highest-$l$ bin used was determined by the formula $l_{\rm
max}=k_{\rm max}D_{A,20}$, where $k_{\rm max}=0.1h\,$Mpc$^{-1}$ is the smallest scale to be fit and $D_{A,20}$ is the distance corresponding to the
20th percentile of the window function for that sample as defined in Appendix~\ref{app:nvsswin}.  We have fit $f_{\rm NVSS}(z)$ with a
$\Gamma$-distribution,
\beq
f_{\rm NVSS}(z) = \frac{\alpha^{\alpha+1}}{z_\star^{\alpha+1}\Gamma(\alpha)} b_{\rm eff}z^{\alpha}e^{-\alpha z/z_\star}.
\eeq
This function has three free parameters, $b_{\rm eff}$, $z_\star$, and $\alpha$.
Of these the normalization $b_{\rm eff}$ may be viewed as an effective bias in the sense that $\int f_{\rm NVSS}(z)\,dz=b_{\rm eff}$; in the absence of
cosmic magnification this would be the bias averaged over the redshift distribution.  The peak of the distribution is at $z_\star$, and $\alpha$
controls the width of the distribution.  The parameter fit gives $b_{\rm eff}=1.98$, $z_\star=0.79$, and $\alpha=1.18$.  


\subsubsection{High-redshift tail}

The above analysis of the NVSS distribution involved cross-correlations against 
several samples at $0<z<2$.  (The QSO0 sample has a small number of
objects at $2.0<z<2.6$, however they have no significant impact on the fitting of
the QSO0$\times$NVSS cross-spectrum.)  Thus it leaves open the issue
of whether there is a tail of objects at high redshift, $z>2$.  Since $f(z)$ is a
product of bias times redshift probability distribution, it need not
be normalized -- $\int f(z)\,dz$ can have any value -- so there is no way to tell
from the cross-correlation analysis alone whether a portion of the
sample is missing.  If we also use the NVSS autopower spectrum then in principle 
one can determine whether an additional source of angular fluctuations
is necessary.  However the angular clustering at fixed angular scale $l$ is much 
stronger at low than high redshift, and the NVSS autopower spectrum is
of low signal-to-noise ratio and possibly contaminated by systematics, so we have
not chosen this strategy.

An alternative approach to the high-$z$ tail is to directly match against optical/
NIR catalogs.  One can then use the $m_K-z$ relation or (if multiband
imaging is available) photometric redshifts.  There are always some radio sources
without optical identifications, however this method enables one to
set an upper limit to the number of NVSS sources that can be at high redshift.  
For our analysis, we have matched against the COSMOS field, which has a
modest solid angle (2 deg$^2$), multiband imaging allowing good photometric redshifts, 
and deep high-resolution coverage with the VLA.  Area is
required due to the low density of NVSS sources (40 deg$^{-2}$), and high-resolution 
radio images are required to uniquely identify an NVSS source with
an optical counterpart due to the large positional uncertainty in the NVSS ($\sim
7$ arcsec for faint sources) \citep{condon98}.  

The COSMOS field contains 87 NVSS sources that pass our cuts.  We began by matching 
these to the VLA-COSMOS observations, which are much deeper and
have typical positional uncertainties of $\sim 0.2$ arcsec \cite{Schinnerer07}.
Of the NVSS sources, 79 have a match within 30 arcsec (we take
the nearest source in the event of multiple matches).  The 79 VLA-COSMOS sources 
that match to NVSS are then matched to the optical catalog
\cite{capak07}; there are 64 successful matches within 1 arcsec.  This
 represents 74\% of the original NVSS catalog.  It is of course
possible that there are some false matches.  By adding up $\bar n\pi\theta^2$ for
each NVSS source, where $\bar n$ is the density of VLA-COSMOS sources
and $\theta$ is the distance to the nearest VLA-COSMOS source (or 30 arcsec if the
 NVSS source had no match), we estimate that there are $\sim 5$ false
NVSS/VLA-COSMOS matches.  A similar argument suggests that $\sim 0.5$ false matches of VLA-COSMOS to the COSMOS optical/NIR catalog.  Thus we expect
that 58.5 of the matches are correct, corresponding to 67\% of the initial NVSS catalog.

\begin{figure}
\includegraphics[angle=-90,width=3.3in]{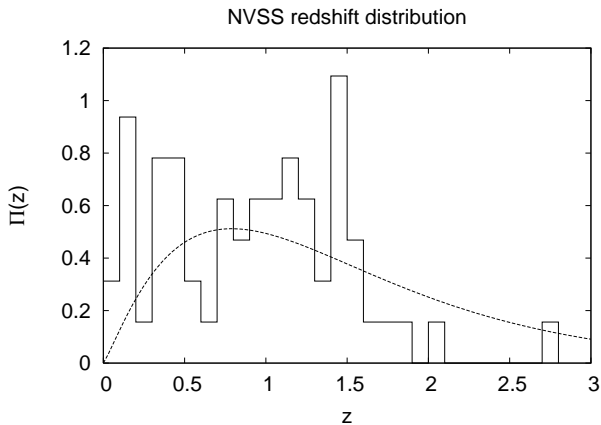}
\caption{\label{fig:nvss-hist}The redshift histogram of NVSS sources matched to COSMOS using the Mobasher et~al. \cite{mobasher07} photometric
redshifts.  The dashed line is the fit three-parameter $f_{\rm NVSS}(z)$, normalized to unity (i.e. the redshift distribution assuming constant bias
and negligible effect from magnification).}
\end{figure}

We show the photometric redshift distribution of the matches (according to Mobasher et~al. \cite{mobasher07}) in Fig.~\ref{fig:nvss-hist}.
Our best-fit $f_{\rm NVSS}(z)$ (with the $\Gamma$ distribution) has 24\% of the bias-weighted source distribution at $z>2$ and 8\% at $z>3$; if the
source bias increases with redshift, as usually found for optical quasars, this number would be lower.  From Fig.~\ref{fig:nvss-hist} we see that only
2 out of 64 matches fall at $z>2$, i.e. the high-redshift tail of the $\Gamma$ distribution can only exist in reality if (i) most of the 26\% of the
sources with failed matches to COSMOS optical/NIR data are actually at $z>2$, or (ii) the sources at $z>2$ have a large bias.  Both (i) and (ii) are
physically plausible but we have no direct evidence for them.

The conservative solution in this case is to consider two limiting cases for the redshift distribution of the sources at $z>2$.  One case, which gives
the minimal lensing signal for all cosmologies, and the minimal (maximal) ISW signal for $\Lambda$CDM (closed) cosmologies, is to set $f_{\rm NVSS}=0$
at $z>2$.  In the opposite limiting case, we have assumed that all failed and incorrect NVSS matches, and all sources with $z_{\rm photo}>2$ (i.e. a
total of 35\%) are at $z>2$, and have four times the clustering amplitude measured for the optical quasars (QSO1 sample), e.g. $b(z)=4\times 1.35/D(z)$
(where $D$ is the growth factor) for the fiducial cosmology; the shape of $f_{\rm NVSS}(z)$ at $z>2$ was left unchanged from the $\Gamma$-distribution
fit.  
In order to understand the change of ISW and CMB-lensing signals due to changes of our assumption 
of the high-z end of the redshift distribution of NVSS, we look at two different redshift distributions,
one with nothing at $z>2$ (minimal model) and the other with a ''maximal'' number of sources (assuming
clustering strength 4 times of the optical quasars and all the failed optical IDs are at $z>2$). 
We find that the signals for both ISW (average: $7.8\%$) and CMB-Lensing change by less than 10\%, therefore, 
one won't expect the unidentified high-z tail of the NVSS sources be a problem in our analysis.

\subsubsection{Constraints, robustness, and alternatives}
\label{sss:cra}

\begin{figure}
\includegraphics[angle=-90,width=3.2in]{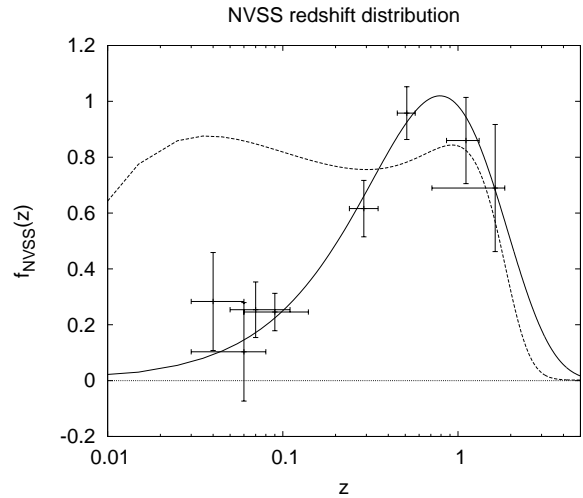}
\caption{\label{fig:fnvss}The constraints on the NVSS redshift distribution from the cross-correlations with the other eight samples.  The horizontal
error bars show the redshift window functions as described in the text.
The dashed line shows the result of using the redshift distribution based on the Dunlop \& Peacock \cite{dunlop90} luminosity function
assuming constant bias and neglecting magnification, as has been done in most ISW studies.}
\end{figure}

While the fit parameters are formally determined by the $\chi^2$, it is useful to graphically display the constraints in order to show what parts of
the distribution are constrained by which data.  This we have done in Fig.~\ref{fig:fnvss}.  For each of the eight samples, we have plotted on the
vertical axis the constant $f_{\rm NVSS}$ value that provides the best fit to cross-correlation with that sample and its $1\sigma$ error bar.  The
horizontal position is determined by the following procedure.  We show in Appendix~\ref{app:nvsswin} that the estimated constant $\hat f_{\rm NVSS}$ is
actually given by an integral over some window function,
\beq
\langle \hat f_{\rm NVSS} \rangle = \int_0^\infty {\cal W}(z) f_{\rm NVSS}(z)\,dz,
\label{eq:fhat}
\eeq
where the window function ${\cal W}(z)$ integrates to unity.  The horizontal position of the data points in Fig.~\ref{fig:fnvss} is the median of the
window function, i.e. the redshift $z$ where $\int_0^z{\cal W}(z')\,dz'=1/2$.  The error bars extend from the 20th to the 80th percentile of the window
function.

Finally we wish to compare the redshift distribution we have obtained to that used in previous ISW studies.  The previous results were based on the
radio luminosity function of Dunlop \& Peacock \cite{dunlop90}.  In each case, it appears that the authors used the luminosity function and
$k$-correction based on the spectral index to infer the redshift distribution, assumed constant bias and negligible magnification, and determined the
one free parameter (the bias) by fitting to the autopower spectrum.  If we do this using the fiducial WMAP cosmology and our autopower spectrum we find
$b=1.7$, and the function $f_{\rm NVSS}(z)=b\Pi(z)$ obtained is shown as the dashed line in Fig.~\ref{fig:fnvss}.  This curve, while roughly consistent
with the NVSS-quasar and NVSS-LRG correlations, badly overpredicts the NVSS-2MASS correlation.  Note that the problem cannot be fixed by changing the
single bias parameter: if $b$ were reduced by a factor of $\sim 3$ to fit the 2MASS data, then the LRG and quasar data would be discrepant.  

There are several possible explanations for this:
\newcounter{dp90}
\begin{list}{\arabic{dp90}. }{\usecounter{dp90}}
\item\label{it:2mag} The shape of $f_{\rm NVSS}(z)$ is being modified by
magnification bias.
\item\label{it:2err} The extrapolation of the luminosity function to faint
sources at high redshift by Dunlop \& Peacock is in error.
\item\label{it:2bias} It is possible that the Dunlop \& Peacock redshift
distribution accurately describes the NVSS sources, but the bias increases
with
redshift so as to produce the shape seen in Fig.~\ref{fig:fnvss}.
\item\label{it:2ext} The cut imposed by us (and by other ISW groups) that
requires NVSS sources to be unresolved is selecting against nearby
objects, and
hence pulling down the low-$z$ part of the $f_{\rm NVSS}(z)$ curve.
\end{list}

Of these, possibility number \ref{it:2mag} is easy to rule out.  Application of
Eq.~(\ref{eq:fi}) implies that $f_{\rm NVSS}(z)$ has a maximum change due
to
magnification bias of $0.09|\alpha-1|$ ($z=0.55$), and a smaller change at
lower redshift ($0.03|\alpha-1|$ at $z=0.1$), where $\alpha=-d\log N/d\log
F$ is
the source count slope.  The NVSS point source counts suggest a slope of
$0.99$ between 2.5 and 5.0 mJy, and $0.95$ between 5 and 10 mJy, which
suggests that
the effect of magnification bias on $\Delta f_{\rm NVSS}(z)$ is at most of
order $0.01$.  In order to accommodate the discrepancy of $\Delta f_{\rm
NVSS}(z)$
between our result and the Dunlop \& Peacock distribution of $\sim 0.6$ at
$z<0.1$, we would need an absurd slope, $\alpha\approx-20$.

Distinguishing among the remaining three possibilities is harder.  We
believe possibility number \ref{it:2err} is unlikely because the discrepancy
between Dunlop
\& Peacock and our work occurs at low redshift where their luminosity
function should be most reliable: this regime is constrained by the local
source counts
rather than by extrapolation.  Redshift-dependent bias (possibility
number \ref{it:2bias}) exists for most samples of objects and there is no reason
to expect it
to be absent for NVSS.  However, based on the Dunlop \& Peacock $dN/dz$
and our $f_{\rm NVSS}(z)$, the bias would have to change from $\sim 0.4$
at $z=0.1$
to $\sim 2$ at $z=0.5$.  Such a large variation, combined with the
unusually low value of the bias at $z=0.1$, suggests that this is not the
full
explanation.  The final possibility (\ref{it:2ext}) is the removal of
extended sources.  This is hard to assess because of the low density of
extended NVSS
sources above our flux cut ($\sim 8\,$deg$^{-2}$).  Of the 20 such sources
in the COSMOS field, 19 match to VLA-COSMOS and 13 of these matches are
found in
the COSMOS optical/NIR catalog.  It is worth noting that 8 of these (62\%)
have $z_{\rm photo}<0.5$, versus 30/64 (30\%) for the unresolved NVSS
sources.
This appears to go in the right direction, however it is difficult to make
quantitative statements about whether the extended sources actually
resolve the
discrepant redshift distributions because of the unknown (but probably
large, especially for the low-$z$ part of the distribution) sampling
variance error
bars.

In summary, while the full explanation for the difference between our
$f_{\rm NVSS}(z)$ and that of Dunlop \& Peacock remains unknown, it seems
likely (based
on process of elimination) that a combination of redshift-dependent bias
and our rejection of the unresolved NVSS sources plays a role.
Magnification bias
is ruled out as the explanation, and the discrepancy occurs in a regime
where the extrapolations used in Dunlop \& Peacock probably do not matter.

\section{\label{sec:sys} Systematics}

We investigate various systematic effects in our correlations utilizing 
a specific multipole range. We choose these multipole bins based on two 
criteria. First, they should not be affected by non-linearities. Second, they should
not be affected by any of the systematic effects in a significant way. We therefore 
only utilize the multipoles corresponding to $k\le 0.05h\,$Mpc$^{-1}$ and we also discard the first $\ell$-bin for all samples since it is affected by 
the 
galactic foreground contamination. The 
specific $l$-bins that are utilized are tabulated in Table~\ref{tab:szcontam}.

\subsection{Dust Extinction} 

Since it is possible that incorrect dust extinction systematically adds signals to our 
ISW cross correlation, 
we cross correlate the reddening maps \cite{schlegel98} in the same 
manner as we cross correlate each of our sample to the cosmic microwave background.
If there is a systematic effect contributed via dust extinction, it will
show up as a correlation, we can then estimate the effect and correct it from 
our tracer-cmb correlation.

In order to the verify that dust extinction does not 
affect our results, we constructed a vector ${\bf f}$ of the estimated spurious cross-spectra $\Delta C_\ell^{gT}$.  The spurious cross-spectra were 
computed by taking the cross-power spectrum of the CMB with the reddening map and multiplying by an estimate of $d\delta_g/dE(B-V)$.  Note that ${\bf 
f}$ has an entry for each $\ell$-bin for each sample, so it has a total length of 42.  We then compute the quantity
(the derivation of this quantity and its relevance to understand
contamination from extinction is detailed in Appendix~\ref{app:foreg}):
\begin{equation}
E_{ext} = {\bf f}^T C^{-1} {\bf f}.
\label{eq:fgderiv}
\end{equation}
Here ${\bf C}$ is the total $42\times 42$ covariance matrix that is generated using looking 
at the covariances of the correlation with each tracer sample and the 
Monte-Carloed CMB temperature map (the MC1 procedure in the terminology of Cabr\'e et~al. \cite{cabre07}; see Sec.~\ref{ss:isw-likelihood} for 
details).

Here $\sqrt{E_{ext}}$ is the maximum number of sigmas at which the effects of dust extinction could be detected if we knew all cosmological and 
redshift distributions 
perfectly; if $E_{ext}\ll 1$ then the dust extinction cannot have any statistically significant effect on any quantity derived from the cross-power 
spectrum, including cosmological parameter estimates.
We estimate that $d\delta_g/dE(B-V)$ = $-0.1$ (all 2MASS samples).  For the SDSS samples we did a Poisson-weighted fit to the LRG and quasar 
overdensities versus $E(B-V)$ (see Fig.~\ref{fig:syst}); this gives
$-0.76$ (low-z LRGs), $-0.18$ (high-z LRGs), $-1.06$ (low-z QSOs), 
and $-0.26$ (high-z QSOs).  (The Poisson error bars are all within $2\sigma$ of zero so there is no evidence that any of these derivatives is nonzero.)  
We ignore extinction for NVSS since it is at radio frequencies.
This gives $\sqrt{E_{ext}} = 0.23$, so the dust extinction is not having a significant effect.

\subsection{Galactic foregrounds}

To test whether galactic foreground contamination is important in our analysis, 
we cross correlate the templates of Galactic emission with the tracer overdensity maps.
The galactic foregrounds that must be considered in
producing a template at higher frequencies are
free-free and thermal dust emission; at lower frequencies
an additional component is present
whose physical origin remains uncertain but which may
include hard synchrotron emission \citep{bennett03} or spinning or
magnetic dust \citep{oliveira-costa04,finkbeiner03}.  We have used Model 8 of
\citet{schlegel98,finkbeiner99} for thermal dust and the H$\alpha$ line radiation
template of \citet{finkbeiner03} rescaled using the conversions of
\citet{bennett03} for free-free radiation (see \cite{hirata04} for further 
details). 
We then construct these maps in the same way as in WMAP temperature maps. 
Cross correlations between these templates with each of the tracer overdensity maps 
are then performed.

To understand the foreground contamination to our result we compute as above
\begin{equation}
E_{fg}= {\bf f}_{fg}^{T} {\bf C}^{-1} {\bf f}_{fg}
\end{equation}
where ${\bf f}_{fg}$ is the vector of cross-power spectra of the LSS and foreground maps, and
${\bf C}$ is the Monte Carlo covariance matrix.
Calculating the $\sqrt{E_{fg}}$ we find that the low multipoles of
some of the low redshift samples correlates with the galactic foreground. 
We investigate this further and realize that there is incidentally a 
low redshift cluster at low latitude, thus correlating with the foreground
map. We therefore restrict our l-range that contributes to our signal 
by leaving out the first  multipole bins for all sample. For the remainder we
get $\sqrt{E_{fg}}= 0.66$. 

\subsection{Thermal SZ effect}

The thermal Sunyaev-Zeldovich (tSZ) effect has a relatively weak frequency dependence compared to the Galactic foregrounds, so we constrain it from 
theoretical models.  We look at the tSZ signal using the
halo model, separating the effect of the tSZ signal into 1-halo term
and 2-halo terms.

The 1-halo term stands for the situation when the flux added towards
the CMB map via tSZ effect comes from the same halo as the one that
hosts the galaxies that we are correlating them with.
The theoretical prediction for the 1-halo term is:
\begin{equation}
\label{eq:1haloa}
C^{tSZ}_{\ell}(1h) = \sum_N \int dF\, \frac{NF}{\bar{n_g}} n_{2D}(N,F)
\end{equation}
where $N$ is the number of galaxies in that halo, $F$ is the flux
from the halo, $\bar{n_g}$ is the average number of galaxies,
$n_{2D}(N,F)$ is the number of halos with $N$ galaxies and flux between $F$ and $F+dF$.
We then turn Eq.~(\ref{eq:1haloa}) into integrals over halo mass and comoving distance:
\begin{equation}
C^{tSZ}_{\ell}(1h) = \int \frac{d\chi}{r^2} \int dM \frac{M}{\rho_0} \phi(M) \frac{N(M)}{\bar{n_g}} F(M,\chi),
\end{equation}
where $\phi(M)$ is the fraction of the mass in haloes between $M$ and $M+dM$, $N(M)$ is the mean number of galaxies in a halo of mass $M$, and $F$ is 
the flux from a halo of mass $M$ at comoving distance $\chi$.

The 2-halo term stands for situation when the flux (from tSZ) comes
from a different halo which hosts galaxies that cross-correlate with
the flux.  It is
\begin{equation}
\label{eq:2haloa}
C^{tSZ}_{\ell}(2h) = \sum_N \int dF \frac{NF}{\bar{n_g}} n_{2D}(N) n_{2D}(F) C_\ell(N;F),
\end{equation}
where $n_{2D}(N)$ is the number of halos with $N$ galaxies
per steradian,  $n_{2D}(F)$ is the number of halos with flux between $F$ and $F+dF$
per steradian and $C_\ell(N;F)$ is the cross-power spectrum between halos
with $N$ galaxies and those with flux $F$.
We then turn the Eq.~(\ref{eq:2haloa}) into integrals over the mass functions and
cosmological distances:
\begin{eqnarray}
C^{tSZ}_{\ell}(2h) &=& \int \frac{d\chi}{r^2} \int dM \frac{M}{\rho_0} \phi(M) b(M)
\nonumber \\ && \times f(\chi) P_{lin}(k) F(M,\chi),
\end{eqnarray}
where $P_{lin}(k)$ is the 3-D linear matter power spectrum.

Now, what is left for us to do is to figure out what the flux $F$ is for
tSZ effect.  One should note that this method is not limited to
the tSZ effect prediction, but any kind of correlations between galaxy number
overdensity and flux of any kind associated with the halos.
For tSZ effect, the flux is
\begin{equation}
F = 2\bar{\tau'} T_{\rm CMB} \frac{f_{\rm ICM}}{f_b} \frac{k_B T_e(M)}{m_e c^2},
\end{equation}
where $\bar{\tau'}$ is the mean Thomson optical depth per unit comoving distance,
$T_{\rm CMB}$ is the observed averaged CMB temperature,
$f_{\rm ICM}$ is the baryon fraction in the intracluster medium,
$f_{b}$ is the cosmic baryon fraction,
$k_B$ is the Boltzmann coefficient,
$T_e(M)$ is the average temperature of electrons inside halos of mass M,
$m_e$ is the mass of electrons, $c$ is the speed of light.

In order to assess the effect of tSZ on the ISW correlation, we
calculate the $C^{tSZ}_{\ell}(1h)$ and $C^{tSZ}_{\ell}(2h)$
with a high $\sigma_8$ (0.92) in order to give a conservative estimate.
We must also estimate $N(M)$.
For the 2MASS samples, we use $N(M)$ of the satellites and the conditional luminosity
function from \cite{lin04} while assuming that there
is 1 BCG per cluster. This is a conservative estimate as some of the
BCGs may fall out of the flux limit.
For the LRGs, we use $N(M)$ from \cite{ho07} for our calculation without modification,
as we use the same galaxy sample.
The quasars and NVSS are both examples of active galactic nuclei, and are generally found in haloes of some mass range with a small probability [i.e. 
$\langle N\rangle(M)<1$] usually interpreted as the duty cycle.
For these cases, we first obtain the redshift distribution ($dN/dz$) and bias.
For NVSS, we assume that bias $\propto 1/D(a)$ where $D(a)$ is the growth
factor of scale factor $a$, as there is no better available information (our determination of $f(z)$ is not capable of separately distinguishing the 
bias from the redshift distribution).
From the bias, we constrain the minimum halo mass that will host a QSO or NVSS object,
and then obtain the duty cycle based on $dN/dz$.
Duty cycles cannot exceed unity, so we cap $f_{duty}$ at 1 and above this use $dN/dz$ to get minimum halo mass. Then, $N(M) = f_{duty}$  if $M> 
M_{min}$ and
0 otherwise.

We assess the level of contaminations by calculating
\begin{equation}
E_{tSZ} = C^{tSZ}_{\ell}(1h+2h) {\bf C}^{-1} C^{tSZ}_{\ell}(1h+2h),
\end{equation}
which is the tSZ analogue to Eq.~(\ref{eq:fgderiv}).
We find that $\sqrt{E_{tSZ}} = 0.109$ using the $\ell$-bins that are tabulated in Table~\ref{tab:szcontam} and thus thermal SZ effect is not a 
significant 
contamination for the ISW effect.

We present our results for the tSZ contamination for the $l$-bins that
we use in our analysis  the cosmological parameter estimation in
Table~\ref{tab:szcontam}.

\begin{table}
\caption{\label{tab:szcontam}The tSZ and point source contamination for each of the samples we used in the analysis.  For tSZ the 1 halo and 2 halo 
terms are shown separately and combined.}
\begin{tabular}{clrlrlrlrlr}
\hline\hline
Sample & & $\ell$ & & \multicolumn{7}{c}{\mbox{$[l(l+1)/2\pi]C_\ell^{gT}$ ($\mu$K)}} \\
 & & & & tSZ 1h & & tSZ 2h & & tSZ 1+2h & & pt src \\
\hline
2MASS0  	& &	6	& & $	-0.0085	$ & & $	-0.0458	$ & & $	-0.0543	$ & & $	-0.4056	$ \\
2MASS1  	& &	6	& & $	-0.0048	$ & & $	-0.0324	$ & & $	-0.0372	$ & & $	-0.0743	$ \\
2MASS1  	& &	11	& & $	-0.0151	$ & & $	-0.0574	$ & & $	-0.0725	$ & & $	0.0070	$ \\
2MASS2  	& &	6	& & $	-0.0027	$ & & $	-0.0241	$ & & $	-0.0268	$ & & $	-0.0875	$ \\
2MASS2  	& &	11	& & $	-0.0086	$ & & $	-0.0458	$ & & $	-0.0544	$ & & $	0.0216	$ \\
2MASS3  	& &	6	& & $	-0.0016	$ & & $	-0.0182	$ & & $	-0.0198	$ & & $	-0.1717	$ \\
2MASS3  	& &	11	& & $	-0.0050	$ & & $	-0.0375	$ & & $	-0.0425	$ & & $	0.0089	$ \\
LRG0  	& &	18	& & $	-0.0045	$ & & $	-0.0196	$ & & $	-0.0241	$ & & $	0.0020	$ \\
LRG0  	& &	31	& & $	-0.0132	$ & & $	-0.0394	$ & & $	-0.0526	$ & & $	0.0261	$ \\
LRG0  	& &	43	& & $	-0.0251	$ & & $	-0.0574	$ & & $	-0.0826	$ & & $	0.0123	$ \\
LRG1  	& &	18	& & $	-0.0017	$ & & $	-0.0064	$ & & $	-0.0081	$ & & $	0.0018	$ \\
LRG1  	& &	31	& & $	-0.0049	$ & & $	-0.0173	$ & & $	-0.0222	$ & & $	-0.0379	$ \\
LRG1  	& &	43	& & $	-0.0094	$ & & $	-0.0269	$ & & $	-0.0363	$ & & $	0.0109	$ \\
LRG1  	& &	56	& & $	-0.0159	$ & & $	-0.0361	$ & & $	-0.0520	$ & & $	-0.0028	$ \\
LRG1  	& &	68	& & $	-0.0240	$ & & $	-0.0460	$ & & $	-0.0700	$ & & $	-0.0332	$ \\
QSO0  	& &	18	& & $	-0.0003	$ & & $	-0.0012	$ & & $	-0.0015	$ & & $	-0.0039	$ \\
QSO0  	& &	31	& & $	-0.0010	$ & & $	-0.0036	$ & & $	-0.0046	$ & & $	0.0058	$ \\
QSO0  	& &	43	& & $	-0.0018	$ & & $	-0.0067	$ & & $	-0.0085	$ & & $	-0.0254	$ \\
QSO0  	& &	56	& & $	-0.0031	$ & & $	-0.0102	$ & & $	-0.0133	$ & & $	0.0097	$ \\
QSO0  	& &	68	& & $	-0.0047	$ & & $	-0.0135	$ & & $	-0.0182	$ & & $	-0.0509	$ \\
QSO0  	& &	81	& & $	-0.0064	$ & & $	-0.0164	$ & & $	-0.0228	$ & & $	0.0660	$ \\
QSO0  	& &	94	& & $	-0.0086	$ & & $	-0.0193	$ & & $	-0.0279	$ & & $	0.0169	$ \\
QSO0  	& &	110	& & $	-0.0118	$ & & $	-0.0230	$ & & $	-0.0347	$ & & $	0.0626	$ \\
QSO0  	& &	130	& & $	-0.0164	$ & & $	-0.0278	$ & & $	-0.0442	$ & & $	0.1854	$ \\
QSO1  	& &	18	& & $	-0.0006	$ & & $	-0.0010	$ & & $	-0.0017	$ & & $	0.0000	$ \\
QSO1  	& &	31	& & $	-0.0018	$ & & $	-0.0027	$ & & $	-0.0045	$ & & $	-0.0169	$ \\
QSO1  	& &	43	& & $	-0.0035	$ & & $	-0.0046	$ & & $	-0.0081	$ & & $	-0.0131	$ \\
QSO1  	& &	56	& & $	-0.0058	$ & & $	-0.0068	$ & & $	-0.0126	$ & & $	0.0030	$ \\
QSO1  	& &	68	& & $	-0.0088	$ & & $	-0.0091	$ & & $	-0.0179	$ & & $	-0.0073	$ \\
QSO1  	& &	81	& & $	-0.0121	$ & & $	-0.0112	$ & & $	-0.0233	$ & & $	0.0332	$ \\
QSO1  	& &	94	& & $	-0.0163	$ & & $	-0.0134	$ & & $	-0.0297	$ & & $	0.0627	$ \\
QSO1  	& &	110	& & $	-0.0223	$ & & $	-0.0158	$ & & $	-0.0381	$ & & $	0.0801	$ \\
QSO1  	& &	130	& & $	-0.0311	$ & & $	-0.0184	$ & & $	-0.0494	$ & & $	0.0794	$ \\
QSO1  	& &	150	& & $	-0.0413	$ & & $	-0.0207	$ & & $	-0.0620	$ & & $	0.0924	$ \\
QSO1  	& &	170	& & $	-0.0530	$ & & $	-0.0232	$ & & $	-0.0763	$ & & $	0.0223	$ \\
NVSS  	& &	6	& & $	-0.0001	$ & & $	-0.0007	$ & & $	-0.0008	$ & & $	-0.0398	$ \\
NVSS  	& &	11	& & $	-0.0003	$ & & $	-0.0020	$ & & $	-0.0023	$ & & $	-0.0124	$ \\
NVSS  	& &	20	& & $	-0.0010	$ & & $	-0.0050	$ & & $	-0.0059	$ & & $	-0.0111	$ \\
NVSS  	& &	31	& & $	-0.0023	$ & & $	-0.0091	$ & & $	-0.0113	$ & & $	0.0014	$ \\
NVSS  	& &	43	& & $	-0.0043	$ & & $	-0.0135	$ & & $	-0.0178	$ & & $	0.0103	$ \\
NVSS  	& &	56	& & $	-0.0073	$ & & $	-0.0179	$ & & $	-0.0252	$ & & $	0.0025	$ \\
NVSS  	& &	68	& & $	-0.0107	$ & & $	-0.0217	$ & & $	-0.0324	$ & & $	-0.0141	$ \\
\hline\hline
\end{tabular}
\end{table}

\subsection{Point source contamination}

Point source contamination is one of the main concerns that we have for
analysis for cross correlation of CMB with large scale structure,
as point sources add to the CMB, while they are probably
correlated with the tracers of large scale matter density field.
Therefore,
we estimate the contamination from the point sources by estimating $C^{ps}_{\ell}(\nu)$
by looking at the differences of cross correlation of the tracer samples with different
frequency maps of WMAP.
We estimate $C^{ps}_{\ell}$ at 61 GHz (V band):
\begin{equation}
C^{ps}_{\ell}(V) = \frac{C_{\ell}(Ka)- C_{\ell}(V)}{r_{Ka}\nu^{-2}_{Ka} - r_{V}\nu^{-2}_{V}} (r_{V}\nu^{-2}_{V}).
\end{equation}
where $r_{X}$ is the ratio of thermodynamic temperature to the antenna temperature of band $X$ and 
we assume that $T(\nu)$ is proportional to $\nu^{-2}$.
We assess the level of contaminations by calculating (similarly as above mentioned foreground analysis):
\begin{equation}
E_{ps} = C^{ps}_{\ell}(\nu) {\bf C}^{-1} C^{ps}_{\ell}(\nu)
\label{eq:eps}
\end{equation}
We find that $\sqrt{E_{ps}} = 0.495$ using the $\ell$-bins that are tabulated in Table~\ref{tab:szcontam} and thus point sources is not a significant 
contamination for the ISW effect.
Note that this includes some effect from Galactic foregrounds (which probably dominate the low $\ell$'s), since 
any foreground effects that have frequency dependence will show up 
in $C^{ps}_{\ell}(V)$.  In particular one would be double-counting the Galactic foreground if one added $E_{ps}$ and $E_{fg}$.

We present the point sources contamination for $\ell$-bins we use for our analysis in the last column of Table~\ref{tab:szcontam}.

\section{Cosmological Implications}
\label{sec:cosmo}

\subsection{Significance of ISW detection after rejecting contaminating bins}

After investigating all the listed systematics and taking into account 
of the non-linearities, 
we decide to only take the $\ell$-bins as 
are listed in Table~\ref{tab:szcontam}.
The high-$\ell$ bins are cut off due to the non-linearities;
we cut off all the bins that at the median redshift for the fiducial cosmology correspond to $k\ge 0.05h\,$Mpc$^{-1}$ using $k=(\ell+1/2)/r$.  This 
is a more conservative cut than the usual $k=0.1h\,$Mpc$^{-1}$ but it must be remembered that in linear theory the ISW effect is sensitive to 
the derivative of $D(a)/a$ which contains a cancellation from the growth of structure in the numerator and the scale factor in the denominator.  
Therefore nonlinear effects could be larger than one naively expects.  We cut off the first $\ell$-bin for all samples as these are most affected by 
Galactic foregrounds. 

We calculate the significance of each of the sample by the standard method. 
First, we compute the amplitude of the signal (Appendix~\ref{app:foreg}, in our case,
fiducial model is based on the WMAP 3-year parameters):
\begin{equation}
A=\frac{C^{data}_{\ell}\cdot {\bf C}^{-1} C^{theory}_{\ell}}{C^{theory}_{\ell}\cdot {\bf C}^{-1} C^{theory}_{\ell}},
\label{eq:ath}
\end{equation}
where $C^{theory}_\ell$ is the vector of predicted cross-power spectra for the fiducial cosmology, $C^{data}_\ell$ is the vector of observed
cross-spectra, and ${\bf C}^{-1}$ is the inverse-covariance matrix.  We obtain ${\bf C}^{-1}$ by Monte Carlo simulation as described in the next 
section.

The error is similarly computed with:
\begin{equation}
{\rm \sigma}=\frac{1}{\sqrt{C^{theory}_{\ell}\cdot {\bf C}^{-1} C^{theory}_{\ell}}},
\label{eq:sth}
\end{equation}
and the significance in sigmas is obtained by the usual calculation, $A/\sigma$.
The result is shown in Table~\ref{tab:significance}.

\begin{figure}
\includegraphics[width=3.0in]{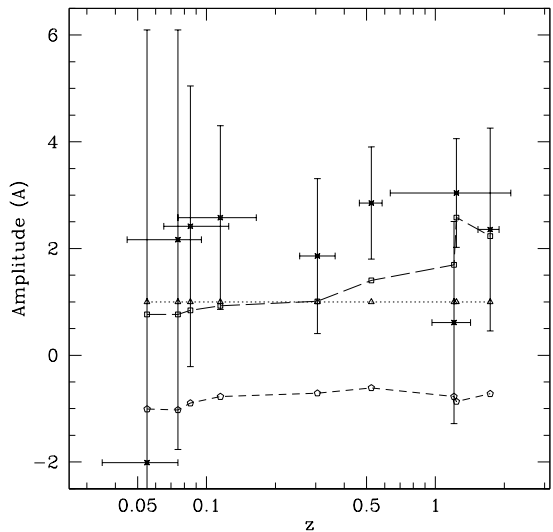}
\caption{The ISW amplitude (A) and errorbars  $\sigma(A)$ for all samples plotted along 
the redshifts compared with predictions of WMAP-3 year parameters. 
The fitting and errors for this figure used the Fisher
matrices from the correlation of the tracer samples with the various WMAP maps. 
We also show the expected amplitude for 3 model Universes
along the angular diameter distance degeneracy curve.
We calculate the expected amplitude by substituting our observed 
correlations with predicted correlations for each of the model 
Universe and proceed in the same manner as described in Eq.\ref{eq:ath}.
The three model Universes are:
$\Lambda$CDM model with the WMAP-3 year parameters 
(open triangle with dotted line);
closed Universe (open pentagons with short dashed line) $\Omega_b$=$0.215$,
$\Omega_m$ = $1.25$, $\Omega_K$ = $-0.29$, $h=0.32$, $\sigma_8 =0.61$; and
Open Universe (open squares with long dashed line) $\Omega_b$=$0.015$,
$\Omega_m$ = $0.089$, $\Omega_K$ = $0.03$, $h=1.20$, $\sigma_8 =0.73$.
Note that the redshift distribution is very broad for NVSS, giving rise to the
jump in the open model prediction, even though the effective
redshift of the sample is nearly the same as for low redshift QSO sample.}
\label{fig:constrain}
\end{figure}

In  Fig.~\ref{fig:constrain} we plot the amplitude ($A$) and its error using 
covariance matrices and fisher matrices from the correlation of the tracer sample with WMAP V-band, computed with angular and redshift weighting 
optimized for WMAP3 model, together with theoretical predictions
for three cosmological models 
(open, closed and flat)
 to illustrate 
the constraining power on $\Omega_K$ from ISW effect. 
Flat model is WMAP3 model and by definition 
its theoretical prediction is $A=1$ (see Eq.~\ref{eq:ath}). 
The other two models were chosen to lie
along the WMAP degeneracy curve (which essentially keeps fixed $\Omega_mh^2$, 
$\Omega_bh^2$ and $\theta$, defined to be 100 times the ratio of the sound horizon to the 
angular diameter distance to recombination), although this does not 
imply they are  
necessarily good fits to the WMAP data: the ISW signal in the CMB power spectrum itself can break the 
degeneracy between the parameters that keep the angular diameter fixed,  
but because ISW is a subdominant contribution to 
primary CMB even on the largest scales its power to discriminate among models is limited. 
We can see that the predicted amplitude of ISW signal for $\Lambda$CDM is positive (using 
the standard sign convention) 
because at late time when cosmological constant becomes important 
growth of structure is decreasing in time relative to Einstein-de Sitter (EdS) 
model
and the associated gravitational potential, constant at high redshift when the Universe
is effectively EdS, begins to decay. The decay is larger if we decrease 
$\Omega_m$ (for which we need to go to a slightly 
open universe to preserve angular diameter distance), which in turn increases ISW. 
On the other hand, a closed universe with $\Omega_m>1$ accelerates the growth of structure 
relative to EdS, so potential is growing and this model 
predicts ISW signal with opposite sign. 
While the sign is essentially 
determined by the growth rate, its amplitude and scale dependence depend on 
other cosmological parameters as well and vary as a function of redshift, 
as shown in Fig.~\ref{fig:constrain}.

As can be seen from Fig.~\ref{fig:constrain} and Table~\ref{tab:significance} we 
have a detection of ISW signal in a number of data sets. Most convincing are 
SDSS LRG1 and NVSS, both at about $3\sigma$, followed by LRG0, QSO1 and 2MASS3
at 1.2--1.5$\sigma$ evidence. Remaining data sets have significance below 1$\sigma$, 
although only one among them has negative signal, opposite to $\Lambda$CDM model 
predictions. The overall significance of detection with $\Lambda$CDM weighting
is 3.7$\sigma$. We emphasize that while we use optimal weighting of data 
to maximize the signal by downweighting the scales and redshifts where we 
do not expect the signal, this depends somewhat on the assumed model, so 
the significance of detection can be somewhat affected by this. For example, 
we could instead of $\Lambda$CDM have used a model that predicts an upward feature at $l=30$ that only 
occurs at redshift around 0.5, therefore taking advantage of the 3 sigma excess power
seen in LRG1 at that scale (Fig~\ref{fig:cross_sdss}). Using this model would give high weight to that 
feature and would lead to a higher significance of the overall detection. 
Of course such aposteriori procedure is not really waranted, but it does 
highlight the difficulty of comparing 
the significance of detection among different analyses, which may have used
different priors. This problem is exacerbated if cross-correlation function analysis 
is used, as in most of the previous work, because in that case 
a narrow feature in Fourier space
would spread out to a broader feature in correlation function. 

While we find a 3.7$\sigma$ detection we also note that 
the observed ISW signal exceeds the predictions of WMAP3 $\Lambda$CDM model by about 
2$\sigma$, since the fit gives $A=2.23 \pm 0.60$ relative to model prediction 
$A=1$. The discrepancy is reduced if we change cosmological parameters somewhat and this is 
explored further in the next subsection using MCMC analysis. 

\begin{table}
\caption{\label{tab:significance}Amplitude of ISW signal and the associate one sigma error 
relative to WMAP3 model and significance of detection for each of the sample and when we combine all samples. 
These are calculated using the covariance matrix that are derived from the correlations with the Monte Carlo 
CMB maps (as described in Eq~\ref{eq:cov_mcmc}.The overall signal is 2 sigma higher than WMAP3 model prediction.}
\begin{tabular}{ccccr}
\hline\hline
Sample & & Amplitude ($A\pm\sigma$) & &  \# sigmas \\
\hline
2MASS0& & $-2.01 \pm 11.41$  & & $-0.18$ \\
2MASS1& & $+3.44 \pm 4.47$ &  &  0.77 \\
2MASS2& & $+2.86 \pm 2.87$ & & 1.00 \\
2MASS3& & $+2.44 \pm 1.73$ & & 1.41 \\
LRG0&  &  $+1.82 \pm 1.46$ & & 1.25 \\
LRG1&  &  $+2.79 \pm 1.14$ &  & 2.46 \\
QSO0&  &  $+0.26 \pm 1.69$ & & 0.16 \\
QSO1&  &  $+2.59 \pm 1.87$ & &  1.38 \\
NVSS&  &  $+2.92 \pm 1.02$ & & 2.86\\
\hline
All Samples & & $+2.23\pm 0.60$ & & 3.69 \\
\hline\hline
\end{tabular}
\end{table}

To show that our results are consistent throughout different bands in WMAP, thus there is no significant 
contamination from frequency dependent systematics, we show the amplitude of ISW signal and associate 
one sigma error relative to the WMAP3 model for each of the sample for all of the WMAP bands (except K band)
in Table~\ref{tab:A_band}.
The differences in frequency $A$(Q)-$A$(V) and $A$(W)-$A$(V) are all $<0.25\sigma$ and most are $<0.15\sigma$,
and there is no consistent sign.  This reassures us that the frequency-dependent foregrounds are subdominant to the statistical
errors in these higher-frequency bands.
The comparison with Ka band, i.e. $A$(Ka)-$A$(V), is worse 
especially for 2MASS0 (the difference is $<0.5\sigma$ for the other samples), probably due to Galactic emission.

\begin{table*}
\caption{\label{tab:A_band}Amplitude of ISW signal and the associated $1\sigma$ error 
relative to WMAP3 model for each of the sample for the WMAP bands (i.e. Ka, Q, V, W). 
The fitting and errors for this table used the Fisher 
matrices from the correlation of the tracer samples with the various WMAP maps.}
\begin{tabular}{ccrrrrrrr}
\hline\hline
Sample & & \multicolumn{7}{c}{\mbox{Amplitude $A$}} \\
 & & \multicolumn{1}{c}{\mbox{Ka}} & & \multicolumn{1}{c}{\mbox{Q}} & & \multicolumn{1}{c}{\mbox{V}} & & \multicolumn{1}{c}{\mbox{W}} \\
\hline
2MASS0& & $-9.04 \pm 8.21$ & &$-3.54 \pm 8.19$ & & $-2.01 \pm 8.11$ & & $-3.38 \pm 7.79$\\
2MASS1& & $1.80  \pm 3.97$ & &$2.73  \pm 3.94$ & & $2.17 \pm 3.93$ & & $1.64 \pm 3.86$\\
2MASS2& & $2.16 \pm 2.66$ & & $2.95  \pm 2.65$ & & $2.42 \pm 2.63$ & & $2.04 \pm 2.61$ \\
2MASS3& & $1.74 \pm 1.72$ & & $2.56  \pm 1.72$ & & $2.58 \pm 1.72$ & & $2.39 \pm 1.69$\\
LRG0&   & $2.00 \pm 1.44$ & & $2.05  \pm 1.44$ & & $1.86 \pm 1.45$ & & $1.92 \pm 1.46$\\
LRG1&   & $2.67 \pm 1.04$ & & $2.59  \pm 1.04$ & & $2.85 \pm 1.05$ & & $2.92 \pm 1.06$ \\
QSO0&   & $0.62 \pm 1.90$ & & $0.39  \pm 1.92$ & & $0.61 \pm 1.89$ & & $0.63 \pm 1.94$\\
QSO1&   & $2.41 \pm 1.90$ & & $2.17  \pm 1.92$ & & $2.36 \pm 1.90$ & & $1.93 \pm 1.90$\\
NVSS&   & $2.56 \pm 1.01$ & & $2.80  \pm 1.01$ & & $3.04 \pm 1.02$ & & $2.88 \pm 1.02$\\
\hline\hline
\end{tabular}
\end{table*}

\subsection{MCMC methodology and Likelihood function}

\subsubsection{MCMC methodology}

A major goal of this paper is to provide a full likelihood function with which 
cosmological models can be compared to each other. Here we describe 
the details of the likelihood function construction
and apply it to some simple cosmological model parametrizations. 
Our goal is not to give an exhaustive parameter estimation analysis, but just to provide 
some characteristic examples of possible applications. We include both ISW
analysis of this paper and the lensing analysis of Paper II. However, the latter 
effect has small statistical significance and does not contribute significantly to 
the likelihood analysis. 
We decided to test the following cosmological models:
flat $\Lambda$CDM model ($\Omega_m h^2$, $\Omega_b h^2$, $\theta$, $\tau$, $n_s$, $A_s$),
$\Lambda$CDM + $\Omega_K$ (not assuming flatness),
flat $\Lambda$CDM + $w$ (assuming flatness, but allowing
dark energy to evolve). Here $\Omega_m$ is the matter density, $\Omega_b$ is the 
baryon density in units of critical density, $\Omega_K=-K/H_0^2$ is the curvature 
$K$ expressed in terms of critical density, $h=H_0/{\rm 100km/s/Mpc}$ is the Hubble parameter, 
$\theta$ is 100 times the ratio of sound horizon to angular diameter distance at recombination,
$\tau$ reionization optical depth and $n_s$ and $A_s$ are the slope and 
amplitude (at $k=0.05/{\rm Mpc}$) of primordial power spectrum. 
We also refit for the bias with the redshift distributions for each of the dataset used
for each of the cosmological parameter sets which we calculate the $\chi^2$ for.
There is a detailed description of the determination of bias and redshift distribution
in Section~\ref{sec:dndz}.
We limit our search to models with scalar fluctuations only with no running of
spectral index, no tensors, and no neutrino masses.
We assume flat priors on all of the parameters defined above.
The priors we use are shown at Table \ref{tab:priors}.
In addition we impose  $40{\rm km/s/Mpc} < H_0 < 100{\rm km/s/Mpc}$ and
that age of the Universe has to be at least $10$ Gyr and at most $20$ Gyr.
These priors are applied to all the chains that we show in the paper (including
those with WMAP alone).

\begin{table}
\caption{\label{tab:priors}The priors applied to the 3 different chains. Note that
all priors are flat.}
\begin{tabular}{ccccc}
\hline\hline
Parameter & & minimum & & maximum  \\
\hline
\multicolumn{5}{c}{\mbox{for all models, 6 parameters}} \\
\hline
$\Omega_bh^2$    & & $0.005$ & & $0.1$ \\
$\Omega_ch^2$    & & $0.01$   & & $0.99$   \\
$\theta$         & & $0.5$   & & $10$   \\
$\tau$           & & $0.01$   & & $0.8$   \\
$n_s$           & & $0.5$   & & $1.5$   \\
$log_e(10^{10}A_s)$ & & $2.7$   & & $4.0$   \\
\hline
\multicolumn{5}{c}{\mbox{for $\Lambda$CDM + $\Omega_K$ only}} \\
\hline
$\Omega_K$ & & $-0.3$   & & $0.3$   \\
\hline
\multicolumn{5}{c}{\mbox{for $\Lambda$CDM + $w$ only}} \\
\hline
$w$ & & $-2.1$   & & $-0.1$   \\
\hline
\end{tabular}
\end{table}

In most cases the intervals are sufficiently broad that the boundaries do not 
matter, with exception of WMAP only case with curvature or dark energy, where 
we apply additional prior with $H_0>40{\rm km/s/Mpc}$. 
We search the parameter space using COSMOMC \citep{lewis02}  with likelihood function from WMAP 3 year analysis \citep{spergel07}.
We discuss the Integrated Sachs Wolfe likelihood function in the following 
section, and leave the discussion of the Weak Lensing of CMB likelihood
function to Paper II.
We test the convergence of our Markov chains following Dunkley et~al. \cite{dunkley05}.

\subsubsection{Integrated Sachs Wolfe likelihood function}
\label{ss:isw-likelihood}

This section describes the ISW likelihood function.
We utilize the amplitude from the galaxy-temperature cross-spectrum $C_\ell^{gT}$ 
from
cross correlating the CMB sky (V-band) with the following samples:
2MASS (0-3), SDSS-LRG (low-$z$ and high-$z$), SDSS-QSO (low-$z$ and high-$z$), NVSS.
When we construct the likelihood function, we need three items: (i) the ``data'', which is $C_\ell^{gT}$ for
each of the sample for each $\ell$-bin (ii) the theoretical prediction; and (iii) the covariance matrix of the $\{C_\ell^{gT}\}$.

The data vector consists of the measured $C_\ell^{gT}$ in each $\ell$-bin and for each LSS sample used.  After our cuts there are 42 such bins 
remaining, when combining all samples, thus the data vector has length 42.

We calculate this covariance matrix by first generating 1000 simulated CMB skies of
WMAP resolution and then cross-correlate each of the samples with these simulated
CMB sky. We call these $C_\ell^{gT_{sim},\mu}$. We then calculate the covariance among
the samples by first calculating the $\langle C_\ell^{gT_{sim},\mu} \rangle$ by averaging
over all the correlations with all the simulated maps, then we find:
\begin{eqnarray}
\label{eq:cov_mcmc}
[{\bf C}]_{\mu\nu} &=& \langle (C_\ell^{gT,\mu} - \langle C_\ell^{gT_{sim},\mu} \rangle) \nonumber \\ && \times
(C_\ell^{gT,\nu} - \langle C_\ell^{gT_{sim},\nu} \rangle) \rangle.
\end{eqnarray}
Note that this is a $42\times 42$ covariance matrix, and that it is {\em not} block-diagonal in the LSS samples because there is some overlap in sample 
volume.  The Monte Carlo procedure, by considering many realizations of the CMB but the actual realization of the galaxies, includes the implied 
correlations between different LSS samples.

The issue of how to construct error bars on estimates of the galaxy-temperature cross-spectrum $C_\ell^{gT}$, or its real-space equivalent
$w_{gT}(\theta)$, has been a contentious issue ever since the first claimed ISW detections were announced.  The methods used have ranged from Gaussian
error estimates based entirely on the theoretical galaxy and CMB spectra, to jack-knife methods that are based entirely on the data.  Among the
intermediate options are the Monte Carlo approach used here (MC1 in the terminology of Cabr\'e et~al. \cite{cabre07}) in which the real
galaxy field is cross-correlated against many random realizations of the CMB.

If we knew the CMB and galaxy power spectra perfectly from theory or observation, we would like to use analytic Gaussian error estimates for
$C_\ell^{gT}$ or do Monte Carlo simulations of random CMB and galaxy fields. Unfortunately, the galaxy maps, particularly QSO0 and NVSS, are subject to
spurious power at large angular scales for which we have no good theory, and for which we cannot measure the power spectrum accurately due to sampling
variance.  However we do know the theoretical CMB power spectrum so we can implement MC1.  It would also have been possible (but 
computationally expensive) to implement a jack-knife;
we chose not to do so because of concerns that at low multipoles the jack-knife regions would not be independent \cite{afshordi04} although we
note that the Cabr\'e et~al. simulations \cite{cabre07} suggest that at least in some cases this is not a significant problem.  The MC1
method is however subject to two biases that could understimate the errors: a ``correlation bias'' due to neglect of the galaxy-temperature correlation
when determining the error bars, and a ``realization bias'' due to the fact that only one realization of the galaxy field is used.  These biases are
discussed in Appendix~\ref{app:errors}, where we find them to be negligible.


We construct the likelihood function as the following:
\beq
\chi^2 = 
 [x^{\mu}({\rm obs})-\langle x^{\mu}\rangle_{({\bf p})}]
[{\bf C}^{-1}]_{\mu\nu}
[x^{\nu}({\rm obs})-\langle x^{\nu}\rangle_{({\bf p})}],
\eeq
where $x^{\mu}$ is simply $C_\ell^{gT}$; the index $\mu$ encodes both the $\ell$-bin and the sample used.
We denote by $x^{\mu}(\rm {obs})$ the observed correlations
$C_\ell^{gT}$, and $\langle x^{\nu}\rangle_{({\bf p})}$ denotes the mean value predicted for cosmological parameters ${\bf p}$.
Note that the vector $x^{\mu}$ is of length 42 and that all LSS samples are included in a single $\chi^2$; we do {\em not} add
the $\chi^2$ values of different samples separately since they are correlated and such an addition would be invalid.
Among the three components of the likelihood function, only the predicted $C_\ell^{gT}$
needs to be re-calculated for each cosmological model.

\subsection{Parameter fits}

\begin{figure}
\includegraphics[width=3.0in]{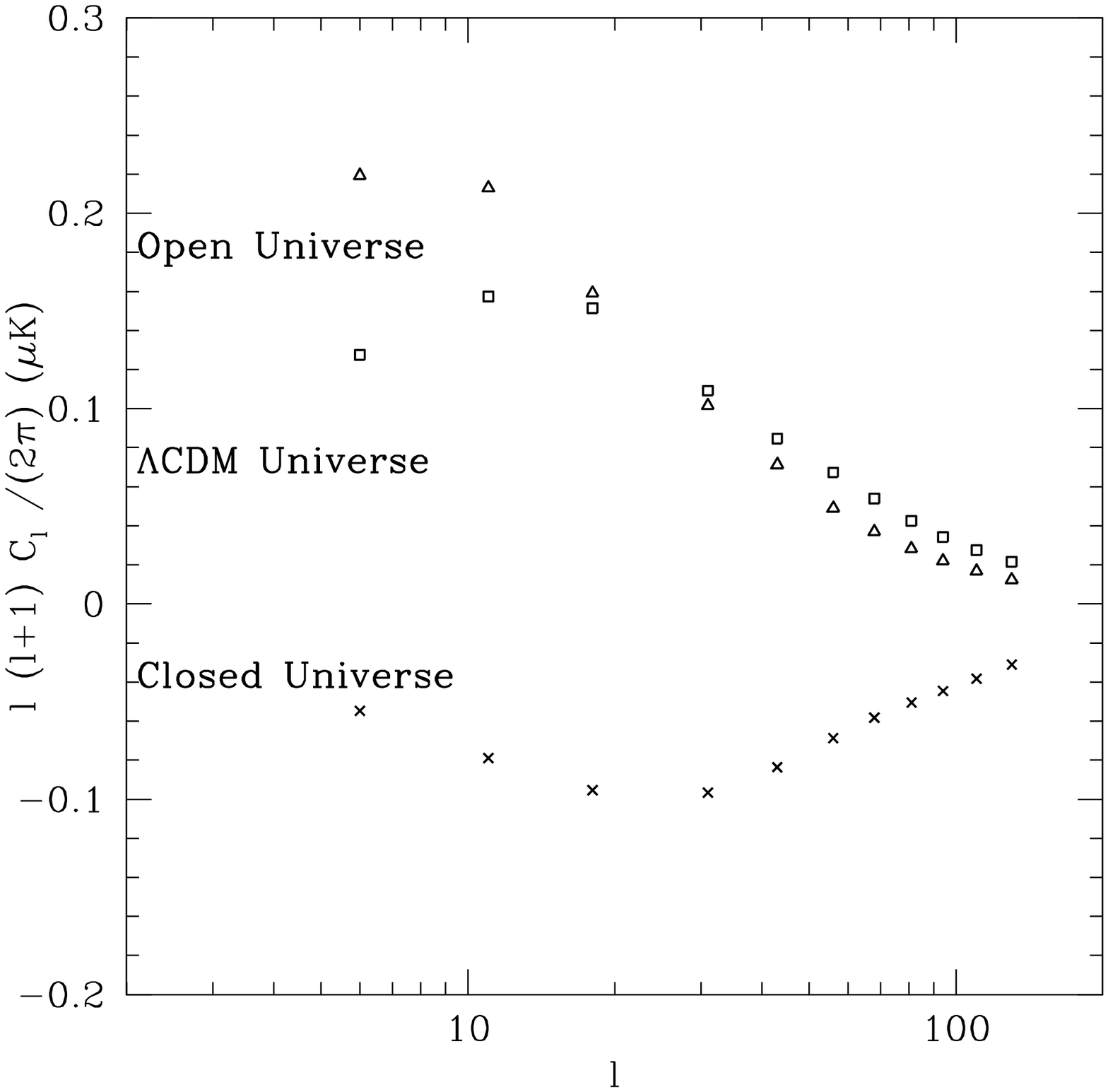}
\includegraphics[width=3.0in]{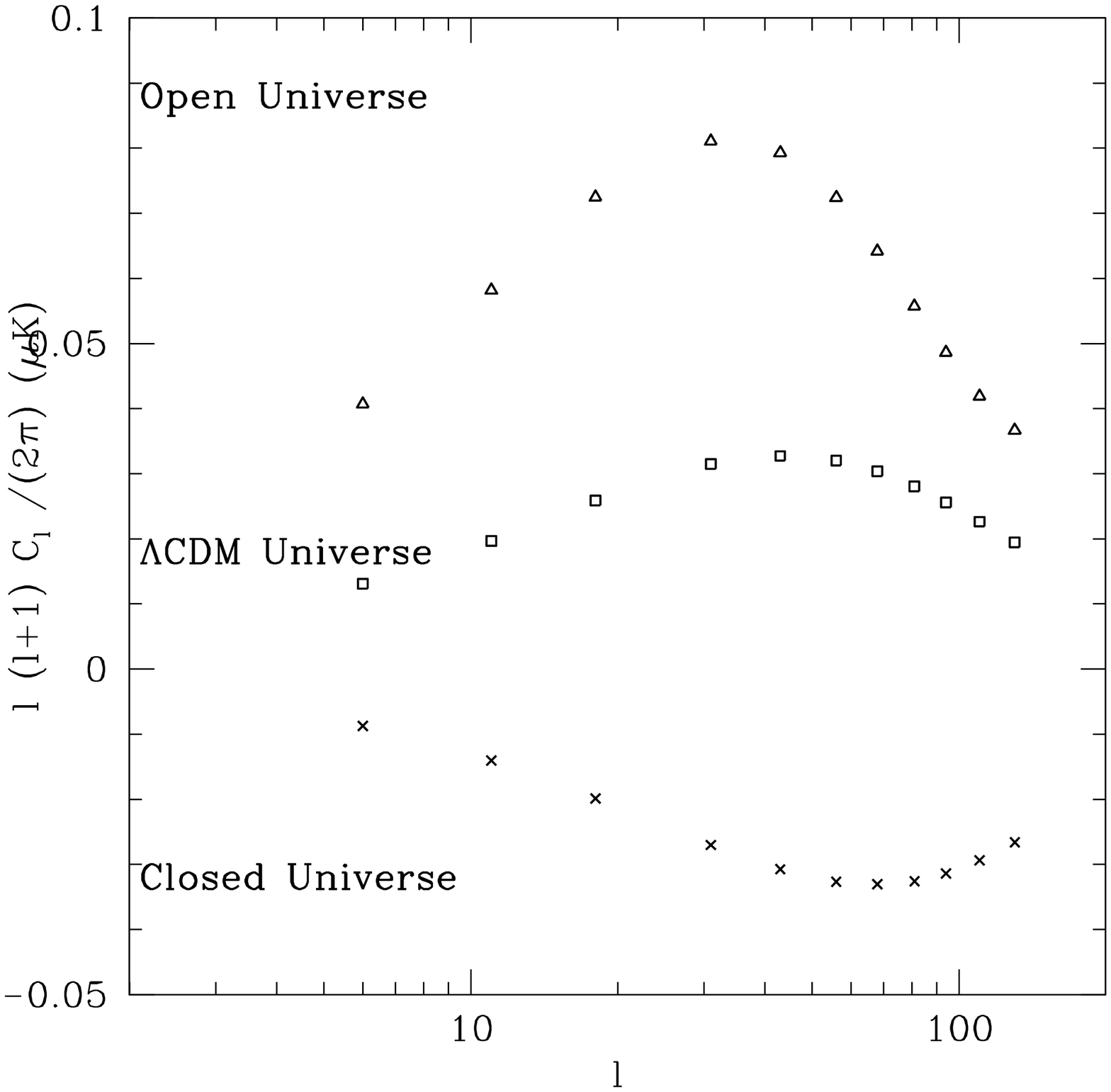}
\caption{The predicted ISW signal for the low-z LRGs (above) and high-z QSOs (below) sample
for sample open, closed, and flat $\Lambda$CDM models.  Parameters are the same as in Fig.~\ref{fig:constrain}.}
\label{fig:prediction}
\end{figure}

We investigate the following cosmological models:
(i) $\Lambda$CDM model ($\Omega_b h^{2}$, $\Omega_{c} h^{2}$, $\theta$, $\tau$, $n_s$, $A_s$);
(ii) $\Lambda$CDM model + $\Omega_K$; and
(iii) $\Lambda$CDM model + $w$.
Note that $\theta$ is the ratio of the sound horizon to the angular 
diameter distance,
while $A_s$ is the the primordial superhorizon power in the curvature perturbation on $0.05/{\rm Mpc}$ scale.
The numerical results are shown in Table~\ref{tab:paramm} for both the full likelihood (CMB+ISW+WL) and CMB alone.
We also looked at the effect of WL (or ISW) separately 
in constraining cosmological parameters
by analyzing a cosmological model ($\Lambda$CDM + $\Omega_K$)
using only CMB+ISW (without lensing). We find the constraints to be similar to the 
full case (CMB+ISW+WL), but with slightly larger errorbars 
(see table~\ref{tab:compare}).
Note that for the CMB-only model including $\Omega_K$, the Markov chain ran up against the $H_0>40$ km/s/Mpc boundary, thus 
artificially tightening the constraints; this did not occur for the full CMB+ISW+WL chains.

For the $\Lambda$CDM model, the combined constraints from WMAP+ISW+WL is only slightly improved over using WMAP alone, but does lead to a decrease in $\Omega_m$ as expected, 
because this is the direction of increase in ISW, which is needed given that 
we find the measured ISW exceeds WMAP3 prediction. The effect is smaller than expected
because moving along the WMAP degeneracy line in the direction of decrease in $\Omega_m$
also requires an increase in $h$ and decrease in $\sigma_8$, both of 
which reduce ISW (see also Fig.~\ref{fig:prediction}). 

\begin{figure}
\includegraphics[width=2.5in]{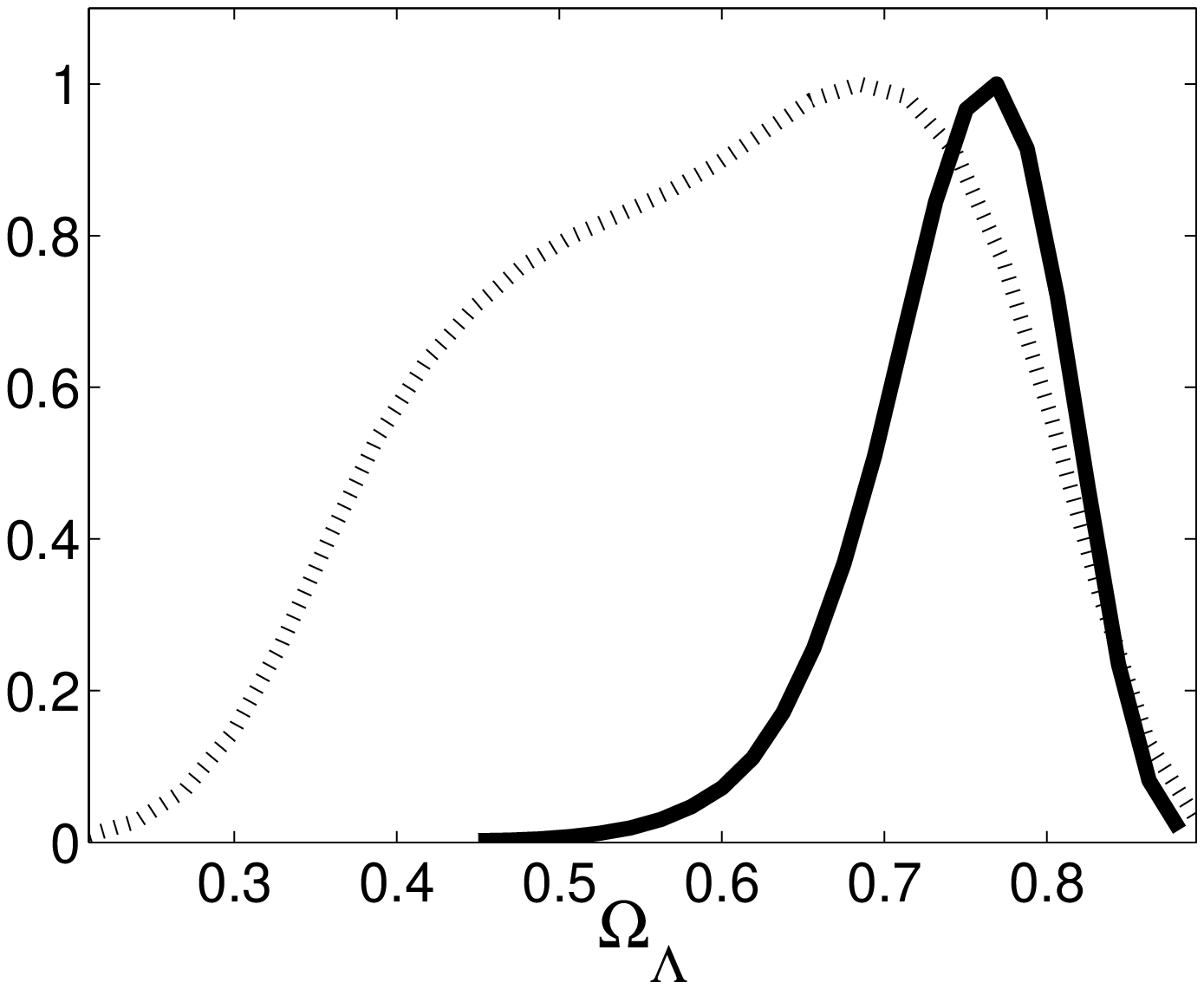}
\includegraphics[width=2.5in]{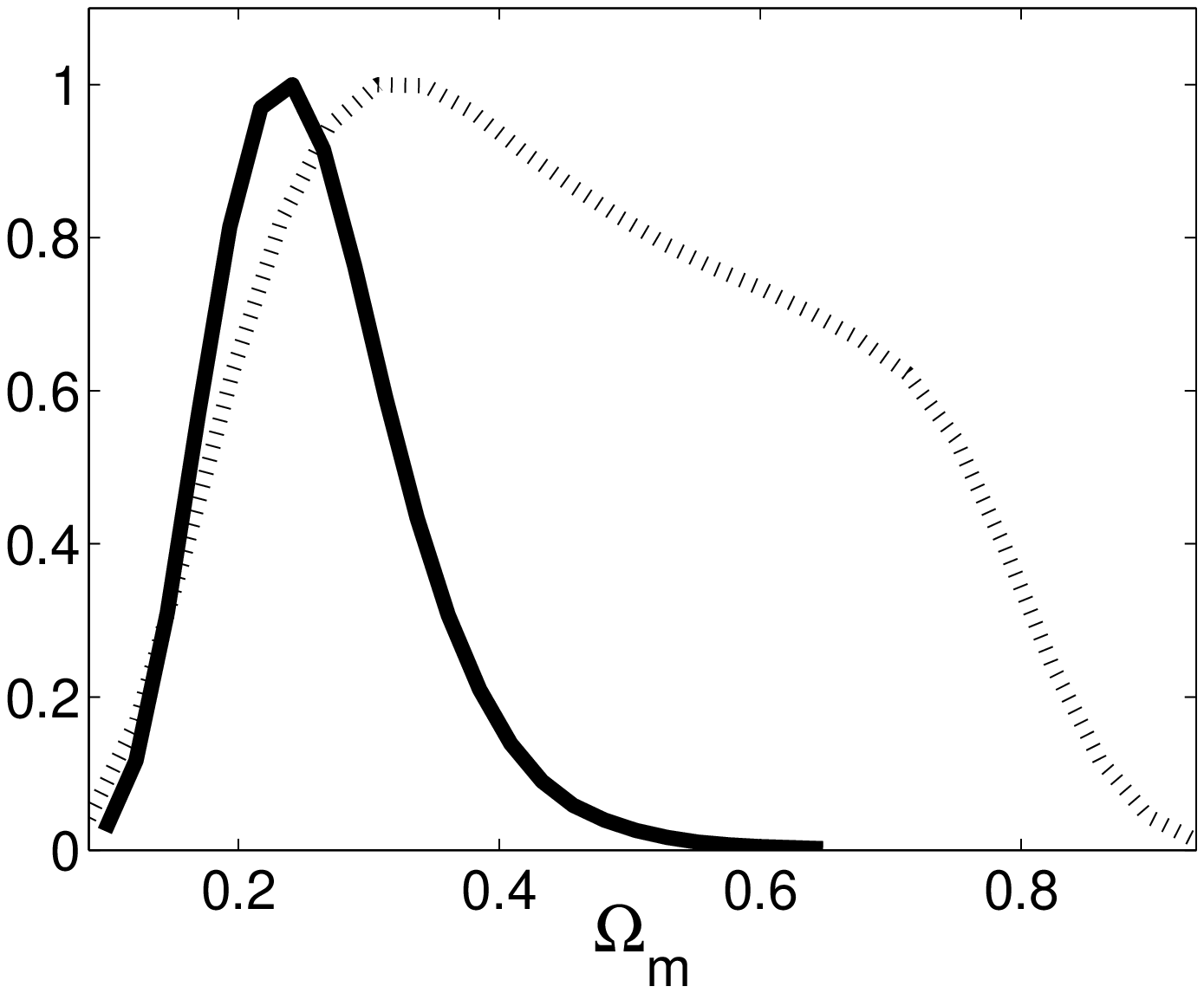}
\caption{\label{fig:ini2_lam}$\Lambda$CDM + $\Omega_K$ model: the 1-D distributions of $\Omega_\Lambda$ and $\Omega_m$. The solid (dashed) 
line represents constraints from 
using WMAP+ISW+WL (WMAP alone).}
\end{figure}

\begin{figure}
\includegraphics[width=2.5in]{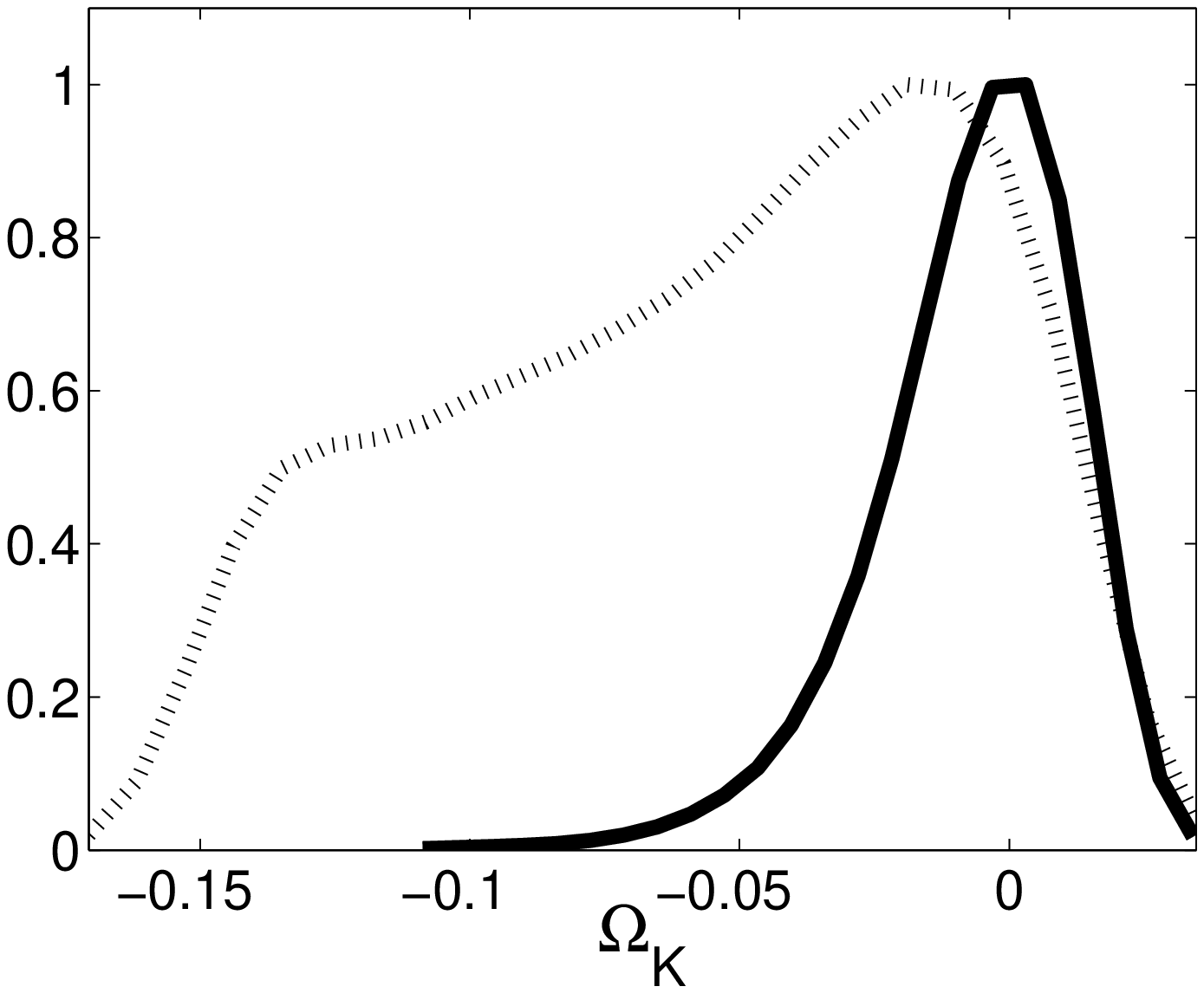}
\includegraphics[width=2.5in]{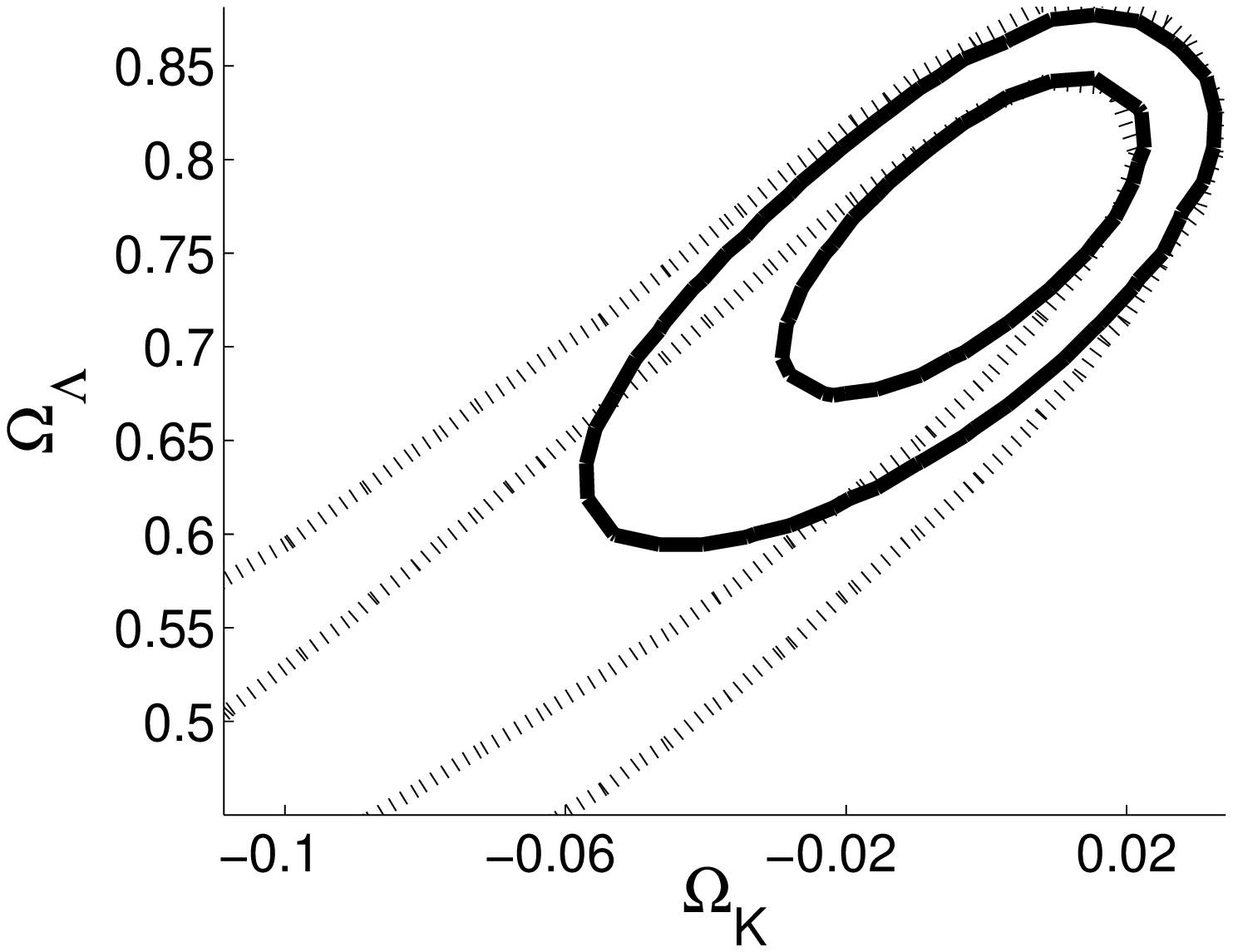}
\caption{\label{fig:ini2_lambda_K}$\Lambda$CDM + $\Omega_K$ model: the 1-D distribution of $\Omega_K$ and the 2-D distribution of $\Omega_\Lambda$ and 
$\Omega_K$ (68\% and 95\% confidence contours shown).  The solid (dot-dashed) line represents constraints from 
using WMAP+ISW+WL (WMAP alone).}
\end{figure}

For $\Lambda$CDM + $\Omega_K$ model, we improve significantly over what using CMB alone can do.
In Fig.~\ref{fig:ini2_lam}
we compare 1-D distributions of the $\Omega_\Lambda$ and $\Omega_{m}$  
when we use WMAP+ISW+WL versus using WMAP alone. 
The ISW effect, as discussed above, can constrain
the change of 
gravitational potential of the Universe as it depends linearly 
on the change of 
growth factor of the potential ($D(a)/a$). 
For example, in the closed Universe model we plotted in Fig~\ref{fig:prediction}, $D(a)/a$ increases as redshift decreases, 
while in the other two models, $D(a)/a$ would decrease as redshift decreases. 
As $\partial\phi/\partial\eta$ has a different sign for the
closed Universe model on WMAP degeneracy curve 
as compared to the open and the flat universe model on the same curve,
the sign of the ISW effect changes too. 
In Fig.~\ref{fig:prediction} we plot the predicted ISW signal using the low-redshift
LRG and the high-redshift quasar distribution for 3 different Universes along 
the WMAP degeneracy curve. 
As expected closed model differs drastically from open and flat models. 
We also see that for LRG there is not much difference between 
flat and open models even though the latter has $\Omega_m=0.088$ compared to 
$\Omega_m=0.24$, but the 
increase in ISW induced by $D(a)/a$ is compensated by the reduction caused by 
other parameters such as $h$ and $\sigma_8$. The differences between the two 
are more significant for the high-$z$ quasar redshift distribution. 
ISW effect breaks the WMAP degneracies between 
$\Omega_K$ and $\Omega_\Lambda$ (or $\Omega_m$). 
In Fig.~\ref{fig:ini2_lambda_K}
we show the 2-D contour plots of this set of parameters 
to demonstrate how our analysis improves
the constraints on these parameters.

%

\begin{figure}
\includegraphics[width=2.5in]{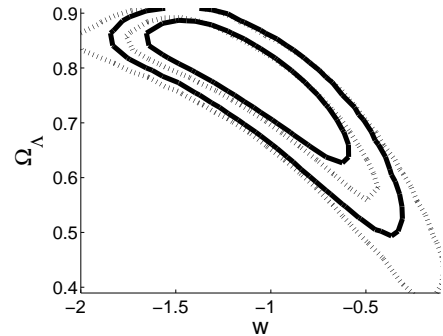}
\caption{$\Lambda$CDM + $w$ model:  the 2-D distribution of $\Omega_\Lambda$ and $w$ (68\% and 95\% confidence contours shown). 
 The solid (dashed) line represents constraints from 
using WMAP+ISW+WL (WMAP alone).}
\label{fig:ini6_lambda_w}
\end{figure}

Finally, we look at the $\Lambda$CDM + $w$ model where we look for better constraints on dark energy equation of state ($w$). The constraint on $w$ is 
modestly improved,
since the dark energy equation of state changes the growth factor along the WMAP 
degeneracy curve, thus the evolution of the gravitational potential. 
We also see that there is a tilt of 
$\Omega_\Lambda$ towards lower value when we combine WMAP with ISW and WL effects.
We also plot the 2-D contours for the $\Omega_\Lambda$ and $w$ in Fig~\ref{fig:ini6_lambda_w}. 

As mentioned above, WMAP 3-year model predicts ISW amplitude that is about
two sigma below our constraints and this is also the case for the best fit
$\Lambda$CDM model (which is almost the same as WMAP 3-year).
Adding curvature or dark energy equation of state
does not reduce this discrepancy either and in both cases these two parameters are not
needed to improve the fit. While reducing matter
density goes in the desired direction of increasing ISW in cross-correlations,
such models also increase
the CMB power at large scales through the ISW auto-correlation power,
which is in disagreement with the
low power observed on large scales in WMAP. For example, we find that there are models
with $\Omega_m=0.18$ which
improve the $\chi^2$ fit to ISW data by 13 relative to the best fit
$\Lambda$CDM + $\Omega_K$ model, but at the same time
make the WMAP $\chi^2$ fit worse by 15. There is thus some mild tension
between low power in WMAP at low $l$ and the high ISW power we measure, but
it is a tension that cannot be removed by
simple extensions of parameter space explored here.
As this is only a two sigma effect there is a considerable probability that it
is just a statistical fluctuation.

\begin{table}
\caption{\label{tab:compare}Comparing the constraints for several 
parameters with and without Weak Lensing
of CMB in $\Lambda$CDM + $\Omega_K$ cosmological model.
The limits shown are mean and standard deviation for each of the parameter.}
\begin{tabular}{ccc}
\hline\hline
Parameter & Limits (CMB+ISW+WL) & Limits (CMB+ISW) \\
\hline
$\Omega_K$       & $-0.068\pm 0.019$    & $-0.0073\pm 0.020$  \\
$\Omega_\Lambda$ & $0.746\pm0.059 $    & $0.745\pm 0.065$  \\
$\Omega_m$       & $ 0.261\pm0.075 $   & $0.263\pm 0.083$  \\
\hline\hline
\end{tabular}
\end{table}



\begin{table*}
\caption{\label{tab:paramm}The percentiles of the posterior distribution (2.5\%, 16\%, 50\%, 84\% and 97.5\%)
  on cosmological parameter for each model with the CMB only (``C'') and also including the ISW and weak 
lensing
likelihood functions (``I'').  $H_0$ is in km$\,$s$^{-1}\,$Mpc$^{-1}$.  For a Gaussian distribution these percentiles correspond approximately to 
$-2\sigma$, $-1\sigma$, central, $+1\sigma$, and $+2\sigma$ values.
Note that for the 7-parameter chains
with CMB only there are significant prior effects in the CMB degeneracy direction.
}
\begin{tabular}{cccccccccccc}
\hline\hline
Parameter &  C($2.5\%$) & C($16\%$) & C($50\%$) & C($84\%$) &C($97.5\%$)& & I($2.5\%$)& I($16\%$) &I($50\%$) & I($84\%$) & I($97.5\%$) \\
\hline
\multicolumn{12}{c}{\mbox{$\Lambda$CDM, 6 parameters}} \\
\hline
$\Omega_bh^2$    & 0.0208& 0.0214 & 0.0222 & 0.0229 & 0.0236 &  & 0.0208 & 0.0215 &0.0222 & 0.0229 & 0.0236\\
$\Omega_ch^2$    & 0.0901 &  0.0976&  0.105 &  0.113 &  0.121 & & 0.0901 &  0.0970 &  0.104 &  0.111 &  0.119 \\
$\tau$           & 0.0312 & 0.0612 & 0.0911 & 0.121 & 0.151 &   & 0.0359 &  0.0662 &  0.0956 &  0.125 &  0.154\\
$n_s$            & 0.929 &  0.943 &  0.959 &  0.976 &  0.993&   & 0.929 &  0.944 &  0.960 &  0.977 &  0.994 \\
$\Omega_\Lambda$ & 0.684 &  0.724 &  0.760 &  0.793 &  0.822&   &0.698 &  0.734 &  0.766 &  0.796 &  0.822  \\
$\Omega_m$       & 0.178 &  0.207&  0.240 &  0.276 &  0.316&    &0.178 &  0.204 &  0.234 &  0.266 &  0.302 \\
$\sigma_8$       & 0.670 &  0.717&  0.767 &  0.816 &  0.863 &   & 0.671 &  0.715 &  0.763 &  0.810 &  0.855  \\
$H_0$            & 67.0 &  69.9 &  72.9 &  76.3 &  79.7 &       & 67.9 &  70.6& 73.5 &  76.6 & 79.8 \\
\hline
\multicolumn{12}{c}{\mbox{$\Lambda$CDM + $\Omega_K$, 7 parameters}} \\
\hline
$\Omega_bh^2$    & 0.0203 &  0.0211 &  0.0218&  0.0226 &  0.0233 & &0.0206 &  0.0213 &  0.0221 &  0.0229 &  0.0236\\
$\Omega_ch^2$    & 0.0916 &  0.0990 &  0.107 &  0.115 &  0.123&    & 0.0900 &  0.0968&  0.104 &  0.112 &  0.120 \\
$\tau$           & 0.0269&  0.0546 &  0.0836 &  0.113 &  0.142 &   & 0.0330&  0.0637&  0.0934&  0.123 &  0.152  \\
$\Omega_K$       & -0.147 & -0.115 & -0.0499 & -0.00574&0.0150&    & -0.0515 &  -0.0235 &  -0.00395 &  0.0103&  0.0201  \\
$n_s$            & 0.917 &  0.932 &  0.948 &  0.966 &  0.984 &     & 0.925 &  0.941 &  0.958 &  0.976 &  0.993 \\
$\Omega_\Lambda$ & 0.332 &  0.437 &  0.606 &  0.745 &  0.821 &     & 0.610 &  0.691 &  0.754 &  0.802 &  0.837\\
$\Omega_m$       &0.166 &  0.262 &  0.445&  0.678 &  0.804&        & 0.148 &  0.190&  0.250 &  0.330 &  0.436   \\
$\sigma_8$       & 0.648&  0.690 &  0.738 &  0.788 &  0.839 &      & 0.663 &  0.709 &  0.758 &  0.807 &  0.857 \\
$H_0$            & 40.5 &  43.6 &  53.8 &  69.5 &  86.6 &          & 54.0 &  62.1 &  71.0&  81.3 &  92.0  \\
\hline
\multicolumn{12}{c}{\mbox{$\Lambda$CDM + $w$, 7 parameters}} \\
\hline
$\Omega_bh^2$    & 0.0208 &  0.0215 &  0.0222 &  0.0231 &  0.0239 & & 0.0207 &  0.0214 &  0.0222 &  0.0230 &  0.237\\
$\Omega_ch^2$    & 0.0900 &  0.0981 &  0.106 &  0.114 &  0.122 &    & 0.0906 &  0.0975 &  0.105 &  0.112 &  0.120 \\
$\tau$           & 0.0294&  0.0600&  0.0894 &  0.119 &  0.149 &     & 0.0347 &  0.0647 &  0.0940 &  0.123 &  0.153 \\
$w$              & -1.731&  -1.457 &  -1.031 &  -0.573&  -0.240 &   & -1.646 &  -1.401 &  -1.006&  -0.704 &  -0.425 \\
$n_s$            & 0.927 &  0.943&  0.960 &  0.981 &  1.010 &       & 0.928 &  0.943 &  0.960&  0.978 &  0.998  \\
$\Omega_\Lambda$ & 0.457 &  0.617 &  0.764 &  0.844 &  0.870&       & 0.546 &  0.672 &  0.778 &  0.845 &  0.871 \\
$\Omega_m$       & 0.130 &  0.156 &  0.235 &  0.383 &  0.543 &      & 0.128 &  0.155 &  0.220 &  0.328 &  0.454 \\
$\sigma_8$       & 0.437 &  0.613 &  0.776 &  0.919 &  1.032 &      & 0.540&  0.659 &  0.781&  0.898 &  1.00  \\
$H_0$            & 47.8 &  57.9 &  73.8 &  90.7 &  98.5 &           & 53.2&  62.4 &  75.6 &  90.2&  98.2  \\
\hline\hline
\end{tabular}
\end{table*}

%
%
%
%
%

\section{Discussion}
\label{sec:discussion}

The main goal of this paper is to perform a full analysis of the integrated 
Sachs-Wolfe effect using the cross-correlations
between WMAP CMB maps and maps of large scale structure. In contrast to
previous work on this subject we place less emphasis on establishing a detection 
of ISW and more emphasis on developing a tool with which 
cosmological models can be compared to the data in a close to optimal fashion.  
For this reason we only select the data sets that can be reliably used towards 
this goal, as discussed in more detail below. The redshift range of the datasets we use is between 
0 and 2.5. 
We use optimal weighting of the data both in angular space and in redshift space
to extract the maximum amount of information, taking into account properly 
the correlations between them. 
Our final product is the likelihood function with which different cosmological 
models can be compared to each other. 

As the ISW effect is both a probe of cosmological parameters and a consistency test of the standard $\Lambda$CDM cosmology, 
there have been significant previous 
efforts made to detect it.  A number of different 
analysis methods have been used and the WMAP data have 
been cross-correlated with several samples.  These include the 2MASS XSC; several SDSS samples including magnitude-limited galaxy samples, LRGs, and 
quasars; the NVSS; and the HEAO hard X-ray map.  Most of these samples (or samples with similar spatial 
coverage and redshift range) are included in the present work, but not all. 
Here we compare our analysis with the previous work and comment on the 
reasons for our choice of data sets. 

\newcounter{oisw}
\begin{list}{\arabic{oisw}. }{\usecounter{oisw}}
\item {\em Near-infrared galaxies (2MASS).}  The 2MASS galaxies are useful for ISW due to high sky coverage and the ability to see closer to the 
Galactic plane in the near-IR than in the optical.  However they can only probe the lowest redshifts ($z<0.2$).
Afshordi et~al. \cite{afshordi04} and Rassat et~al. \cite{rassat07} have measured the ISW signal using the 2MASS sample and we deliberately 
cut our 2MASS sample into brightness bins 
such as theirs so that we can compare the results. We find that our measured signal from 2MASS is very similar.  
We do however derive cosmological 
constraints using a Markov chain (which fits all the cosmological parameters instead of just $\Omega_\Lambda$) from these samples. 
We also take into account (albeit in a crude way) the redshift dependence of the bias resulting from seeing all nearby galaxies but 
only the brightest and most biased galaxies at $z\ge 0.1$.

\item {\em Optical galaxies (SDSS, APM).} Wide-angle multicolor galaxy surveys such as SDSS open almost limitless possibilities for constructing galaxy 
samples, and many of these samples have been used in previous ISW work.  Most work so far has been on either flux-limited samples \cite{fosalba03, 
cabre06}, which have a broad redshift distribution, or photometric LRGs \cite{fosalba03, scranton03, padmanabhan05ISW, 
cabre06}, which can be seen 
to larger distances and for which it is easier to construct reliable photo-$z$ cuts.  In SDSS, photometric LRG samples oversample the linear 
density field in the redshift range $0.2<z<0.6$ and the lower redshifts are covered by 2MASS, so the flux-limited galaxy samples would be redundant in 
terms of volume for our study; we therefore did not include them.  Our LRG samples cover the largest solid angle to date of any SDSS ISW analysis (6641 
deg$^2$) and for the purposes of cosmological analysis are split into two photo-$z$ slices.
Fosalba \& Gazta\~naga \cite{fosalba04} have also used galaxies from the Automated Plate Measuring (APM) survey \cite{maddox90}, 
which adds $\sim 4300$ 
deg$^2$ in the Southern Hemisphere inaccessible to SDSS.  Their APM sample has a typical redshift $\bar z\approx 0.15$ and thus would add some 
information beyond the most distant of our 2MASS samples.  Considering that APM area is 
only 16\% of 2MASS and that it only marginally 
extends the redshift range we have not used APM in our analysis. 
However adding a deeper galaxy survey in the South, comparable to or 
deeper than SDSS, would be valuable for improving ISW 
constraints. Overall signal to noise from SDSS LRG galaxies is about 3$\sigma$, most 
of which comes from the higher redshift sample around $z \sim 0.5$. 

\item {\em Optical quasars (SDSS).}  Photometrically selected quasars can probe large-scale structure at much higher redshifts than ``normal'' galaxies 
because they are bright enough to be seen in wide-angle surveys (such as SDSS) even at $z\sim 2$.
The only ISW analysis with quasars so far has been that of Giannantonio et~al. \cite{giannantonio06}, who cross-correlated WMAP with a sample of 
photometric quasars from the SDSS.  Our analysis uses similar selection criteria, but we have used photo-$z$ cuts to eliminate most of the 
lower-redshift objects, and used a combination of spectroscopic data and angular clustering to constrain $b*dN/dz$ taking into account the multimodal 
nature of the photo-$z$ failures.  We also slice our quasars into two photo-$z$ bins.
Despite these improvements we find that the significance is only 1.3$\sigma$ (1.24$\sigma$ 
for the high redshift sample with $z>1$), and we therefore
do not confirm that the 2.1$\sigma$ signal seen in \cite{giannantonio06} comes from $z>1$.

\item {\em Radio sources (NVSS).}  There have been several past WMAP$\times$NVSS ISW analyses \cite{boughn04, nolta04, pietrobon06, 
mcewen07}, 
taking advantage of the high redshift (compared to most optical samples) and wide sky coverage of the NVSS.
We have used the angular power spectrum whereas the previous works have used correlation functions or wavelet coefficients.
However, the most important difference between our analysis and the previous result is that we fit $b*dN/dz$ from cross-correlations rather than
using the 
Dunlop \& Peacock model \cite{dunlop90} for the redshift distribution and assuming constant bias.  This is important as we find the fit $b*dN/dz$ looks 
very different (see Fig.~\ref{fig:fnvss}).  All of these studies, including ours, 
have found positive cross correlations at the $\sim 3\sigma$ level. However, the 
interpretation of this result depends sensitively on one's ability to measure $b*dN/dz$ 
and this is where we believe our analysis is an improvement upon previous efforts.

\item {\em Hard X-ray background (HEAO).}  Boughn \& Crittenden \cite{boughn04}
have used the HEAO hard X-ray map \cite{boldt87} for ISW 
cross-correlation.  The 
background is due mainly to unresolved (by HEAO) active galactic nuclei and hence traces large scale structure at redshifts of order unity.  This, 
combined with the all-sky nature of HEAO, is beneficial for ISW projects.
However, we decided not to add in HEAO sample to our analysis for several 
reasons. First is the difficulty
in understanding the $b(z)*dN/dz$ of the sample (we use the general notation $dN/dz$ here even though for unresolved X-ray flux it would be 
more accurate to write $dF/dz$).
Only $\sim 75$\% of the background is resolved by
Chandra into sources with measured redshifts \cite{cowie03, boughn04b}, and we have little guidance on where to place the other 
25\%.  Even if we knew $dN/dz$ 
perfectly, this does not tell us $b*dN/dz$: the modeled $dN/dz$ spans the range $0<z<3$ and it is unlikely that the bias would be even 
approximately constant over this range.
An alternative is to fit for their bias and redshift distributions
up to high $z$ using a cross-correlation method similar to that done for NVSS in Sec.~\ref{sec:dndz}.
Unfortunately HEAO has FWHM of $\sim 2^\circ$ and does not resolve individual sources, so we would have to fit
the data to the model without small-scale information, which loses signal-to-noise
on the cross-correlation very rapidly.
A secondary reason is that there is considerable overlap between HEAO and NVSS, so it is 
likely that the two trace partly the same structure, and thus the improvement in ISW 
constraint is not as large as adding two independent data sets.
We note that it may make sense to include the hard X-ray maps in parameter estimation
in the future if a robust determination of
$b*dN/dz$ becomes available.
\end{list}

In summary, we believe we used most of the available large scale structure
data useful for ISW analysis. This not only updates previous work on 
ISW effect \cite{boughn98,fosalba03,scranton03,afshordi04,boughn04,fosalba04,nolta04,padmanabhan05ISW,gaztanaga05,cabre06,giannantonio06,vielva06, 
pietrobon06,mcewen07,rassat07}, but is also the first one that attempts to do the tomography of ISW, in the sense of
encompassing a wide redshift range via our mass tracers going from the local Universe to 
$z\sim 2.5$, while reducing the amount of overlap in area and redshift as much as 
possible. We have argued that many of the previous measurements have a considerable
overlap in redshift and area, which means that they cannot be combined independently
and that the effective redshift of the sample is not necessarily the redshift from
where most of the ISW signal is coming from.
Our analysis, while attempting to minimize the overlap in the first place, 
takes the residual correlations into account explicitly via the construction of the 
full covariance matrix.  We note that Giannantonio et~al. \footnote{T. Giannantonio et~al., in preparation.} are also pursuing an ISW tomography 
analysis, with somewhat different choices of LSS samples and cross-correlation methodologies.

We spend a significant fraction of our analysis obtaining the correct 
redshift distributions for all of the samples. To be more accurate, 
it is the $b*dN/dz$ that we constrain for all samples. 
The signature of ISW effect is highly affected by the redshift distribution 
of the tracer, and thus one would need to have an accurate idea of 
what the redshift distribution is in order to interpret the correlation. 
Apart from employing spectroscopic datasets that overlap in magnitude range 
and sky coverage, we correlate the tracer samples with one another so 
as to obtain the $b*dN/dz$ for some of the samples. 
This is mainly possible because LRGs have relatively good photometric redshifts 
and so we correlate the LRGs with other overlapping datasets to determine
what are the $b*dN/dz$ at the redshift range that LRGs cover. 
In addition, we account for redshift-dependent bias in 2MASS and for
the multimodal error distributions for the quasars. 
We also made the first determination of $b*dN/dz$ for NVSS sample which is 
not based simply on a theoretical model fitting the luminosity function.

Correlations of mass tracers with the CMB sky can be caused not 
only by the ISW effect, but also by other cosmological effect such as thermal SZ, 
Galactic foregrounds and extinction, and extragalactic point sources. 
We provide an estimate for all these effects and only include the scales
deemed reliable, where the contamination is subdominant or negligible. 

We report a detection of $3.7\sigma$
of the ISW effect combining 2MASS, SDSS, and NVSS with WMAP data.
We make a joint analysis of all samples by constructing a reliable covariance matrix 
including cross-correlations of different samples, which is 
needed for cosmological parameter fitting.
We combine our ISW correlation functions with weak lensing of the CMB (Paper II) to derive cosmological constraints on three different 
cosmological 
models: (i) the ``vanilla'' $\Lambda$CDM model, (ii) $\Lambda$CDM+$\Omega_K$, and (iii) $\Lambda$CDM+$w$.
We find a slight improvement of our measurement of $w$ in model (iii) 
over the measurement made by CMB alone: $w=-1.01^{+0.30}_{-0.40}$
instead of $-1.03^{+0.46}_{-0.43}$.
The constraining power of our analysis is however most prominent in 
determining that curvature of the Universe: for CMB+ISW+WL we find
$\Omega_K=-0.004^{+0.014}_{-0.020}$ instead of $-0.050^{+0.044}_{-0.065}$ for CMB alone.
These constraints are
not as tight as that obtained by some other methods, such as combining the CMB with baryonic
oscillations or with supernovae \cite{eisenstein05,seljak06,spergel07}, but it is 
subject to very different systematics.  It is thus reassuring that all of them are consistent with each other. Even more importantly, there are other 
models where ISW 
can be crucial in distinguishing them from standard $\Lambda$CDM, such as $f(R)$
models in which the growth of structure is not fixed by the background geometry \cite{song07}.
Some of these models may already be inconsistent with our ISW signal;
we plan to present such constraints in a future paper.
These constraints should improve further in the 
future with deeper galaxy surveys that should reach the cosmic variance limit out to $z\sim 1-2$, and 
future CMB data that enables lower-noise lensing reconstruction.

Finally, we would like to note that we plan to release a package for calculating ISW likelihood function given the datasets and cosmological 
parameters. This will be described further in the documentation for the package. 

\begin{acknowledgments}
We would like to thank Joanna Dunkley for her extensive help on the discussion of chain convergences,
and Lucas Lombriser and An\v{z}e Slosar for bringing an error in an earlier version of this paper to our attention.
We would also like to thank David Spergel, Kendrick Smith and Jim Gunn for helpful conversations.
C.H. was a John Bahcall fellow at the Institute for Advanced Study during most of the preparation of this paper.
N.P. is supported by a Hubble Fellowship
HST.HF-01200.01 awarded by the Space Telescope Science Institute, 
which is operated by the Association of Universities for Research in 
Astronomy, Inc., for NASA, under contract NAS 5-26555. Part of this work was supported by the 
Director, Office of Science, of the U.S. 
Department of Energy under Contract No. DE-AC02-05CH11231.
U.S. is supported by the Packard Foundation and NSF
CAREER-0132953. 

Funding for the SDSS and SDSS-II has been provided by the Alfred P. Sloan Foundation, the Participating Institutions, the National Science Foundation, 
the U.S. Department of Energy, the National Aeronautics and Space Administration, the Japanese Monbukagakusho, the Max Planck Society, and the Higher 
Education Funding Council for England. The SDSS Web Site is http://www.sdss.org/.

The SDSS is managed by the Astrophysical Research Consortium for the Participating Institutions. The Participating Institutions are the American Museum 
of Natural History, Astrophysical Institute Potsdam, University of Basel, University of Cambridge, Case Western Reserve University, University of 
Chicago, Drexel University, Fermilab, the Institute for Advanced Study, the Japan Participation Group, Johns Hopkins University, the Joint Institute 
for Nuclear Astrophysics, the Kavli Institute for Particle Astrophysics and Cosmology, the Korean Scientist Group, the Chinese Academy of Sciences 
(LAMOST), Los Alamos National Laboratory, the Max-Planck-Institute for Astronomy (MPIA), the Max-Planck-Institute for Astrophysics (MPA), New Mexico 
State University, Ohio State University, University of Pittsburgh, University of Portsmouth, Princeton University, the United States Naval Observatory, 
and the University of Washington.

The 2dF QSO Redshift Survey (2QZ) was compiled by the 2QZ survey team
from observations made with the 2-degree Field on the Anglo-Australian
Telescope.
The 2dF-SDSS LRG and QSO (2SLAQ) Survey was compiled by the 2SLAQ team
from SDSS data and observations made with the 2-degree Field on the
Anglo-Australian Telescope.
This publication makes use of data products from the Two Micron All Sky Survey, which is a joint project of the University of Massachusetts and the 
Infrared Processing and Analysis Center/California Institute of Technology, funded by the National Aeronautics and Space Administration and the 
National Science Foundation.

\end{acknowledgments}

\appendix

\section{NVSS redshift window functions}
\label{app:nvsswin}

In Sec.~\ref{sss:cra}, we fit a constant $f_{\rm NVSS}$ to the cross-spectra of NVSS and the other LSS samples.  We argued that this procedure gave an
estimator $\hat f_{\rm NVSS}$ whose expectation value was given by Eq.~(\ref{eq:fhat}).  The purpose of this appendix is to prove this equation and
construct the functional form ${\cal W}(z)$.

We begin by noting that we have measured cross-spectra $C_\ell^{i,\rm NVSS}$ and their covariance matrix $\Sigma^i_{ll'}$.  The theoretical 
cross-spectrum
is on the other hand simply the Limber result,
\beqa
C_\ell^{i,\rm NVSS}({\rm th}) &=& \int_0^\infty f_{\rm NVSS}(z) f_i(z)
\nonumber \\ && \times
P\left(k=\frac{\ell+1/2}r\right) \frac{dz}{[r(z)]^2H(z)}
\nonumber \\ &\equiv& \int_0^\infty f_{\rm NVSS}(z) \kappa^i_\ell(z)\,dz,
\label{eq:cli}
\eeqa
where $r(z)$ is the comoving angular diameter distance and $\kappa^i_\ell(z)$ is defined by the equivalence in the second line.
The $\chi^2$ fitting procedure for the constant $f_{\rm NVSS}$ is to minimize
\beqa
\chi^2 &=& \sum_{\ell\ell'} [\Sigma^{i\,-1}]_{\ell\ell'}
(\hat C_\ell^{i,\rm NVSS} - f_{\rm NVSS}K^i_\ell)
\nonumber \\ && \times
(\hat C_{\ell'}^{i,\rm NVSS} - f_{\rm NVSS}K^i_{\ell\ell'}),
\eeqa
where $K_\ell=\int_0^\infty\kappa^i_\ell(z)\,dz$ and $\hat C_\ell^{i,\rm NVSS}$ are the measured cross-spectra.  The minimum value of $\chi^2$ is 
obtained 
for
\beq
\hat f_{\rm NVSS} = \frac{
\sum_{\ell\ell'} [\Sigma^{i\,-1}]_{\ell\ell'}\hat C_\ell^{i,\rm NVSS}K^i_{\ell'}
}{
\sum_{\ell\ell'} [\Sigma^{i\,-1}]_{\ell\ell'}K^i_\ell K^i_{\ell'}
}.
\eeq
Since the $\hat C_l^{i,\rm NVSS}$ have expectation value given by Eq.~(\ref{eq:cli}), we have
\beq
\langle \hat f_{\rm NVSS} \rangle = \frac{
\sum_{\ell\ell'} [\Sigma^{i\,-1}]_{\ell\ell'}  K^i_{\ell'} \int_0^\infty f_{\rm NVSS}(z) \kappa^i_\ell(z)\,dz
}{
\sum_{\ell\ell'} [\Sigma^{i\,-1}]_{\ell\ell'}K^i_\ell K^i_{\ell'}
}.
\eeq
This proves Eq.~(\ref{eq:fhat}) and shows that the window function is
\beq
{\cal W}(z) = \frac{
\sum_{\ell\ell'} [\Sigma^{i\,-1}]_{\ell\ell'}  K^i_{\ell'} \kappa^i_\ell(z)
}{
\sum_{\ell\ell'} [\Sigma^{i\,-1}]_{\ell\ell'}K^i_\ell K^i_{\ell'}
}.
\eeq

\section{Error bars on galaxy-CMB correlations}
\label{app:errors}

The purpose of this appendix is to discuss our choice of the Monte Carlo (``MC1'') estimator for the error bars on the $C_\ell^{gT}$ estimator, and
then give a crude estimate for the possible biases that are induced by its use.  As mentioned in the main text there are two biases: the correlation
bias (because the galaxies and CMB are correlated and MC1 does not take this into account) and a realization bias (since we have only one realization
of the galaxies).  The correlation bias is deterministic in the sense that the error bar is always underestimated in every $l$-bin.  The realization
bias is more subtle: if $C_\ell^{gT}=0$, then the MC1 estimator returns an unbiased estimate of $\sigma^2(C_\ell^{gT})$.  However the $\ell$-bins where
the error is underestimated are weighted more heavily than those where it is overestimated, resulting in a final error bar on cosmological parameters
that is biased low.

We consider each of these issues separately in a toy model.  The toy model has the following assumptions:
\newcounter{assumptions}
\begin{list}{\arabic{assumptions}. }{\usecounter{assumptions}}
\item We are computing cross-spectra $C_\ell^{gT}$ in $M$ $\ell$-bins (call these cross-spectra $x_1...x_N$).
\item The galaxies and CMB temperature are Gaussian random fields.  (We are at linear scales where large scale structure is Gaussian; the systematics
may not be.)
\item In the $i$th $\ell$-bin, there are $N_i$ galaxy modes, and all of the CMB modes in this region are observed.  (This is true except that NVSS goes
slightly closer to the Galactic plane than WMAP.)  We ignore mode coupling at the boundaries, i.e. each mode is ascribed to a single $\ell$-bin.
\item We are fitting the cross-correlation data to some parameter, say an amplitude $A$ of some template $t_i$.  More generally, when the ISW effect is
essentially constraining one direction in parameter space with all others constrained by the CMB alone (the case with $\Lambda$CDM+$\Omega_K$ and
$\Lambda$CDM+$w$ models here) the template would be $dx_i/dp$ where $p$ parameterizes the CMB-degenerate direction.  The fit is done using the Monte
Carlo covariance matrix.
\item The objective is to determine what is the ratio of the ``true'' error bar on $A$ to that derived from the fitting procedure.
\end{list}

Within these assumptions, we evaluate the correlation bias $R_1$ and realization bias $R_2$, which we define to be the ratio of true to estimated
variance.  We find, using correlation coefficients and numbers of modes for the worst-case bins, that $R_1\approx 1.02$ and $\langle R_2\rangle=1.11$.  
This corresponds to $\sim 6$\% underestimation of the error bars in the worst case, 
which is negligible.

\subsection{Correlation bias}

We will introduce the notation $\tilde C_\ell^{gg}=C_\ell^{gg}+\bar n^{-1}$ for the galaxy power spectrum including Poisson noise, and for a
matrix Cov we will write Cov$^{-1}_{ij}$ to mean the $ij$ element of Cov$^{-1}$ rather than the reciprocal of Cov$_{ij}$.  (In the
cases considered in this appendix the covariance matrices are diagonal so this distinction will not matter.)  We will also use the shorthand $C_i^{gT}$
for the galaxy-temperature cross-spectrum in the $i$th bin.

The estimator for the cross correlation is
\begin{equation}
x_i \equiv \hat C_{\ell_i}^{gT} = \frac 1{N_i}\sum_\alpha g_\alpha T_\alpha,
\label{eq:est}
\end{equation}
where $\alpha = 1...N_i$ is a mode index.

The true uncertainty in Gaussian theory, using independence of modes, is
\begin{equation}
{\rm Cov}_{ij}\equiv {\rm Cov}(x_i, x_j) = \frac {\delta_{ij}}{N_i} [\tilde C_i^{gg} C_i^{TT} + (C_i^{gT})^2].
\end{equation}
However the MC1 procedure gives
\begin{equation}
\widehat{\rm Cov}_{ij} = \frac 1{N_iN_j}\sum_{\alpha,\beta} g_\alpha g_\beta
          \langle T_\alpha T_\beta\rangle_{\rm Monte Carlo},
\end{equation}
where the $\alpha$ modes are in bin $i$, the beta modes are in bin $j$, and $g$ is the actual realization of the galaxies.  Simplifying with CMB
covariance matrix gives
\begin{equation}
\widehat{\rm Cov}_{ij} = \frac{\delta_{ij}}{N_i^2} \sum_\alpha g_\alpha^2 C_i^{TT}.
\label{eq:covij-act}
\end{equation}
Note that this is diagonal, even though we have only Monte-Carloed one of the data sets.

In the presence of a nonzero cross-correlation, the MC1 covariance matrix is biased:
\begin{equation}
R_1 \equiv \frac{{\rm Cov}_{ij}}{\widehat{\rm Cov}_{ij}} = 1 + \rho_i^2,
\end{equation}
where the correlation coefficient is
\begin{equation}
\rho_i = \frac{C_i^{gT}}{\sqrt{\tilde C_i^{gg} C_i^{TT}}}.
\end{equation}
For the fiducial cosmology and the bins that we used, the maximum predicted correlation coefficient is 0.067 (LRG1, $\ell=18$).  This would suggest an 
underestimate of the error bar by a factor of $R_1 = 1.0044$.  For some cosmological models, such as those with lower $\Omega_m$ or higher 
$\sigma_8$, the correlation coefficient could be larger.  Indeed there is some evidence for this: we observe an overall ISW amplitude of $2.2\pm 0.6$ 
times the prediction.  If we multiply the correlation coefficient $\rho$ by 2.2 then the underestimate of the error bar grows to $R_1=1.02$; even this 
is negligible.

\subsection{Realization bias}

Having taken into account the correlation bias, we now consider the case where the cross-correlation coefficient is small ($\rho\ll 1$).  In this case,
the covariance matrix of the $C_\ell^{gT}$ that we obtain from the CMB Monte Carlos is unbiased.  The realization bias comes from the fact that we
invert the covariance matrix, and unbiased Cov does not imply unbiased Cov$^-1$.

The true covariance matrix of the estimator Eq.~(\ref{eq:est}) for $x_i$ is
\begin{equation}
{\rm Cov}_{ij} = {\rm Cov}(x_i,x_j) = \frac{\delta_{ij}}{N_i} \tilde C_i^{gg} C_i^{TT},
\end{equation}
where $\tilde C_i^{gg}$ and $C_i^{TT}$ are the {\em true} (ensemble-averaged) galaxy and CMB power spectra, including Poisson noise for the galaxies.
The estimated covariance matrix is instead given by Eq.~(\ref{eq:covij-act}).  Now define the number
\begin{equation}
y_i \equiv \frac { \widehat{\rm Cov}_{ii} }{ {\rm Cov}_{ii} },
\end{equation}
which is the ratio of the estimated to true variance in a given bin.  This simplifies to
\begin{equation}
y_i = \frac{\sum_\alpha g_\alpha^2}{\tilde C_i^{gg}N_i},
\end{equation}
i.e. it is a $\chi^2$ distribution with $N_i$ degrees of freedom, divided by the number of degrees of freedom.  In particular $\langle y_i\rangle=1$:
the covariance matrix is
unbiased, but we have from $\chi^2$ distribution theory $\langle y_i^{-1}\rangle=(1-2/N)^{-1}$ and $\langle y_i^{-2}\rangle =(1-2/N)^{-1}(1-4/N)^{-1}$.

However what we really want to compare are the true and estimated errors on the parameter $A$.  The estimate $\hat A$ of the amplitude $A$ is
\begin{equation}
\hat A = \frac{ \widehat{\rm Cov}^{-1}_{ij}t_ix_j }{ \widehat{\rm Cov}^{-1}_{ij}t_it_j }.
\end{equation}
Its estimated variance is
\begin{equation}
\widehat{\rm Var}(\hat A) = \frac 1{ \widehat{\rm Cov}^{-1}_{ij}t_it_j }.
\end{equation}
Its true variance is
\begin{equation}
{\rm Var}(\hat A) = \frac{ t_h \widehat{\rm Cov}^{-1}_{hi} {\rm Cov}_{ij} \widehat{\rm Cov}^{-1}_{jk} t_k }
  {\left(\widehat{\rm Cov}^{-1}_{ij}t_it_j\right)^2}.
\end{equation}
(Note that $x_i$ and $\widehat{\rm Cov}_{ij}$ are uncorrelated because the probability distribution is symmetric in $T_\alpha\rightarrow -T_\alpha$,
under which $x_i$ changes sign but $\widehat{\rm Cov}_{ij}$ does not.)  The ratio is
\begin{equation}
R_2 \equiv \frac{{\rm Var}(\hat A)}{\widehat{\rm Var}(\hat A)}
 = \frac{ t_h \widehat{\rm Cov}^{-1}_{hi} {\rm Cov}_{ij} \widehat{\rm Cov}^{-1}_{jk} t_k }
  {\widehat{\rm Cov}^{-1}_{ij}t_it_j}.
\end{equation}
Using the definition of $y_i$ and diagonality of the matrices Cov and $\widehat{\rm Cov}$,
\begin{equation}
R_2 = \frac{ \sum_i t_i^2 {\rm Cov}^{-1}_{ii} y_i^{-2} }
    { \sum_i t_i^2 {\rm Cov}^{-1}_{ii} y_i^{-1} }.
\end{equation}
We now consider two limiting cases.  If we have a single $\ell$-bin, then the ratio is $R=y_1^{-1}$ and
\begin{equation}
\langle R_2\rangle = \frac1{1-2/N_i}.
\label{eq:onebin}
\end{equation}
If we have many $\ell$-bins contributing then the sums go to their mean values and we get
\begin{equation}
\langle R_2\rangle = \frac1{1-4/N_i},
\label{eq:manybins}
\end{equation}
if there were the same number of modes in each $\ell$-bin.  This is larger than Eq.~(\ref{eq:onebin}) because with only a single bin there is then no
possibility for the amplitude estimator to re-weight the bins to take advantage of the ones with smaller estimated variance.

The number of modes per bin is, in the limit of negligible mode coupling,
\begin{equation}
N_i = [(\ell_{\rm max}+1)^2-\ell_{\rm min}^2]f_{\rm sky}.
\end{equation}
This is 40 for the lowest-$\ell$ 2MASS bin that we use in parameter fits, 77 for the lowest-$\ell$ LRGs, 70
for the quasars, and 40 for NVSS.  To be pessimistic, if we take Eq.~(\ref{eq:manybins}) for all these cases the worst number we get is $\langle
R_2\rangle=1.11$, which means that in this
pessimistic case we have underestimated the error bar ($\sigma$) on the cross-correlation by 5\%.  In reality much of the constraint comes from
higher-$\ell$ bins where $N_i$ is greater, so this should be taken as an upper limit.

\section{Signal to noise estimate and upper limit on foreground contamination}
\label{app:foreg}

To assess the statistical signal to noise we
look at correlation between the galaxy overdensity and the temperature
of the CMB ($C^{gT}_{\ell,obs}$), which is the data vector called $\vec{d}$. We also
need its
inverse covariance matrix ($C^{-1}$) and the theoretical prediction, which we
can model as amplitude $A$ times a fiducial model $\vec{t}$.
To assess the possible contamination from foregrounds, tSZ, point sources etc.
to our signal we need to estimate the associated cross-correlation
contamination $(\vec{f})$.
For example,
for foregrounds we look at the correlation between the galaxy overdensity
and the foreground temperature $C^{g(fg)}_{\ell}$ (which is what we calculated
using models such as described by Eq.~\ref{eq:eps}).

Consider the usual $\chi^2$ analysis, where we are trying to fit for $A$ given
$\vec{d}$, $\vec{t}$ and ${\bf C}^{-1}$:
\begin{equation}
\chi^2 = (\vec{d} - A\vec{t})\cdot {\bf C}^{-1}(\vec{d} - A\vec{t}).
\end{equation}
We minimize $\chi^2$ and get
\begin{equation}
A = \frac{\vec{d}\cdot{\bf C}^{-1}\vec{t}}{\vec{t}\cdot {\bf C}^{-1} \vec{t}}.
\end{equation}
This is Eq.~(\ref{eq:ath}) and the associated variance is given by Eq.~(\ref{eq:sth}).
The ratio of estimated amplitude to its variance is the estimated signal to noise.

Since the total signal is a sum of the true signal and contamination such as
foreground, tSZ or point sources, the latter contribute to the
signal to noise,
\begin{equation}
\frac{\Delta A}{\sigma(A)}
= \frac{\vec{f}\cdot{\bf C}^{-1}\vec{t}}{\sqrt{\vec{t}\cdot{\bf C}^{-1} \vec{t}}}
= \frac{\vec{f}\cdot{\bf C}^{-1}\vec{t}}{\sqrt{\vec{t}\cdot{\bf C}^{-1} \vec{t}}} \times 
\frac{\sqrt{\vec{f}\cdot{\bf C}^{-1}\vec{f}}}{\sqrt{\vec{f}\cdot{\bf C}^{-1}\vec{f}}}.
\label{eq:cont}
\end{equation}
While we could use this expression to estimate the possible contamination we
can make it less dependent on the weighting by theoretical model by using
the Cauchy inequality, here written in the (primed) diagonal basis with eigenvectors
normalized to eigenvalue,
$\vec{f'}\cdot \vec{t'} < \sqrt{\vec{t'}\cdot \vec{t'}}\sqrt{\vec{f'}\cdot \vec{f'}}$,
to derive from Eq.~(\ref{eq:cont}):
\begin{equation}
\frac{\Delta A}{\sigma(A)} \le
\sqrt{E_{\rm cont}}\equiv \sqrt{\vec{f}\cdot{\bf C}^{-1}\vec{f}}.
\end{equation}
We use this expression in our estimates of contamination; it represents an upper limit on the number of sigmas of contamination introduced by the 
foreground $\vec{f}$.

\bibliography{ISWI}

\end{document}